\documentclass[iop]{emulateapj}

\usepackage{amsmath}

\begin{document}

\title{The Hunt for Red Quasars: Luminous Obscured Black Hole Growth Unveiled in the Stripe 82 X-ray Survey}

\author{Stephanie M. LaMassa$^{1}$, Eilat Glikman$^{2}$, Marcella Brusa$^{3,4}$, Jane R. Rigby$^{5}$, Tonima Tasnim Ananna$^{6,7}$,  Daniel Stern$^{8}$, Paulina Lira$^{9}$, C. Megan Urry$^{6,7}$, Mara Salvato$^{10}$, Rachael Alexandroff$^{11}$, Viola Allevato$^{12,13}$, Carolin Cardamone$^{14}$, Francesca Civano$^{15}$, Paolo Coppi$^{6,7}$, Duncan Farrah$^{16}$, S. Komossa$^{17}$, Giorgio Lanzuisi$^{3,4}$, Stefano Marchesi$^{18}$, Gordon Richards$^{19}$, Benny Trakhtenbrot$^{20}$, Ezequiel Treister$^{21}$}

\affil{$^1$Space Telescope Science Institute, 3700 San Martin Drive, Baltimore MD, 21218, USA;
  $^2$Department of Physics, Middlebury College, Middlebury, VT 05753, USA;
  $^3$INAF-Osservatorio Astronomico di Bologna, via Ranzani 1, I-40127 Bologna, Italy;
  $^4$Dipartimento di Fisica e Astronomia, Universita' di Bologna, viale Berti Pichat 6/2, I-40127 Bologna, Italy;
  $^5$Observational Cosmology Lab, NASA Goddard Space Flight Center, Greenbelt, MD 20771, USA;
  $^6$Yale Center for Astronomy \& Astrophysics, Physics Department, P.O. Box 208120, New Haven, CT 06520, USA;
  $^7$Department of Physics, Yale University, P.O Box 208121, New Haven, CT 06520, USA;
  $^8$Jet Propulsion Laboratory, California Institute of Technology, 4800 Oak Grove Drive, Mail Stop 169-221, Pasadena, CA 91109, USA;
  $^9$Departamento de Astronom\'ia, Universidad de Chile, Santiago, Chile
  $^{10}$Max-Planck-Institute f{\"u}r extraterrestriche Physik, D-85748 Garching, Germany;
  $^{11}$Center for Astrophysical Sciences, Department of Physics and Astronomy, Johns Hopkins University, Baltimore, MD, 21218;
  $^{12}$Department of Physics, University of Helsinki, Gustaf H\"allstr\"omin katu 2a, FI-00014 Helsinki, Finland;
  $^{13}$University of Maryland, Baltimore County, 1000 Hilltop Circle, Baltimore, MD 21250, USA;
  $^{14}$Department of Math \& Science, Wheelock College, 200 Riverway, Boston, MA 02215, USA;
  $^{15}$Smithsonian Astrophysical Observatory, 60 Garden Street, Cambridge, MA 02138, USA;
  $^{16}$Department of Physics MC 0435, Virginia Polytechnic Institute and State University, 850 West Campus Drive, Blacksburg, VA 24061, USA;
  $^{17}$National Astronomical Observatories, Chinese Academy of Sciences, 20A Datun Road, Chaoyang District, Beijing 100012, China
  $^{18}$Department of Physics \& Astronomy, Clemson University, Clemson, SC 29634, USA;
  $^{19}$Department of Physics, Drexel University, 3141 Chestnute Street, Philadelpha, PA 19104, USA;
  $^{20}$Institute for Astronomy, Department of Physics, ETH Zurich, Wolfgang-Pauli-Strasse 27, CH-8093 Zurich, Switzerland;
  $^{21}$Instituto de Astrof\'{\i}sica, Facultad de F\'{i}sica, Pontificia Universidad Cat\'{o}lica de Chile, Casilla 306, Santiago 22, Chile
}

\begin{abstract}
We present results of a ground-based near-infrared campaign with Palomar TripleSpec, Keck NIRSPEC, and Gemini GNIRS to target two samples of reddened active galactic nucleus (AGN) candidates from the 31 deg$^2$ Stripe 82 X-ray survey. One sample, which is $\sim$89\% complete to $K<16$ (Vega), consists of eight confirmed AGNs, four of which were identified with our follow-up program, and is selected to have red $R-K$ colors ($>4$, Vega). The fainter sample ($K>17$, Vega) represents a pilot program to follow-up four sources from a parent sample of 34 that are not detected in the single-epoch SDSS catalog and have {\it WISE} quasar colors. All twelve sources are broad-line AGNs (at least one permitted emission line has a FWHM exceeding 1300 km s$^{-1}$) and span a redshift range $0.59 < z < 2.5$. Half the ($R-K$)-selected AGNs have features in their spectra suggestive of outflows. When comparing these sources to a matched sample of blue Type 1 AGNs, we find the reddened AGNs are more distant ($z > 0.5$) and a greater percentage have high X-ray luminosities ($L_{\rm X,full} > 10^{44}$ erg s$^{-1}$). Such outflows and high luminosities may be consistent with the paradigm that reddened broad-line AGNs represent a transitory phase in AGN evolution as described by the major merger model for black hole growth. Results from our pilot program demonstrate proof-of-concept that our selection technique is successful in discovering reddened quasars at $z > 1$ missed by optical surveys.
\end{abstract}

\keywords{galaxies: active --- quasars: general --- X-rays: galaxies --- infrared: galaxies --- quasars: supermassive black holes}

\section{Introduction}
To understand the growth of supermassive black holes over cosmic time, it is crucial to identify and study samples of active galactic nuclei (AGNs) with diverse properties. Powerful emission from the accreting black hole imprints signatures on its surroundings which enables these sources to be detected across the electromagnetic spectrum. Optical emission from the accretion disk and gas photoionized by the AGN can be prominent. Indeed, hundreds of thousands of AGNs have been detected by large ground-based optical surveys, like the Sloan Digital Sky Survey \citep{york}, with almost 300,000 Type 1 quasars in the most recent release of the SDSS Quasar Catalog \citep{paris}. These Type 1 quasars are sources where we have a direct view of the growing black holes, allowing them to be easily identified at optical wavelengths due to their blue color, which imparts a power law slope to the optical spectra, and broad emission lines, from gas rapidly orbiting near the black hole.

However, obscured AGNs, where direct view of the central engine is blocked by the circumnuclear torus of the AGN unification scheme \citep[Type 2 AGN;][]{antonucci,urry} and/or large amounts of dust from the host galaxy, are less efficiently detected based on their optical emission alone. In apparent defiance of the canonical AGN unification scheme, red quasars are typically broad-line AGNs, yet are enshrouded by large amounts of dust that reddens the spectrum and attenuates optical emission. Studies of this extreme segment of the obscured AGN population indicate that their reddening is due to a stage of AGN evolution in the major merger model of black hole growth, rather than orientation of the putative torus with respect to the line-of-sight \citep[e.g.,][]{georgakakis,glikman2012}.

According to this model \citep{sanders,hopkins}, when galaxies of comparable mass collide and coalesce, gas is funneled to the central supermassive black hole, which ignites AGN activity and circumnuclear star formation. During this phase, the AGN is cocooned within large amounts of dust and gas associated with on-going star formation, potentially reaching Compton-thick levels \citep[$N_{\rm H} > 1.25 \times 10^{24}$ cm$^{-2}$;][]{kocevski2015,ricci}, causing the AGN to appear heavily reddened while it is intrinsically luminous. According to this major merger evolution model, powerful winds from the AGN eventually clear out the obscuring material, revealing a blue Type 1 quasar, and potentially regulating galaxy growth by shutting down star formation and/or evacuating gas from the host \citep{hopkins2006}. While the Compton-thick phase lasts 10$^{7}$-10$^{8}$ years \citep{hopkins2005,treister2010}, the reddened AGN phase, when the quasar begins to clear out its surroundings, is shorter-lived, $\sim 5\times10^{6}$ years \citep{hopkins2005,glikman2012}, making these sources rare. This pathway for black hole growth only pertains to a portion of the AGN population \citep{treister2012,hopkins2014}, with secular processes apparently responsible for triggering moderate luminosity AGNs \citep[e.g.,][]{schawinski,kocevski,villforth2014} and some high luminosity AGNs \citep{villforth2017}. Identifying reddened AGNs that may be in this evolutionary phase provides us an opportunity to test whether there is a causal or coincidental connection between mergers and black hole growth \citep{mechtley,farrah2017} and learn about how black holes can shape their environment.

Many previous red quasar samples were identified by their radio, optical-to-near-infrared, near-infrared, and/or mid-infrared emission \citep[e.g.,][]{glikman2007,glikman2012,glikman2013,banerji2012,banerji2013,banerji2015,eisenhardt,assef2015,ross,hamann}. The traits of some of these sources are consistent with being in the transitory reddened phase in the AGN evolution model: they are intrinsically luminous, after correcting for extinction \citep{glikman2007,glikman2012,glikman2013,banerji2015,assef2015}; they host outflows that may impart feedback onto the host galaxy \citep{farrah,urrutia2009,zakamska}; and their host galaxies have morphologies consistent with having recently undergone a merger \citep{urrutia2008,glikman2015}. As X-ray emission provides a direct probe of black hole fueling and can pierce through optically obscuring dust, honing in on this emission provides a complementary method for identifying this population. Indeed, reddened AGNs selected from the $\sim$2 deg$^2$ {\it XMM}-COSMOS survey \citep{hasinger,cappelluti,brusa} are launching powerful outflows, suggesting that they may be in the ``clear-out'' phase in the AGN evolution paradigm \citep{brusa2015a,brusa2015b,brusa2016,perna2015,perna2017}.

Before the launch of focusing X-ray instruments with sensitivity beyond a few keV, there had not been wide-area X-ray surveys with sufficient depth to identify heavily obscured AGN beyond the local Universe, limiting our census of these rare, reddened sources. Stripe 82X is a $\sim$31 deg$^2$ X-ray survey with {\it XMM-Newton} and {\it Chandra} \citep{s82x1,s82x2,s82x3,ananna} that overlaps the legacy SDSS Stripe 82 field \citep{frieman}. About 20 deg$^2$ of the Stripe 82X survey is from a dedicated {\it XMM-Newton} observing program from AO10 and AO13 (PI: Urry), reaching depths of 5-7.5 ks, while the rest of the survey is composed of archival {\it XMM-Newton} and {\it Chandra} observations in the field. The flux limit at half the survey area is $\sim 5.4 \times 10^{-15}$ erg s$^{-1}$ cm$^{-2}$ in the soft band (0.5-2 keV). Due to the relatively wide area covered in X-rays plus rich multiwavelenth data, this survey provides an ideal dataset to identify reddened AGNs, building on the census from smaller-area X-ray surveys and complementing samples selected at other wavelengths.

We used a combination of optical and infrared clues to identify signatures of such obscured black hole growth. The full Stripe 82 region is covered at relatively homogeneous depths in the optical by SDSS and in the near-infrared by the UKIRT Infrared Deep Sky Survey \citep[UKIDSS;][]{lawrence} and the VISTA Hemisphere Survey \citep[VHS;][]{mcmahon}. Though, in general, reddened extragalactic sources could be dusty starbursts that may not necessarily host an accreting black hole (e.g., some (ultra) luminous infrared galaxies), X-ray emission from Stripe 82X sources is a clear indicator of supermassive black hole accretion: at the relatively bright X-ray flux limits of Stripe 82X, faint X-ray emission from star formation processes in galaxies beyond the local Universe are not detectable. Optical faintness in tandem with relatively brighter infrared emission is thus consistent with heavily reddened AGNs.

We present two samples of objects that we targeted for follow-up with ground-based near-infrared spectroscopy using Palomar TripleSpec \citep{herter}, Keck NIRSPEC \citep{mclean}, and Gemini GNIRS \citep{elias1,elias2}. Our ``bright NIR'' sample ($K < 16$, Vega) consists of objects selected on the basis of red $R-K$ colors, with an X-ray to optical flux ($X/O$) cut to mitigate contamination from stars; these sources were followed up with Palomar TripleSpec. We used Keck NIRSPEC and Gemini GNIRS to follow-up our ``faint'' sample ($K > 17$, Vega), which are sources that are not detected in the single-epoch imaging of the SDSS survey, yet have {\it WISE} colors consistent with quasars. From our follow-up campaign, the bright NIR sample is nearly complete (i.e., 8 of the 9 sources from the parent sample have secure spectroscopic redshifts) while the faint NIR sample represents a pilot program of a larger sample. In Sections \ref{targ} and \ref{obs}, we describe the target selection and follow-up observations. We discuss the results of our spectroscopic campaigns, multiwavelength properties of the samples, and insight from spectral energy distribution (SED) analysis in Section \ref{results}. In Section \ref{disc}, we compare the properties of the $R-K$ sample with a matched sample of blue Type 1 AGNs and compare both samples with reddened quasars from the literature selected at other wavelengths. We assume a cosmology of H$_{0}$=67.8 km s$^{-1}$ Mpc$^{-1}$, $\Omega_{M}$=0.31, $\Omega_{\Lambda}$=0.69 \citep{planck}

\section{Target Selection}\label{targ}
We focused on photometric signatures of reddening for both samples we targeted for near-infrared spectroscopy. Such red colors can be induced by large amounts of dust (in the host galaxy and/or circumnuclear region) or by radio synchrotron emission \citep{serjeant}. Two of our 12 sources are detected in the radio by the 1.5 GHz FIRST survey \citep{helfand} and are discussed in more detail below. Thus the red colors for most of these sources are likely due to obscuration.

The sources that we followed up are reported in Tables \ref{bright_samp} and \ref{faint_samp}, where we list the full Stripe 82X name based on the X-ray coordinates. For clarity, we use an abbreviated version of the source name in the main text and subsequent tables. 

\subsection{Bright NIR Stripe 82X Sample: $R-K$ versus $X/O$ Selection}
To unveil the brighter end of the reddened AGN population in Stripe 82X, we focus on the 551 sources (9\% of the 6181 unique X-ray sources in Stripe 82X) that are detected in the X-ray full band\footnote{The full band is defined from 0.5-10 keV for {\it XMM-Newton} and from 0.5-7 keV for {\it Chandra}.} and have UKIDSS $K$-band magnitudes brighter than 16 (Vega). We retain the sources where the SDSS $r$ band and $i$ band magnitudes are well-measured (i.e., error is below 0.5) to avoid artificially reddened colors from poor photometric measurements, leaving us with 373 sources.\footnote{We note that only one source had $r$ and $i$ band magnitude errors exceeding 0.5 and otherwise met our selection criteria. This source, selected as a counterpart from the SDSS coadded catalog of \citet{jiang}, is confused with a nearby source and a stellar spike, and hence has unreliable photometry.} For a straightforward comparison to reddened populations from other studies that use $R-K$ (Vega) to identify obscured AGNs \citep[e.g.,][]{banerji2012,brusa2015b}, we convert the SDSS $r$ magnitude from the AB system to the Bessell $R$ bandpass \citep{bessell} in the Vega system using the formulae in \citet{blanton}, which was calibrated on galaxies ranging in redshifts $0 < z < 1.5$ from SDSS, {\it GALEX} \citep{martin}, DEEP2 \citep{davis,faber} and GOODS \citep{giavalisco} surveys:
\begin{equation}
  R_{\rm AB} = r - 0.0576 -0.3718((r - i) - 0.2589)
\end{equation}

\begin{equation}    
  R_{\rm Vega} = R_{\rm AB} - 0.21,
\end{equation}
where $r$ and $i$ are the SDSS pipeline ``modelMag'' magnitudes, which is a PSF model for point sources, and the better of a de Vauculeurs or exponential profile fit for extended sources.

To calculate $X/O$, the ratio of X-ray to optical flux, we use the following \citep[see][]{brandt2005}:
\begin{equation}
  X/O = {\rm Log}(f_{\rm x}/f_{\rm opt}) = {\rm log}(f_{\rm x}) + C + 0.4 \times m_{r},
  \end{equation}
where $C$, a constant that depends on the optical filter, is 5.67 for the SDSS $r$ band \citep[AB;][]{green}, and $m_{r}$ is the ``modelMag'' reported by the SDSS pipeline. Here, the X-ray flux is in the full band.

\begin{figure}[ht]
  \centering
      {\includegraphics[angle=90,scale=0.4]{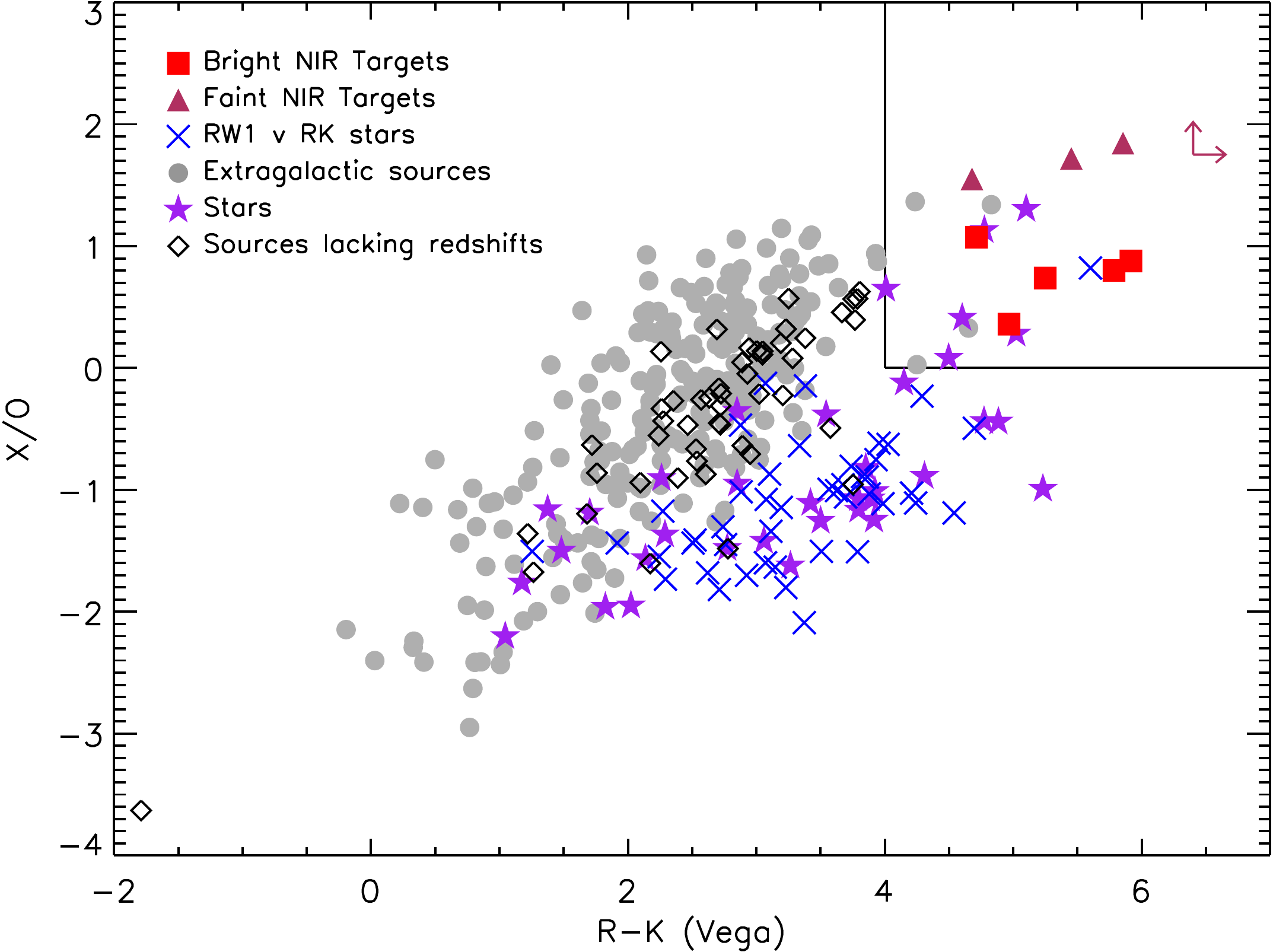}}
      \caption{\label{rkxo_selection} $R-K$ versus $X/O$ colors of Stripe 82X sources brighter than $K$=16 (Vega) with significant detection in the full X-ray band (0.5-10 keV) and well measured $r$ and $i$ band magnitudes (magnitude errors less than 0.5). The boxed region ($R/K > 4$ and $X/O > 0$) indicates the locus for reddened AGN candidates \citep[see][]{brusa} where we defined a sample for follow-up. This sample excludes sources lacking spectra that lie along the $R-K$ versus $R-W1$ stellar locus presented in \citet{rw1} since they are likely not AGNs ({\it blue Xs}). The sources with pre-existing spectroscopic redshifts (from SDSS or our optical follow-up programs) that are extragalactic and stellar are shown by the {\it grey circles} and {\it purple stars}, respectively, while the sources lacking identifications via spectroscopic redshifts are indicated by the {\it black diamonds}. The reddened AGN candidates we targeted with TripleSpec on Palomar are shown by the {\it red squares}. For reference, we show where the faint NIR sample ($K > 17$) lie in this parameter space with the {\it maroon triangles} and {\it lower limits}.}
      \end{figure}

We applied a modified version of the color cuts presented in \citet{brusa} to select our sample: $R-K > 4$ (Vega) and $X/O > 0$. We find 17 sources that meet these cuts (boxed region in Figure \ref{rkxo_selection}), of which seven are spectroscopically confirmed as stars and four are Type 1 AGN with existing SDSS spectra (see Section \ref{supp_samp}). Of the sources lacking spectra, we removed the one object that lies along the stellar locus of $R-K$ versus $R-W1$ color space presented in \citet{rw1}, i.e., $R-W1 = 0.998(\pm0.02) \times (R - K) + 0.18$, leaving us with five reddened AGN candidates lacking spectra (red filled squares).

We note that though the $X/O > 0$ cut is designed to mitigate contamination from stars, this restriction in principle can also omit sources where the galaxy light dominates \citep[e.g.,][]{cardamone}, including low luminosity AGNs or heavily obscured to Compton-thick AGNs where the observed X-ray emission appears weak \citep[e.g.,][]{heckman,me2009,me2011}. The X-ray to optical flux cut selects AGN where the X-ray emission dominates over the host galaxy, implicitly favoring AGN with high X-ray luminosities. However, as Figure \ref{rkxo_selection} shows, the sources at $R-K > 4$ with $X/O < 0$ are either spectroscopically confirmed as stars or are likely stars based on their $R-K$ and $R-W1$ colors. 

We targeted the five reddened AGN candidates lacking spectra with Palomar TripleSpec, as discussed below, and summarize their properties in Table \ref{bright_samp}, where the magnitudes are given in the native units from their parent catalogs, while the $R-K$ color is in the Vega system, following the derivation above. The corresponding X-ray identification numbers refer to those published in \citet{evans}, for {\it Chandra} sources identified in the {\it Chandra} Source Catalog, and \citet{s82x1,s82x2,s82x3}, for Stripe 82X sources detected by {\it XMM-Newton}. We discuss the four extragalactic sources that have SDSS spectra in Section \ref{supp_samp} and include them in our subsequent analysis.

\begin{deluxetable*}{lrllll}
\tablewidth{0pt}
\tablecaption{\label{bright_samp}Bright NIR Stripe 82X Targets: $R-K$ versus $X/O$ Selection}
\tablehead{\colhead{Stripe 82X Name} & \colhead{X-ray ID\tablenotemark{1}} & \colhead{$r$ (AB)} &\colhead{$K$ (Vega)} & \colhead{$R-K$ (Vega)} & \colhead{$X/O$}}

\startdata

S82X 013245.41$-$000835.5 & 4150 & 22.39 & 15.85 & 5.92 & 0.88 \\

S82X 024219.20+000511.9 & 618 (108774C\tablenotemark{2})  & 22.32 & 15.94 & 5.78 & 0.80 \\

S82X 030215.39$-$000335.5 & 783  & 21.37 & 15.49 & 5.25 & 0.74 \\

S82X 030324.58$-$011508.3 & 855  & 21.43 & 15.89 & 4.97 & 0.36 \\

S82X 232801.91$-$002822.9 & 1859 & 21.22 & 16.00 & 4.71 & 1.07

\enddata
\tablenotetext{1}{{\it XMM-Newton} record number introduced in the Stripe 82X survey \citep{s82x2}.}
\tablenotetext{2}{Source also detected by archival {\it Chandra} observations in Stripe 82 \citep{s82x1}. The {\it Chandra} Source Catalog MSID identifying number \citep{evans} for this object is noted in parentheses.}
\end{deluxetable*}

\subsection{Faint NIR Stripe 82X Sample: Optical Drop-outs Recovered by \textit{WISE}}
An interesting population are X-ray sources that lack an optical counterpart in the single-epoch SDSS imaging, but which are detected at infrared wavelengths. The depths of the single-epoch SDSS imaging,\footnote{$r \leq 22.2$ and $i \leq 21.3$ (AB) for 95\% completeness of point sources} NIR imaging,\footnote{$K < 18.1$ (Vega) for 5$\sigma$ point source detection} and MIR imaging\footnote{5$\sigma$ limit at $W1 < 17.30$ and $W2 < 15.84$ (Vega) for 95\% sky coverage} in Stripe 82 are comparable for an AGN SED: $R - K \sim 3.6$ at the flux limits of these surveys, indicating that the infrared coverage is not systematically deeper than the optical. Hence a non-detection in SDSS in conjunction with an NIR detection selects for reddening.

\begin{figure}
  \centering
      {\includegraphics[angle=90,scale=0.45]{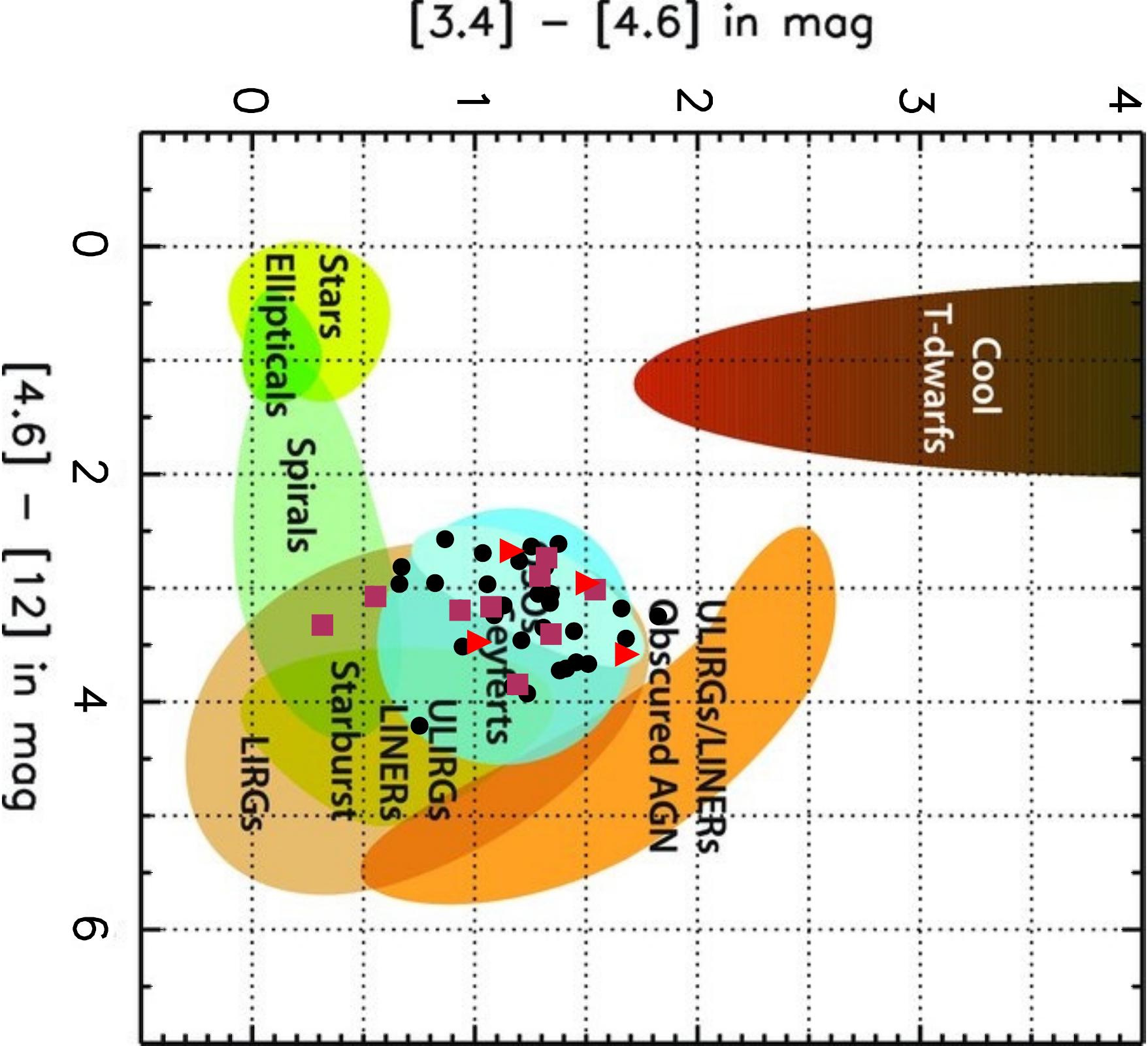}}
      \caption{\label{wiseqso_selection} {\it WISE} color-color plot \citep{wright} for our parent sample of reddened AGN candidates which are optical drop-outs in single-epoch SDSS imaging and have NIR detections in at the least the VHS $K$ band and {\it WISE} colors in the QSO and Seyfert locus (cyan circle). The {\it red triangles} indicate the objects from our pilot program where we identified emission lines and were thus able to determine redshifts and confirm these sources as quasars; the {\it black circles} represent sources from the parent sample currently lacking identifications. We show the $R-K$ versus $X/O$ selected sources from the bright NIR sample as {\it maroon squares} in this color space for reference.}
      \end{figure}

We creates a target list of 47 such optical drop-outs that have mid-infrared colors in the quasar locus of the {\it WISE} color-color diagram \citep[Figure \ref{wiseqso_selection};][]{wright}, and have $K$-band detections in VHS, which is deeper than UKIDSS. These optical drop-outs have no SDSS source within the nominal search radius we used to identify counterparts to the X-ray source (5$^{\prime\prime}$ for {\it Chandra} and 7$^{\prime\prime}$ for {\it XMM-Newton}). We vet each potential target by eye to remove any sources that fall out of the SDSS footprint\footnote{Though the X-ray observations are designed to overlap the Stripe 82 region, some of the archival X-ray observations partially overlap the Stripe while the rest of the field-of-view is outside of the SDSS footprint.} or where visual inspection shows a clear source that failed to be detected by the SDSS pipeline, leaving us 37 objects. 

Thirty-two of these sources are detected in the deeper coadded SDSS catalogs, while five remain undetected \citep{fliri, jiang}. We further vet our target list to preserve the reddening criterion of our selection using the information from the coadded SDSS catalogs. If the source is detected in the $r$ and $i$ bands in the coadded SDSS catalog such that we can calculate $R$, we only retain the sources where $R - K > 4$. We also retain all sources that remain undetected in at least the $r$ or $i$ band in even this deeper imaging, which implies colors of $R - K > 5.9$.

In total, we have 34 such reddened AGN candidates in our faint sample, including the five sources not detected in any SDSS band in the coadded catalogs. We targeted six with NIRSPEC on Keck (two in 2014 September and four in 2015 October; Section 3.2), and three with GNIRS on Gemini (Section 3.3). Though all sources were detected in the NIRSPEC $K$-band spectra, we were only able to identify an emission line in one object. It is likely that the remaining sources were unidentified because of the limited wavelength range in the $K$-band order. Indeed, when using the cross-dispersed mode on Gemini GNIRS, which yields simultaneous $J$, $H$, and $K$-band coverage, we detected emission lines in all three targets.

For the remainder of this work, we focus on the four objects for which we were able to identify emission lines and thus derive redshifts. We list this sample in Table \ref{faint_samp}, where we note whether a reliable optical counterpart was found in the deeper coadded SDSS catalog, and provide the $R-K$ colors or lower limit (if undetected in the coadded SDSS catalogs).

\begin{deluxetable*}{llllllcll}
\tablewidth{0pt}
\tablecaption{\label{faint_samp} Faint NIR Stripe 82X Targets: {\it WISE}-Selected Optical Drop-outs}
\tablehead{\colhead{Stripe 82X Name} & \colhead{X-ray ID\tablenotemark{1}} & \colhead{$W1$ (Vega)} & \colhead{$W2$ (Vega)} & \colhead{$W3$ (Vega)} & \colhead{$K$ (Vega)} & \colhead{SDSS Coadd?\tablenotemark{2}} & \colhead{$r$ (AB)} & \colhead{$R-K$ (Vega)} }

\startdata

\multicolumn{7}{c}{{\textit{Keck NIRSPEC}}}\\

S82X 022723.51+004253.3 & 129832C & 16.86 & 15.84 & 12.36 & 18.75 & Y & 23.44 & 4.68 \\

\multicolumn{7}{c}{{\textit{Gemini GNIRS}}}\\

S82X 010019.25+000844.8 & 2589X & 16.79 & 15.28 & 12.32 & 18.36 & Y & 24.37 & 5.45 \\

S82X 011840.06+001806.0 & 3692X & 16.27 & 15.10 & 12.42 & 18.04 & Y & 24.40 & 5.85 \\

S82X 014152.06$-$001749.5 & 4583X & 16.14 & 14.46 & 10.87 & 17.98 & N & $>$24.7 & $>$6.40 

\enddata
\tablenotetext{1}{If the X-ray ID number is followed by a ``C'', this indicates the {\it Chandra} MSID number from the {\it Chandra} Source Catalog \citep{evans}. If the X-ray ID number is appended by an ``X'', this denotes the {\it XMM-Newton} record number introduced in the Stripe 82X survey \citep{s82x2,s82x3}.}
\tablenotetext{2}{Flag to indicate if optical dropout was recovered in the coadded SDSS catalog.}
\end{deluxetable*}

\section{Observations and Data Analysis}\label{obs}

\subsection{Palomar TripleSpec}
TripleSpec simultaneously covers wavelengths 1 - 2.4 $\mu$m in 4 orders, with an approximate resolution of 120 km/s (based on the instrument specifications\footnote{\url{http://www.astro.caltech.edu/palomar/observer/200inchResources/ tspecspecs.html}}). All five sources from the bright NIR Stripe 82X sample were observed using the standard $ABBA$ nodding sequence, where we integrated for 300s per exposure. Four of the sources were observed on 2015 October 27 (S82X 0242+0005, S82X 0302-0003, S82X 0303-0015, S82X 2328-0028) with 3 $ABBA$ sequences. The fifth source, S82X 0132-0008, was observed on 12 December 2016 with 4 $ABBA$ sequences. At the end of the science exposures for each target, we observed a standard A0V or A1V star for telluric correction. All sources were observed at an airmass below 1.5.

Data were reduced with the \textsc{IDL} program \textsc{Spextool} \citep{cushing} which creates normalized flat field images, performs a wavelength calibration based on sky lines, and extracts the spectra from each order. We note that for S82X 2338-0028, we only extract the spectrum from the $B$ position due to a bad column affecting the emission feature in the $A$ nod. Telluric correction is performed on each order with \textsc{xtellcor} \citep{vacca}, after which the spectra are merged into a continuous spectrum with the \textsc{Spextool} \textsc{xmergerorders} routine. Finally, the spectrum is smoothed with task \textsc{xcleanspec} using the Savitzky-Golay routine, which preserves the average resolving power using a smoothing window that is two times the slit width.

\subsection{Keck NIRSPEC}
Keck NIRSPEC is a near-infrared spectrograph on Keck II, with different filter wheels limiting the wavelength range of a given spectrum to a single waveband. The approximate resolution, found from measuring resolved sky lines, is 200 km/s. We observed S82X 0227+0042 at an airmass of $\sim$1.1 with the K$^{\prime}$ filter (1.950 - 2.295 $\mu$m) on 2014 Sep 7 with the 42 $\times$ 0$\farcs$79 slit. We acquired 4 $ABBA$ exposures at 600s per exposure. Due to an apparent error with target acquisition or the dithering script, the source was only in the slit in the $A$ nod. We observed A0V standard star HD 18571.

We reduced the data with \textsc{IRAF} routine \textsc{WMKONSPEC}\footnote{http://www2.keck.hawaii.edu/inst/nirspec/wmkonspec.html} which corrects the distortion in the $x$ and $y$ directions before spectral extraction. Wavelength calibration was performed using sky lines. We corrected the source spectrum for telluric features using \textsc{xtellcor\_general} \citep{vacca}, part of the \textsc{Spextool} package \citep{cushing}. 

\subsection{Gemini GNIRS}
Our Gemini program, GN-2015B-Q-80 (PI: LaMassa), made use of GNIRS in cross-dispersed mode, with simultaneous coverage from 0.85 - 2.5 $\mu$m, on Gemini North. We used a slit width of 1.0$^{\prime\prime}$ to maximize signal throughput from our sources, with a 32 l/mm grating and short blue camera. With this instrumental set-up, the approximate resolution is 550 km/s.\footnote{\url{https://www.gemini.edu/node/1046?q=node/10543}}

In total, we were awarded 12 hours of queue time in Band 3. Each target was observed for three hours, including acquisition from offset stars and standard star observations, with 24-26 $ABBA$ exposures at 300s and 4 $ABBA$ exposures at 270s. Due to varying sky conditions from changes in cloud cover, not all science exposures were included in the analysis. Thus, we discarded observations that added more noise than signal, with the resulting net exposure times listed in Table \ref{gnirs_sample} for each source.

\begin{deluxetable}{lcc}
\tablewidth{0pt}
\tablecaption{\label{gnirs_sample}Gemini GNIRS Observing Log}
\tablehead{\colhead{Stripe 82X Name} & \colhead{Observation Dates} & \colhead{Net Exposure\tablenotemark{1}} \\
 & \colhead{(year-month-date)} & \colhead{(s)}}

\startdata

S82X 0100+0008 & 2016-01-07 & 3600 \\
          & 2016-01-08 & 3540 \\

S82X 0111+0018 & 2015-12-10 & 2400 \\
          & 2016-01-02 & 6480 \\

S82X 0141-0017 & 2016-01-05 & 5280 

\enddata
\tablenotetext{1}{Net exposure time after discarding observations with poor sky conditions.}
\end{deluxetable}

We reduced the spectra with the \textsc{XDGNIRS} pipeline which calls Gemini GNIRS \textsc{IRAF} routines to clean pattern noise, flat field the data, remove spikes from the data, correct the $S$-distortion, perform the wavelength calibration based on arc lamps, and extract a spectrum from the combined $A$ and $B$ exposures \citep{cooke}. We used \textsc{xtellcor\_general} \citep{vacca} to perform the telluric correction. Spectra from separate nights were averaged, weighted by the number of exposures contributing to each spectrum.

\section{Results}\label{results}

\subsection{Near Infrared Spectroscopy}
For all objects, we used photometry to flux calibrate the spectra to obtain estimates of the emission line fluxes: we interpolated the $K$-band filter response onto the wavelength grid of our spectra, using the filter curve from UKIDSS \citep{hewett} for the bright NIR sample (Section 2.1) and from VHS for the faint NIR sample (Section 2.2). The integrated flux is then measured from this folded spectrum. The ratio of the $K$-band flux, derived from the observed $K$-band catalog photometry, and this pseudo-flux gives us the scale factor by which we adjust the spectrum for an absolute flux calibration. We note that variability between the photometric and spectroscopic observations induce uncertainty into this calibration beyond the statistical errors we report on the emission line fluxes.

\begin{figure*}[ht]
  \centering
      {\includegraphics[scale=0.4]{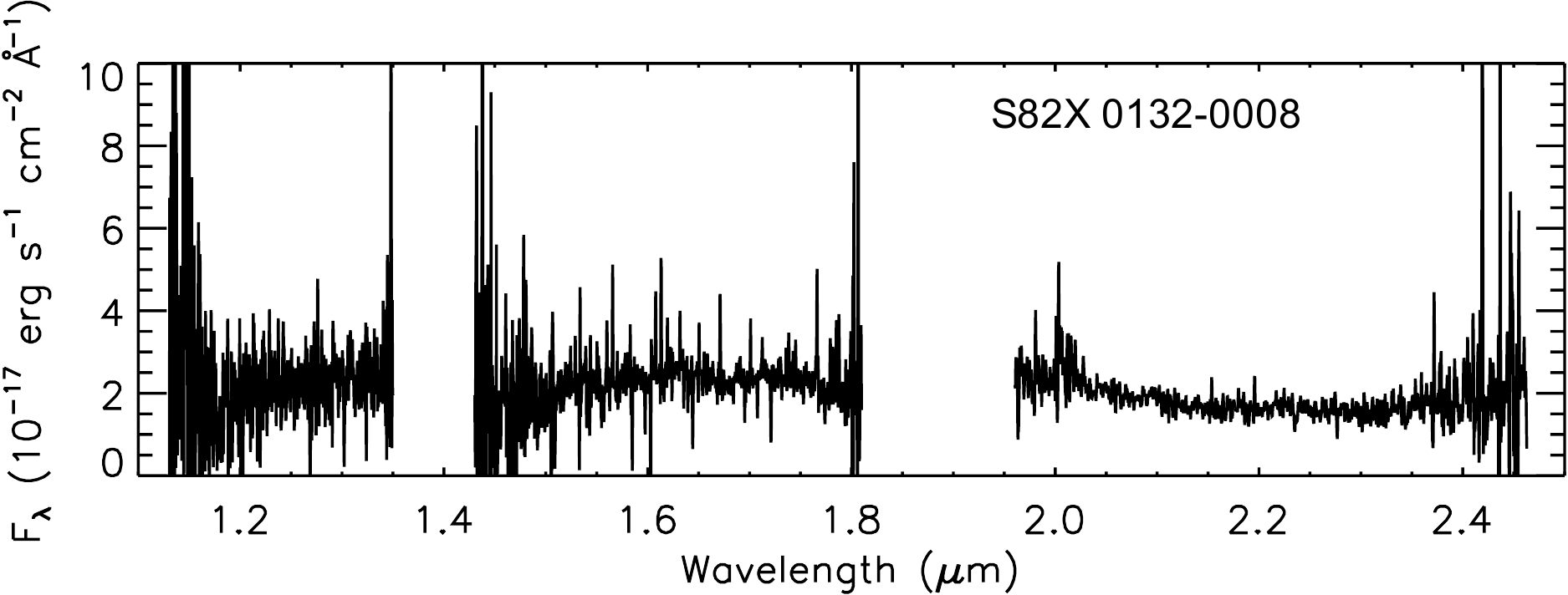}}~
      {\includegraphics[scale=0.4]{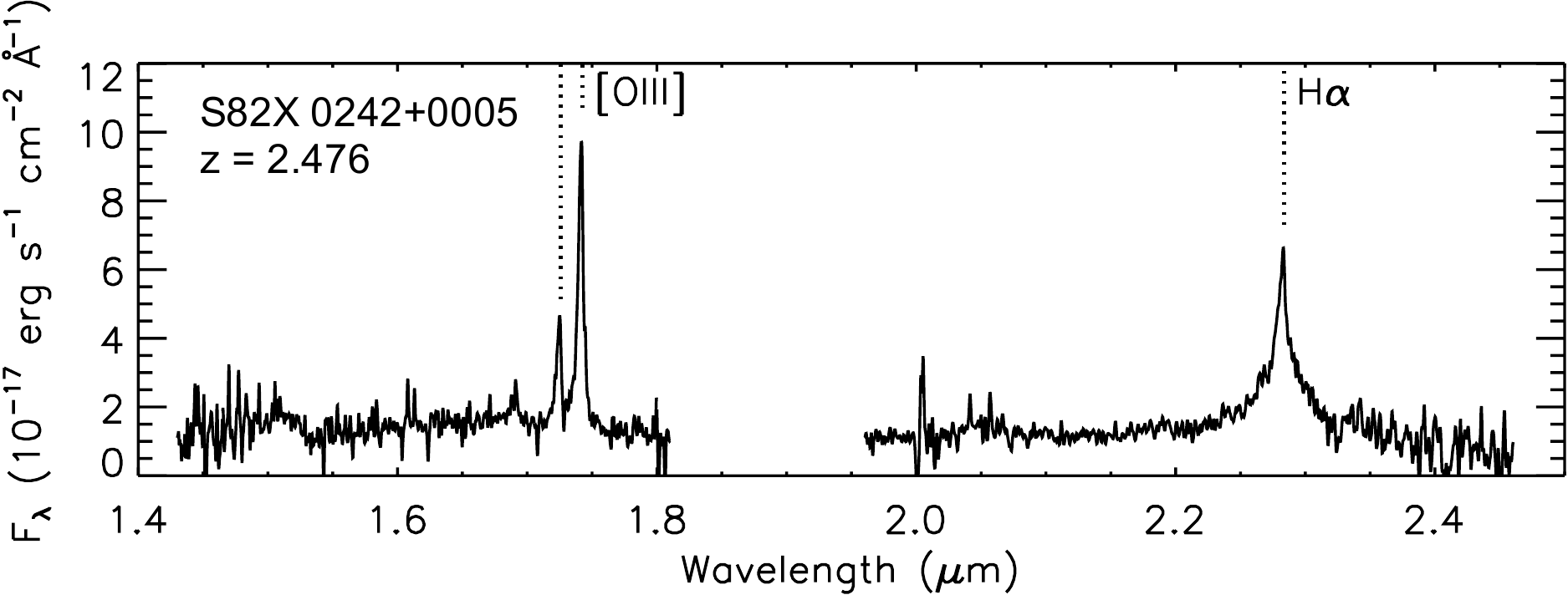}}
      {\includegraphics[scale=0.4]{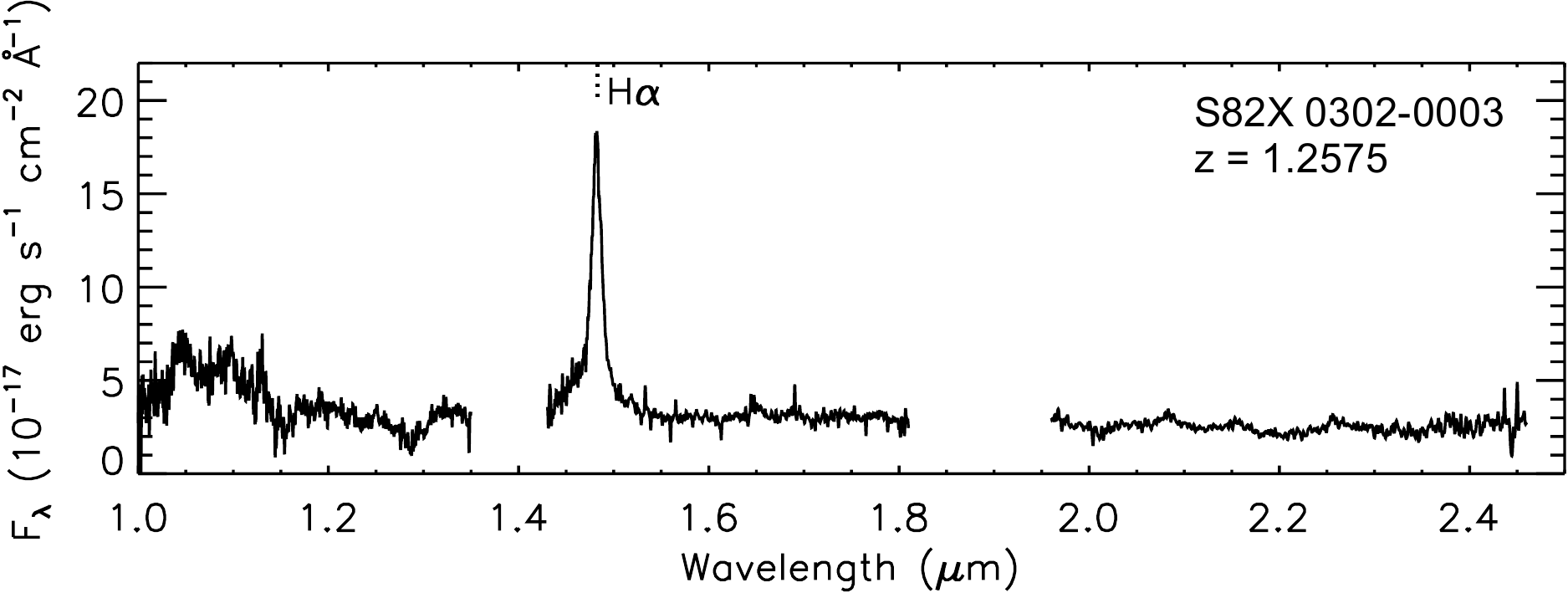}}~
      {\includegraphics[scale=0.4]{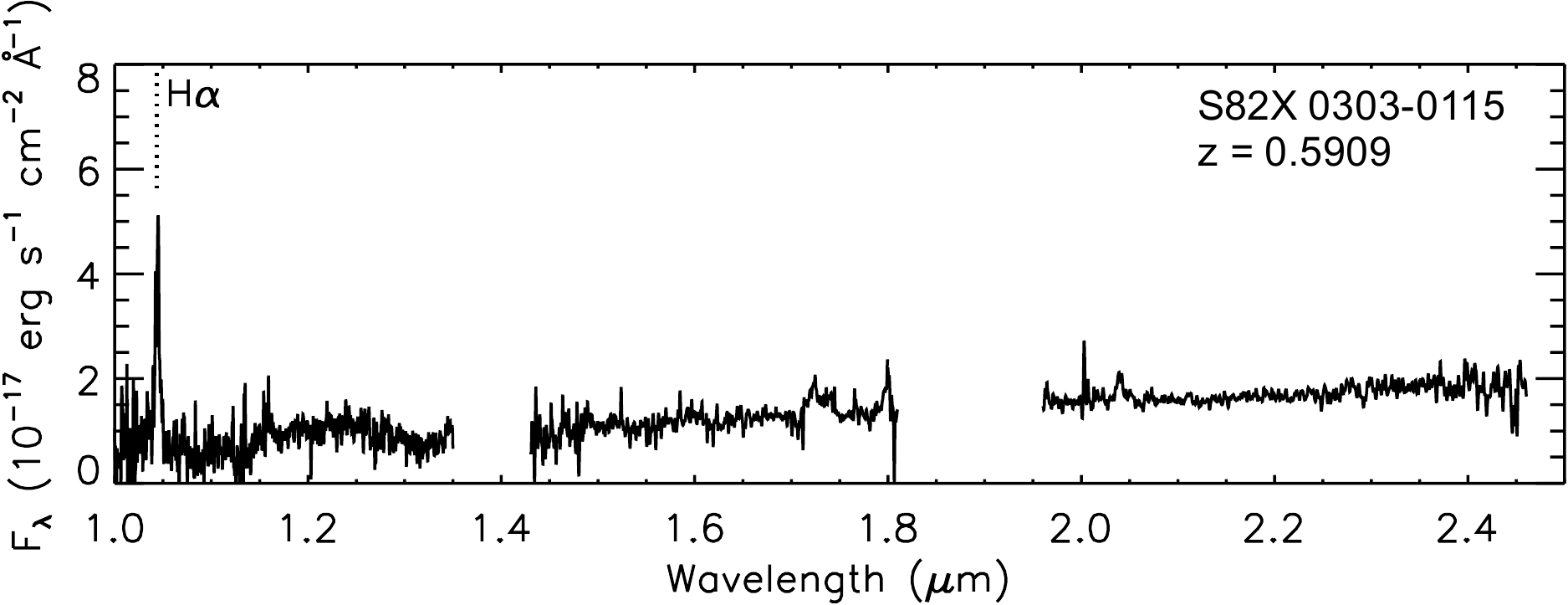}}
      {\includegraphics[scale=0.4]{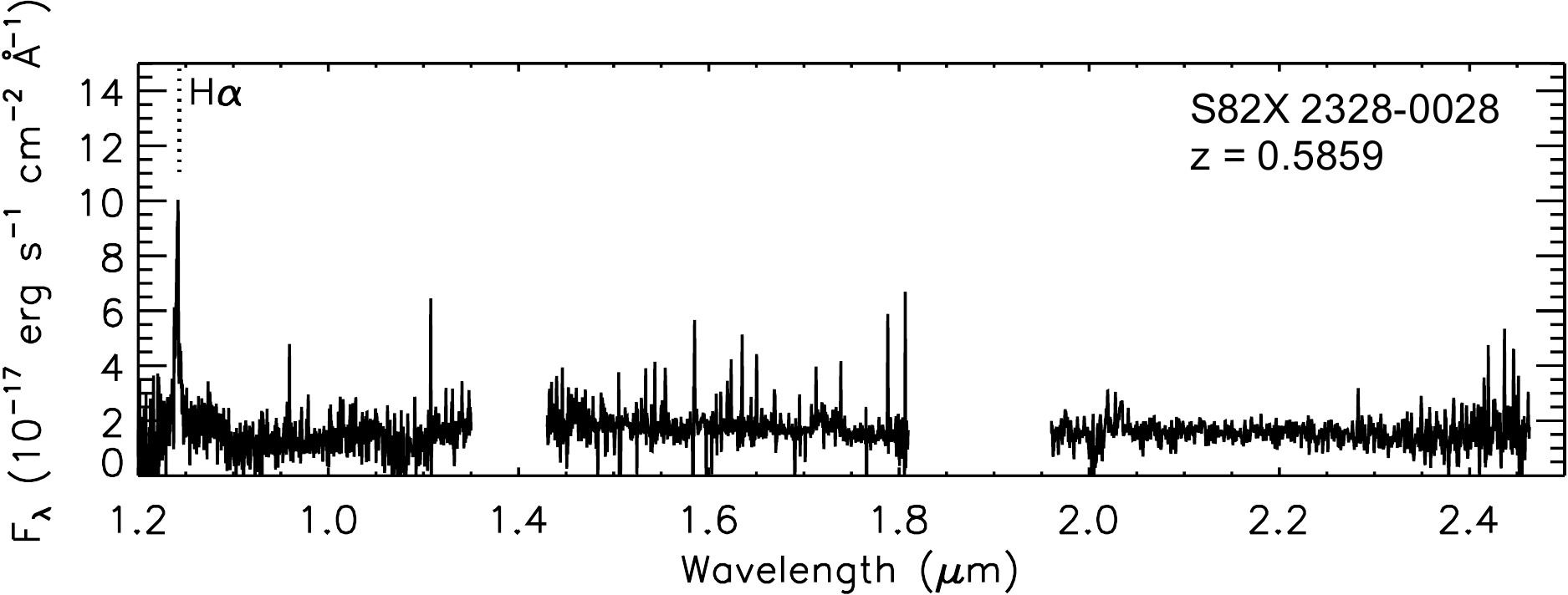}}
      \caption{\label{tspec_spec} Palomar TripleSpec spectra of our bright NIR $R-K$ versus $X/O$ selected sample. ({\it Top left}): Spectrum of S82X 0132-0008; though continuum is detected, no emission lines are present. ({\it Top right}): Spectrum of S82X 0242+0005 with H$\alpha$ and the [\ion{O}{3}] doublet marked. ({\it Middle left}): Spectrum of S82X 0302-0003 with H$\alpha$ marked. ({\it Middle right}): Extracted spectrum of S82X 0303-0115 with H$\alpha$ marked. Due to a bad column that overlaps the H$\alpha$ emission feature at the $A$ position, the spectrum was extracted from the $B$ position only. ({\it Bottom}):  Extracted spectrum of S82X 2328-0028 with H$\alpha$ marked. Marked transitions indicate the emission lines visible in the two-dimensional spectral images.}
\end{figure*}

\begin{figure*}[ht]
  \centering
      {\includegraphics[scale=0.4]{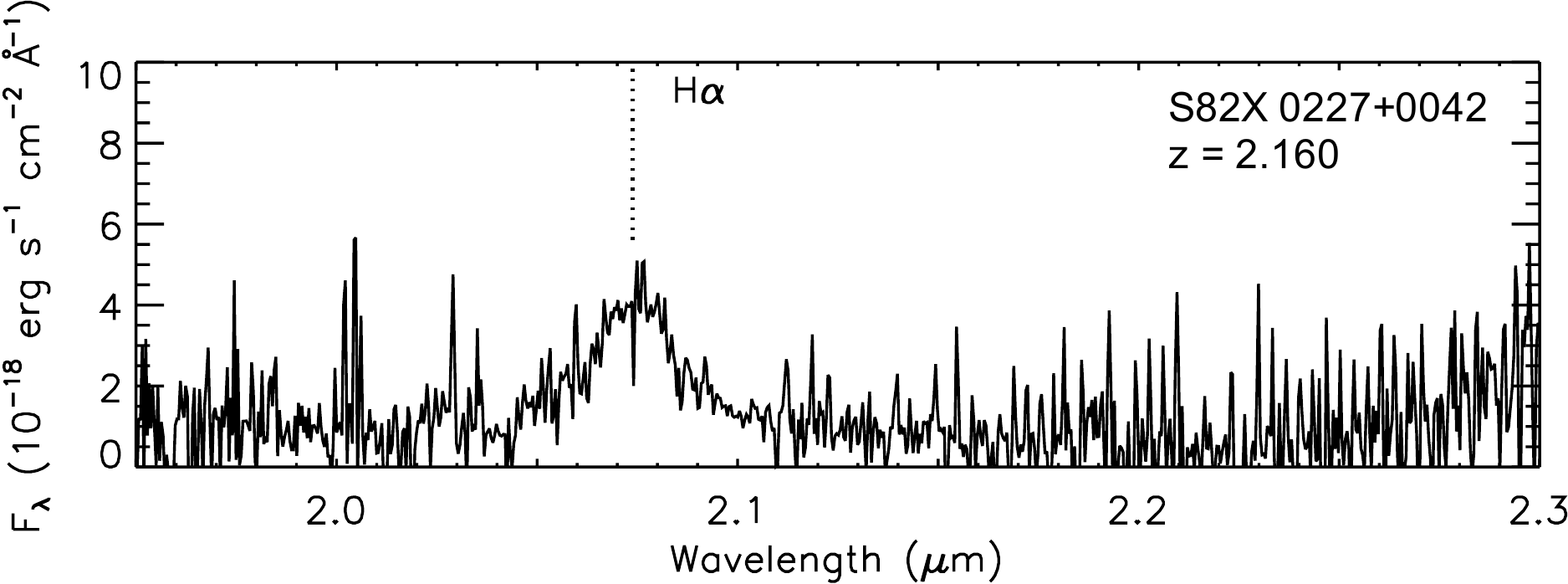}}~
      {\includegraphics[scale=0.4]{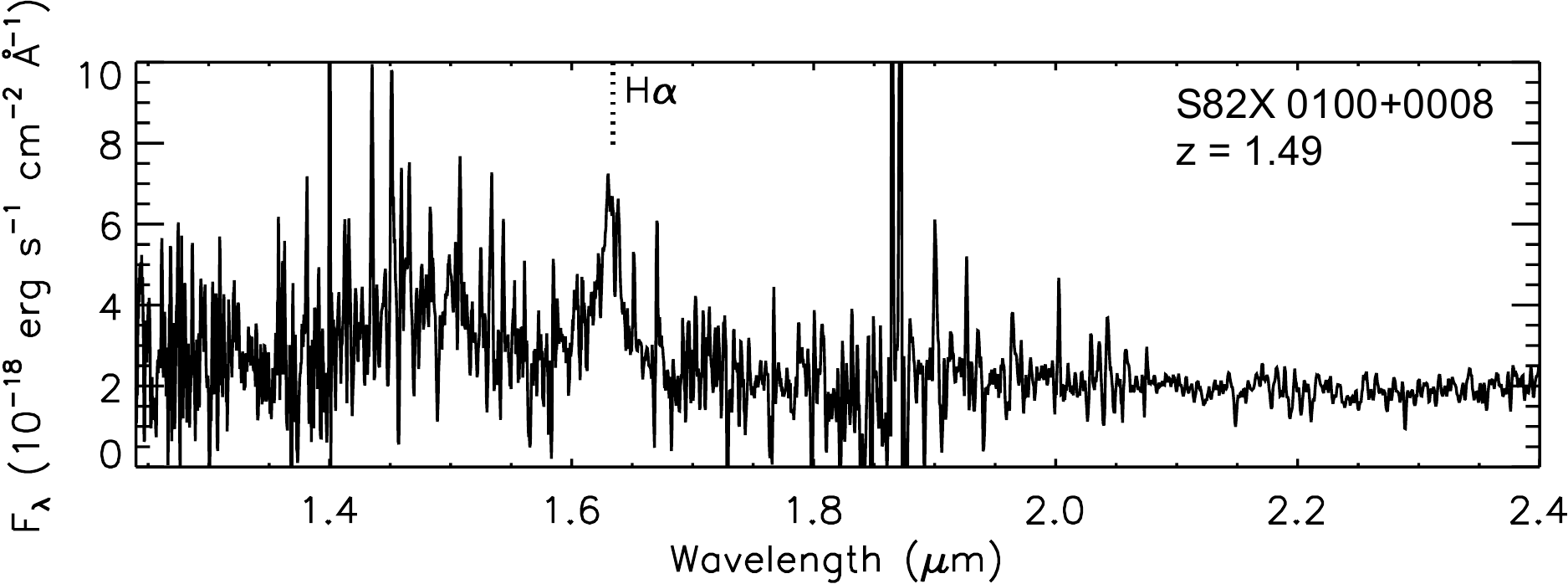}}
      {\includegraphics[scale=0.4]{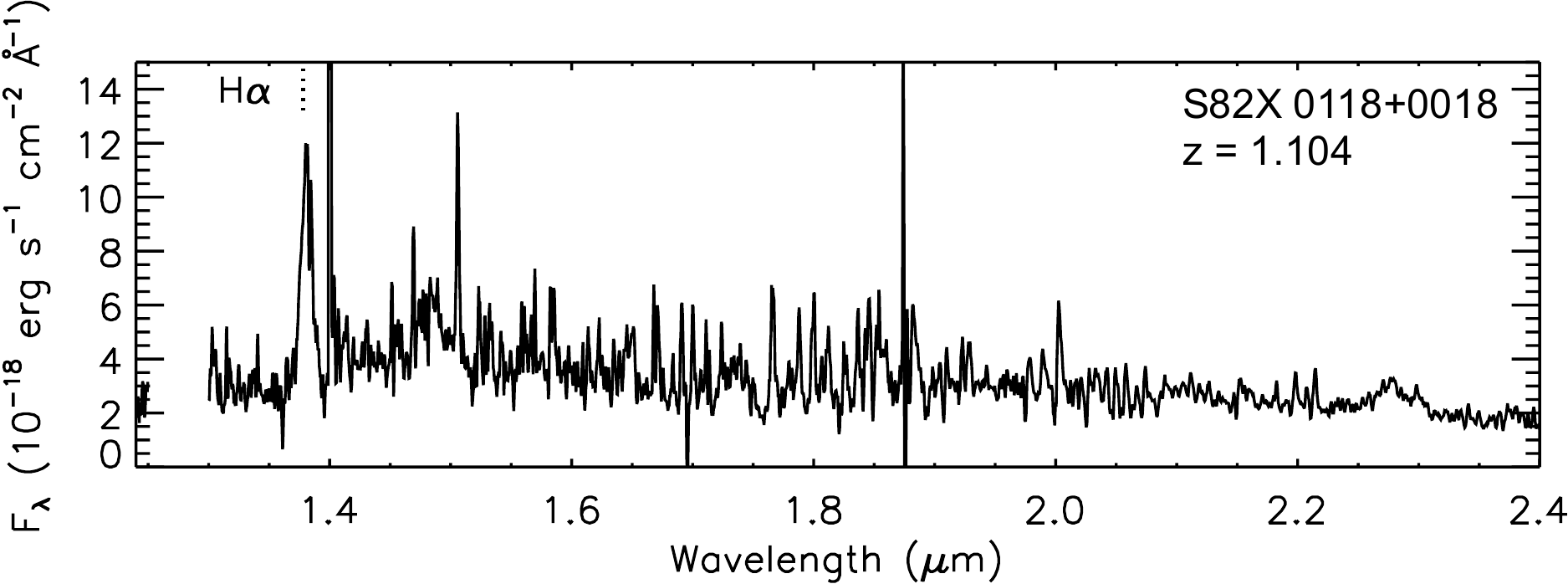}}~
      {\includegraphics[scale=0.4]{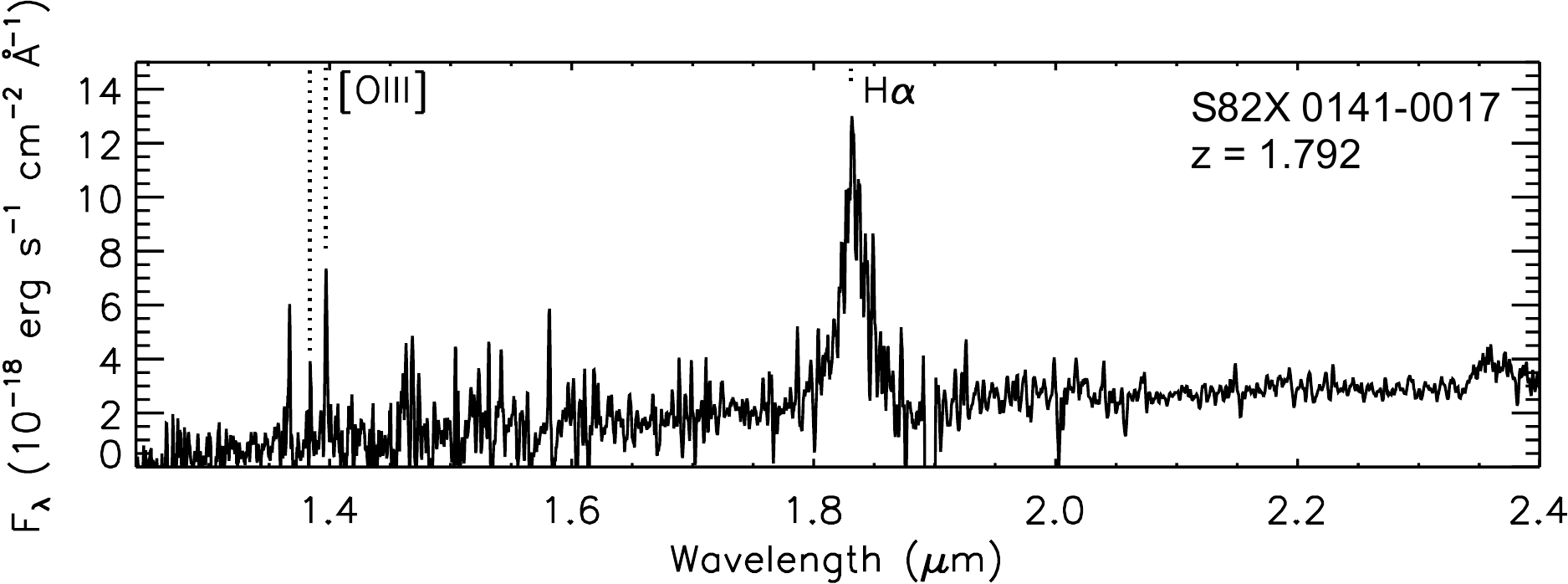}}
      \caption{\label{keck_gem_spec} Spectra of our faint NIR,  SDSS dropout sample. The spectrum in the {\it top left} is from Keck NIRSPEC while the others are from Gemini GNIRS. Marked transitions indicate the emission lines visible in the two-dimensional spectral images. We detected H$\alpha$ emission in each source and [\ion{O}{3}] in S82X 0141-0017. }
\end{figure*}

The spectra for the sources from the bright and faint samples are shown in Figures \ref{tspec_spec} and \ref{keck_gem_spec}, respectively. In the Palomar TripleSpec spectrum of source S82X 0132-0008, we clearly detect continuum, yet we find no emission lines, precluding us from including this source in the analysis and discussion below. We note that the photometric redshift for this source is $z_{\rm phot} \sim 1.74$ \citep{ananna}, such that H$\alpha$ would fall between the TripleSpec spectral orders, consistent with our lack of emission line detections. For the remaining eight sources, we detected H$\alpha$ emission and in two objects (S82X 0242+0005 and S82X 0141-0017), we also detect [\ion{O}{3}] emission. We indicate these emission features in the extracted one-dimensional spectra in Figure \ref{tspec_spec}. We verified that these emission lines are visible in the two-dimensional spectral images.

To obtain precise redshift measurements of these sources, as well as to calculate emission line fluxes and full-width half maxima (FWHMs) of the emission lines, we analyzed the spectra in IDL. To start, we interactively fit a first-order polynomial to the regions of the continuum free of emission and sky lines which was subsequently subtracted from the spectrum. We then used the IDL tool \textsc{mpfitfun} to fit a Gaussian model to the emission lines \citep{markwardt}. In two cases (S82X 0242+0005 and S82X 0302-0003), two broad Gaussian components were required to adequately fit the H$\alpha$ emission. When fitting the [\ion{O}{3}] doublet, the amplitude of the 4959\AA\ line was fixed to 1/3 the amplitude of the 5007\AA\ line, and the width of the lines were tied together. We note that in S82X 0242+0005, the [\ion{O}{3}] doublet has a blue wing to the narrow profile, which we accommodated with additional Gaussian components; we comment more on this feature in Section \ref{outflows}. While the redshifts were not tied when fitting the H$\alpha$ and [\ion{O}{3}] lines, we obtained consistent redshifts when fitting these features independently, indicating no systematic [\ion{O}{3}] blueshift. 

\begin{figure*}[ht]
  \centering
  \includegraphics[angle=90,scale=0.35]{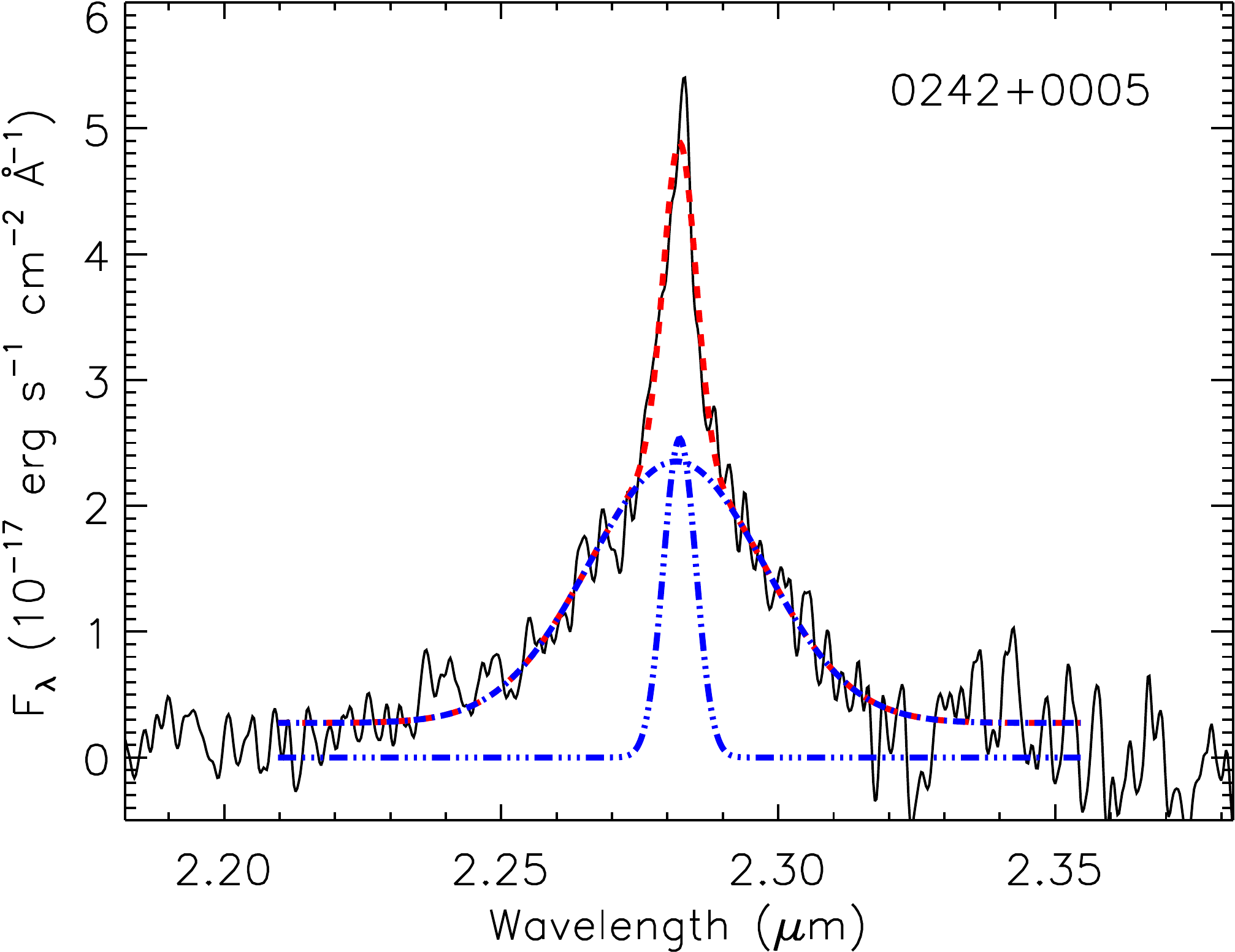}
  \includegraphics[angle=90,scale=0.35]{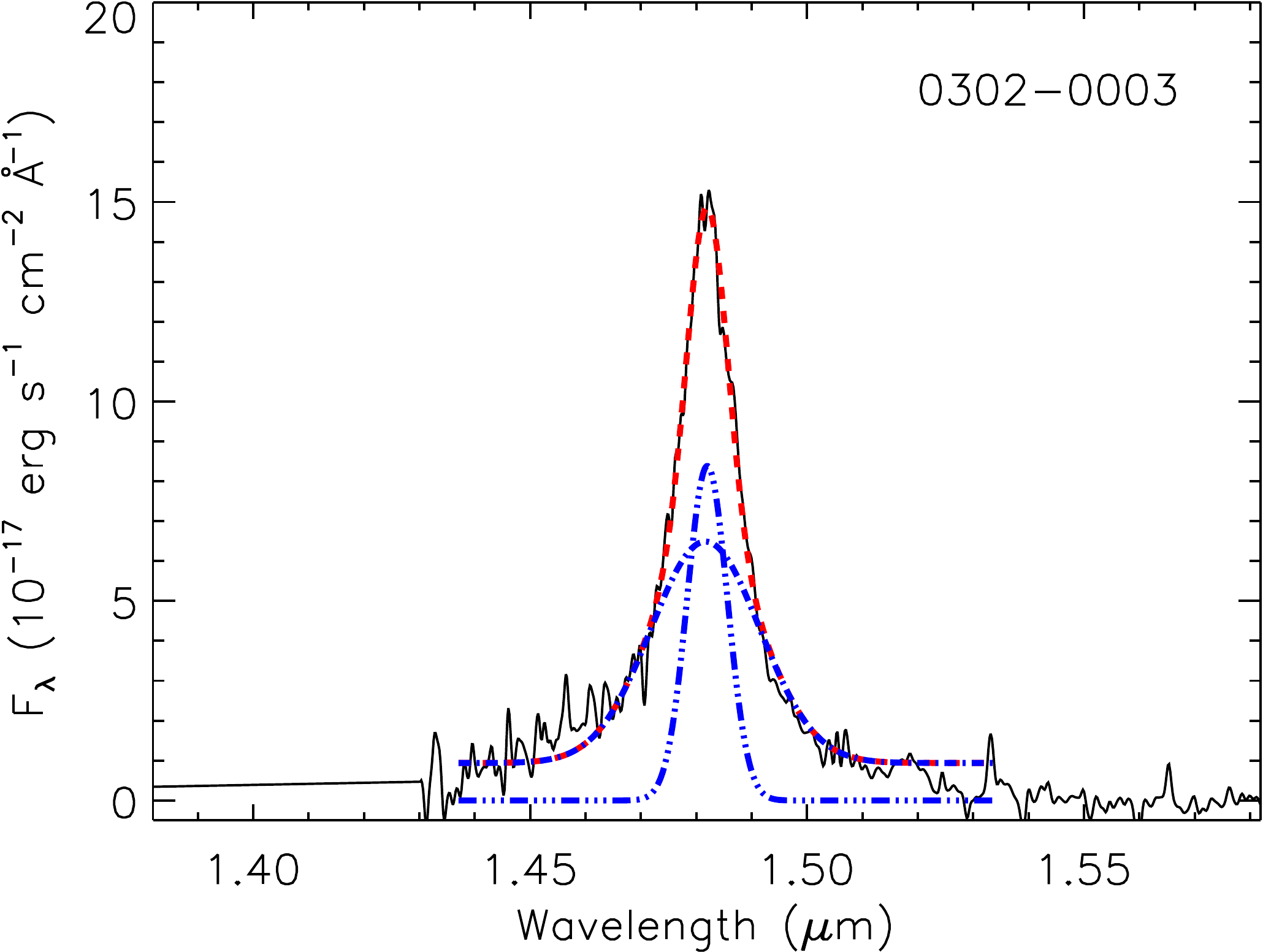}
  \includegraphics[angle=90,scale=0.35]{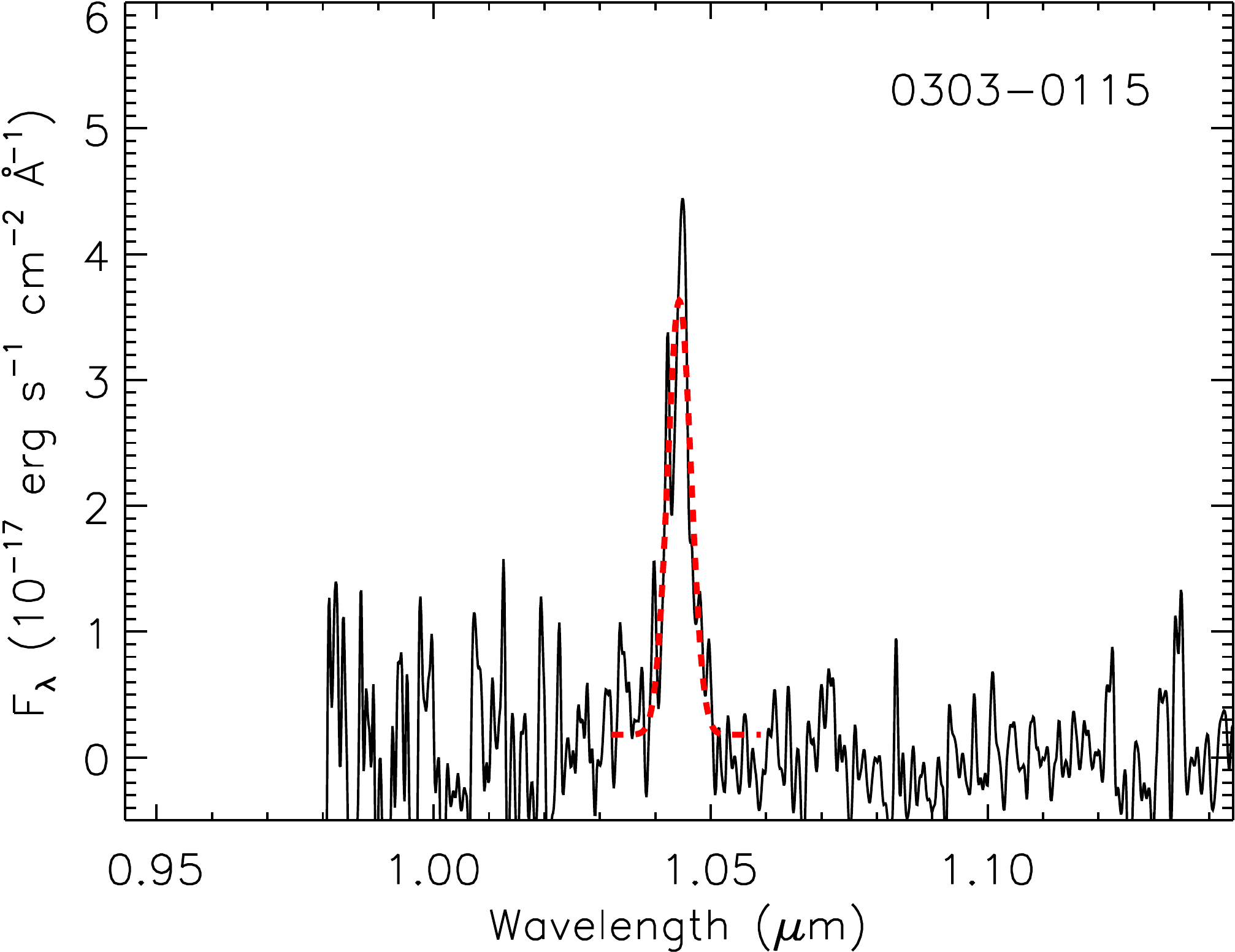}
  \includegraphics[angle=90,scale=0.35]{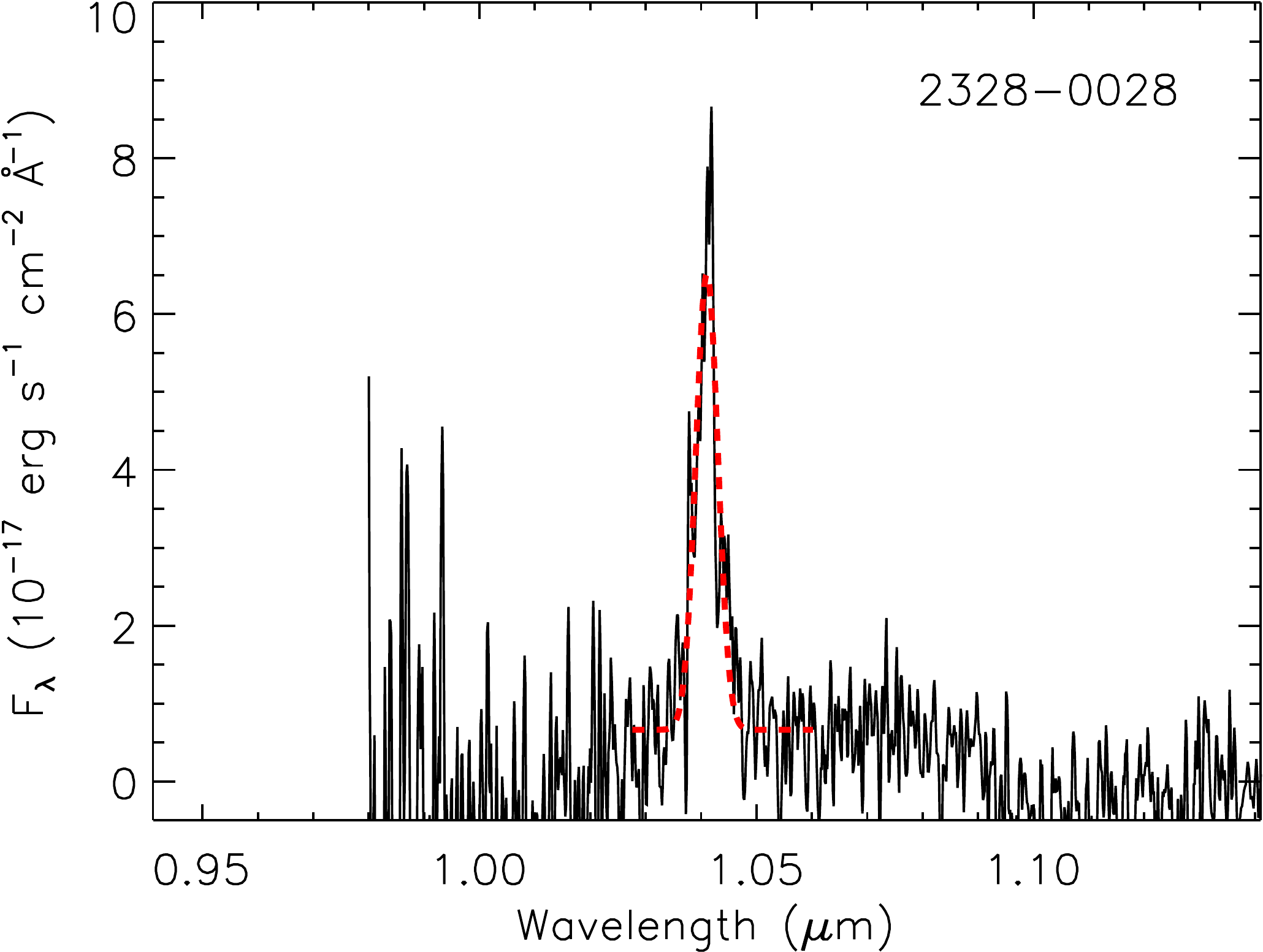}
  \includegraphics[angle=90,scale=0.35]{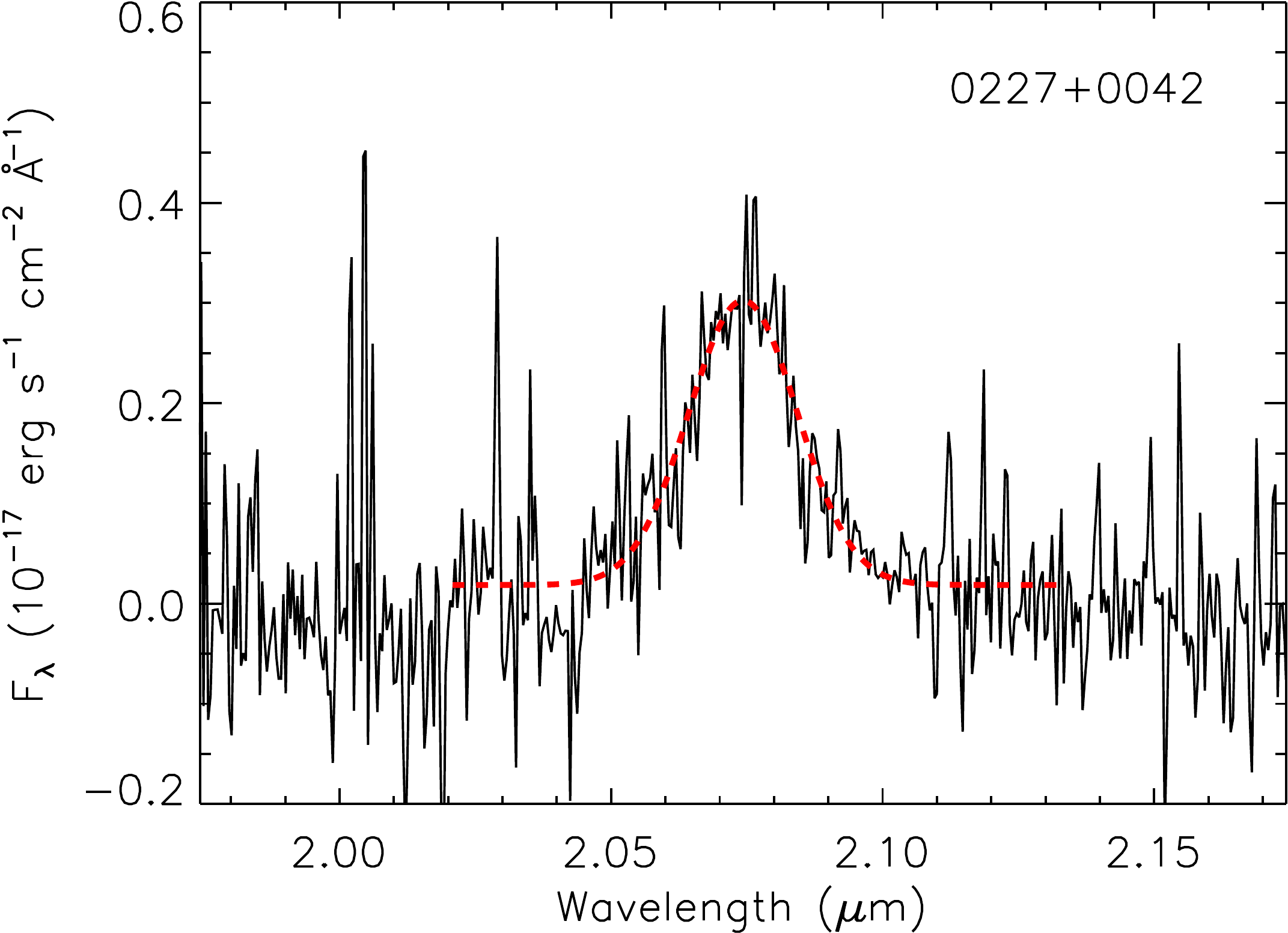}
  \includegraphics[angle=90,scale=0.35]{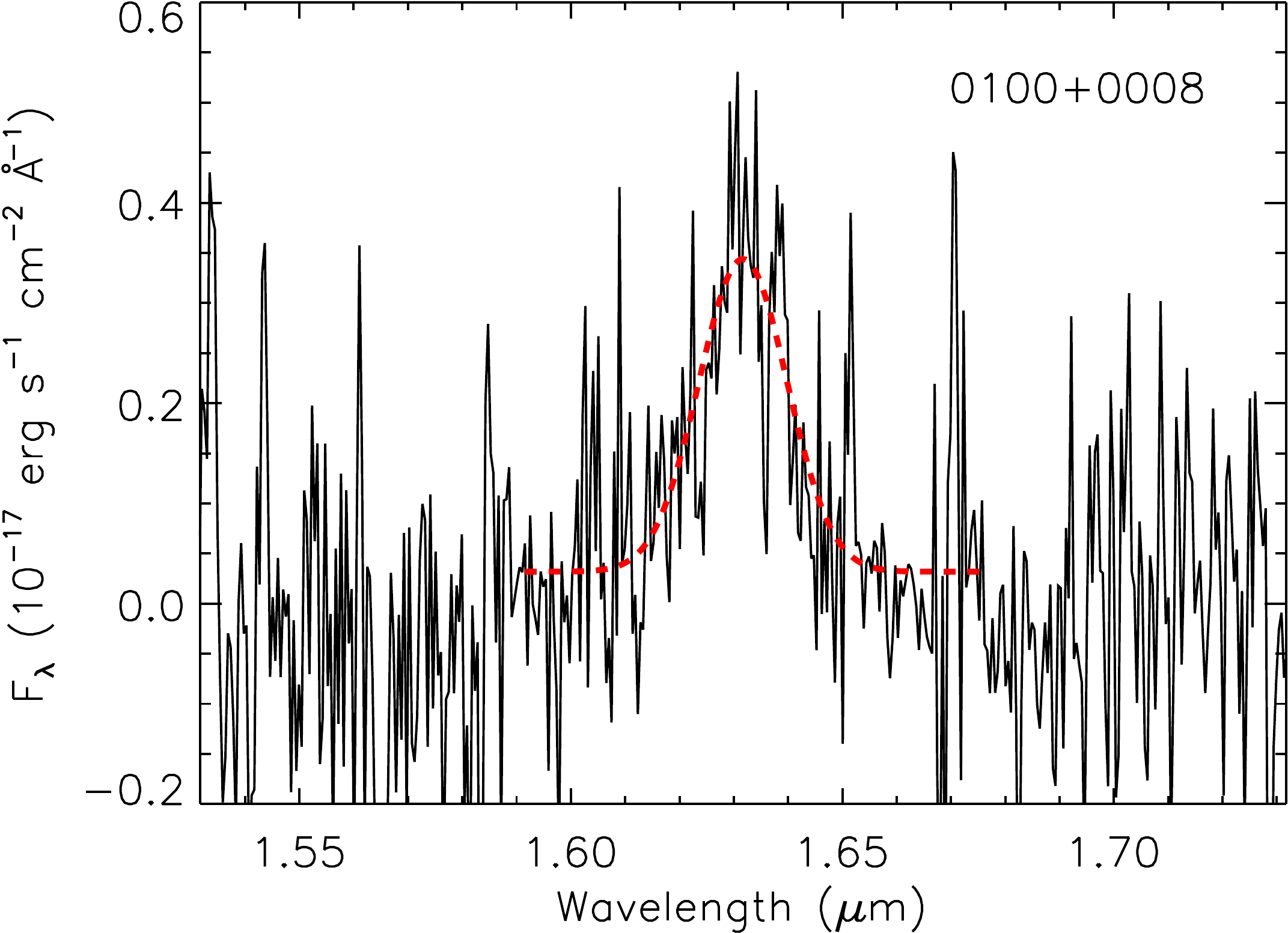}
  \includegraphics[angle=90,scale=0.35]{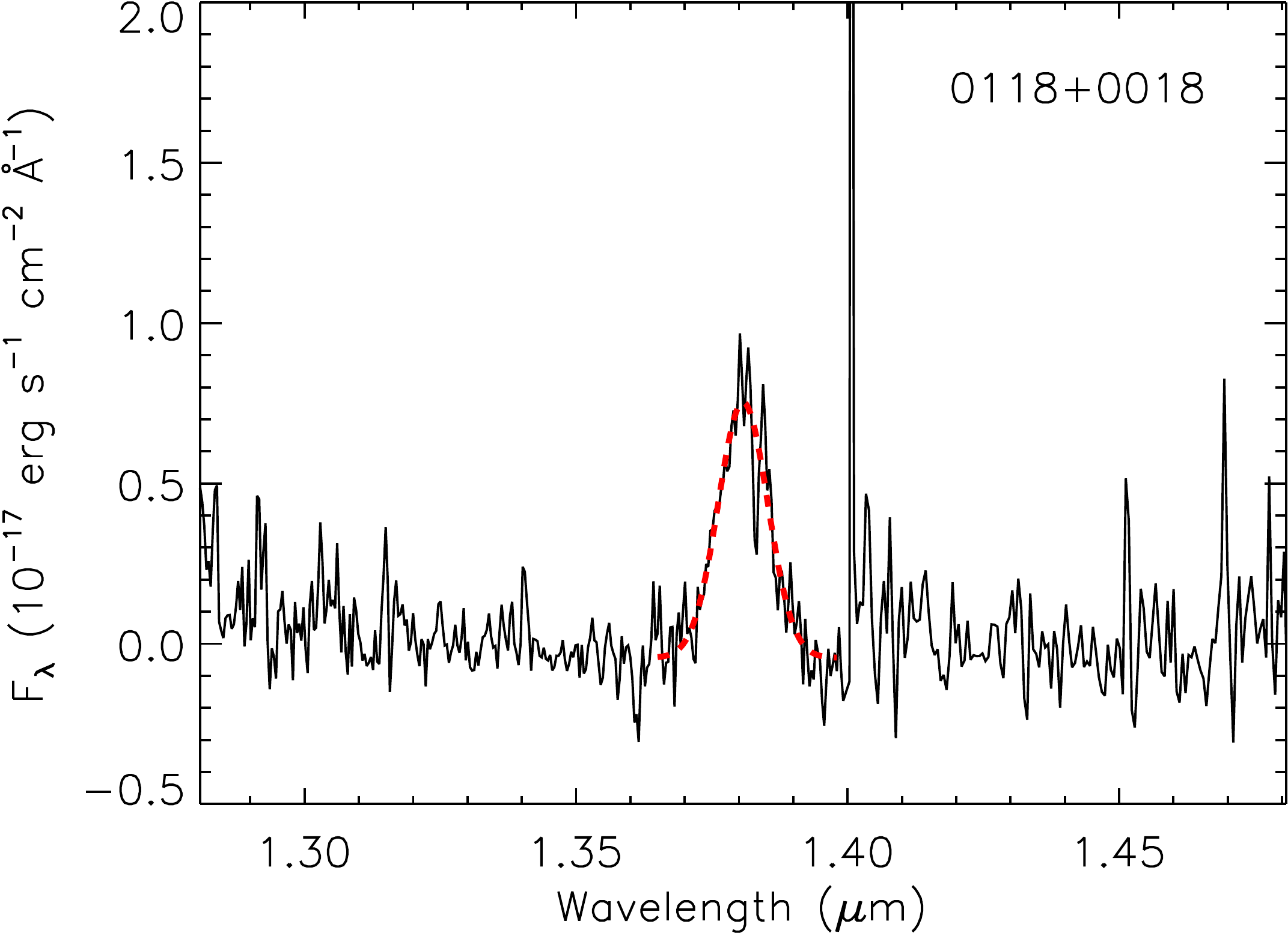}
  \includegraphics[angle=90,scale=0.35]{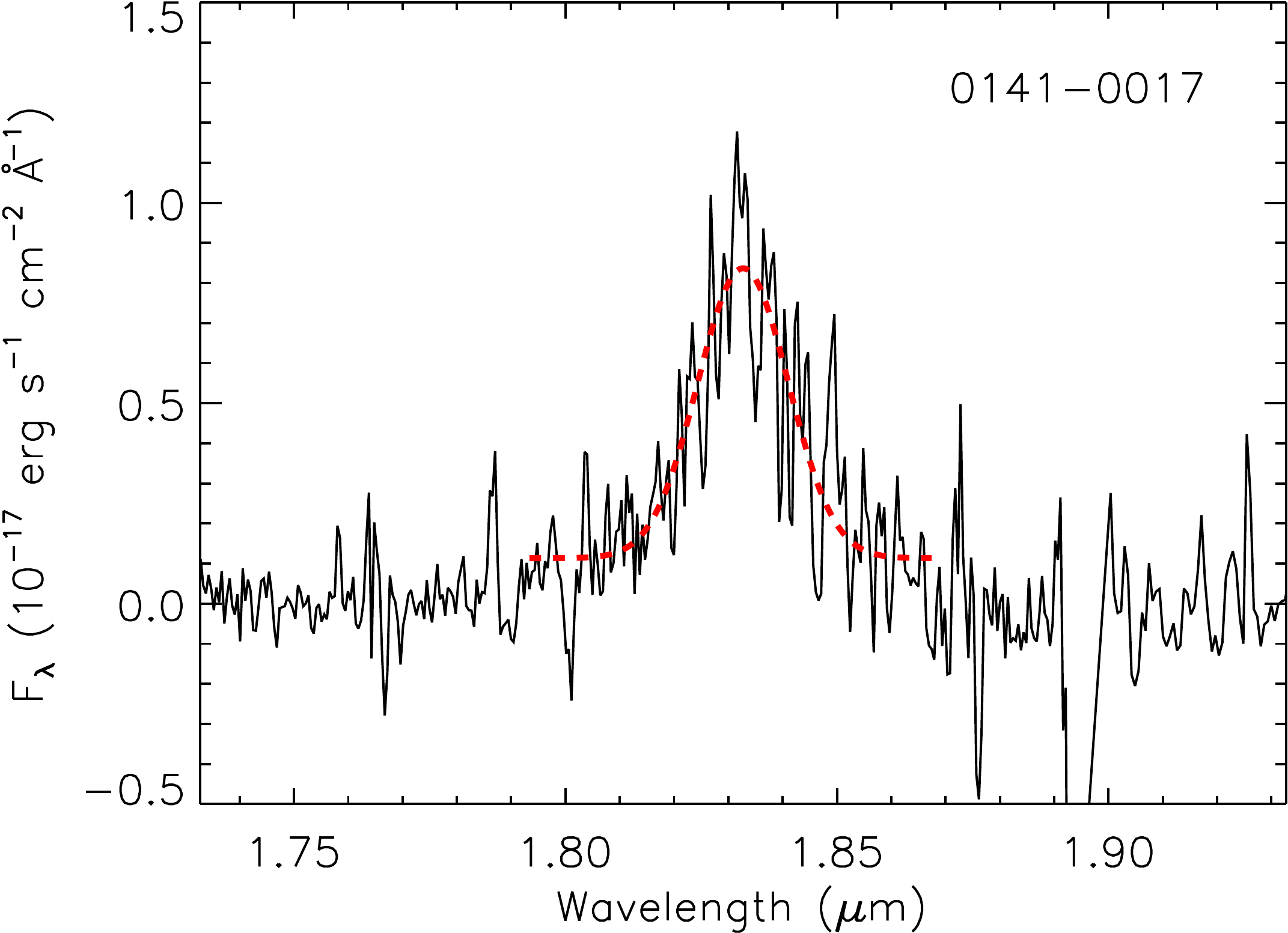}
  \caption{\label{ha_fits} Close-up of H$\alpha$ emission line region, fit with a Gaussian model (dashed red line). For 0242+0005 and 0302-0003, two Gaussian models, with a broad (blue dot-dash line) and narrow (blue dot-dot-dash line) component, were needed to adequately fit the emission feature. Spectra are in the observed frame.}
\end{figure*}

\begin{figure*}[ht]
  \centering
  \includegraphics[angle=90,scale=0.35]{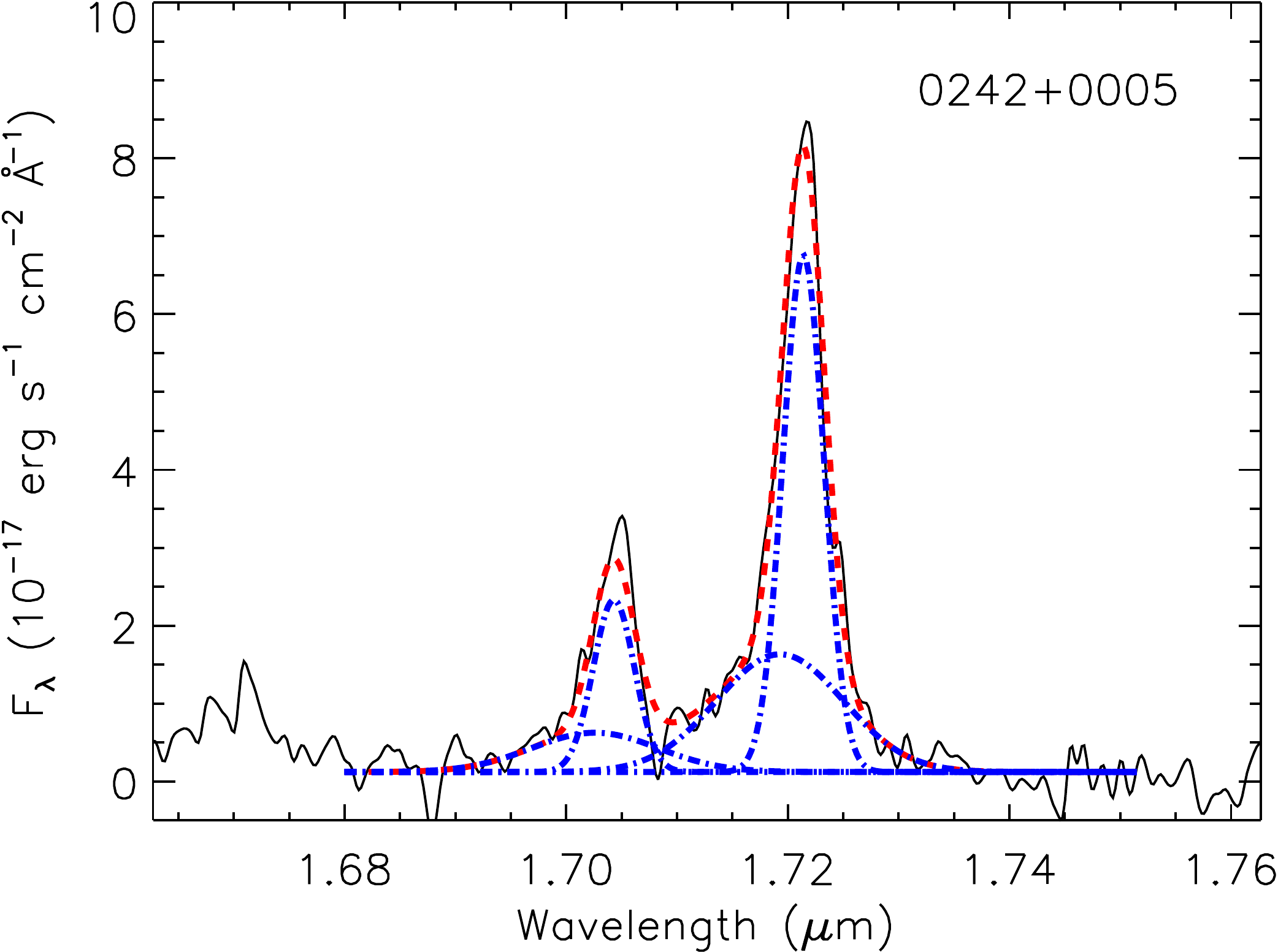}
  \includegraphics[angle=90,scale=0.35]{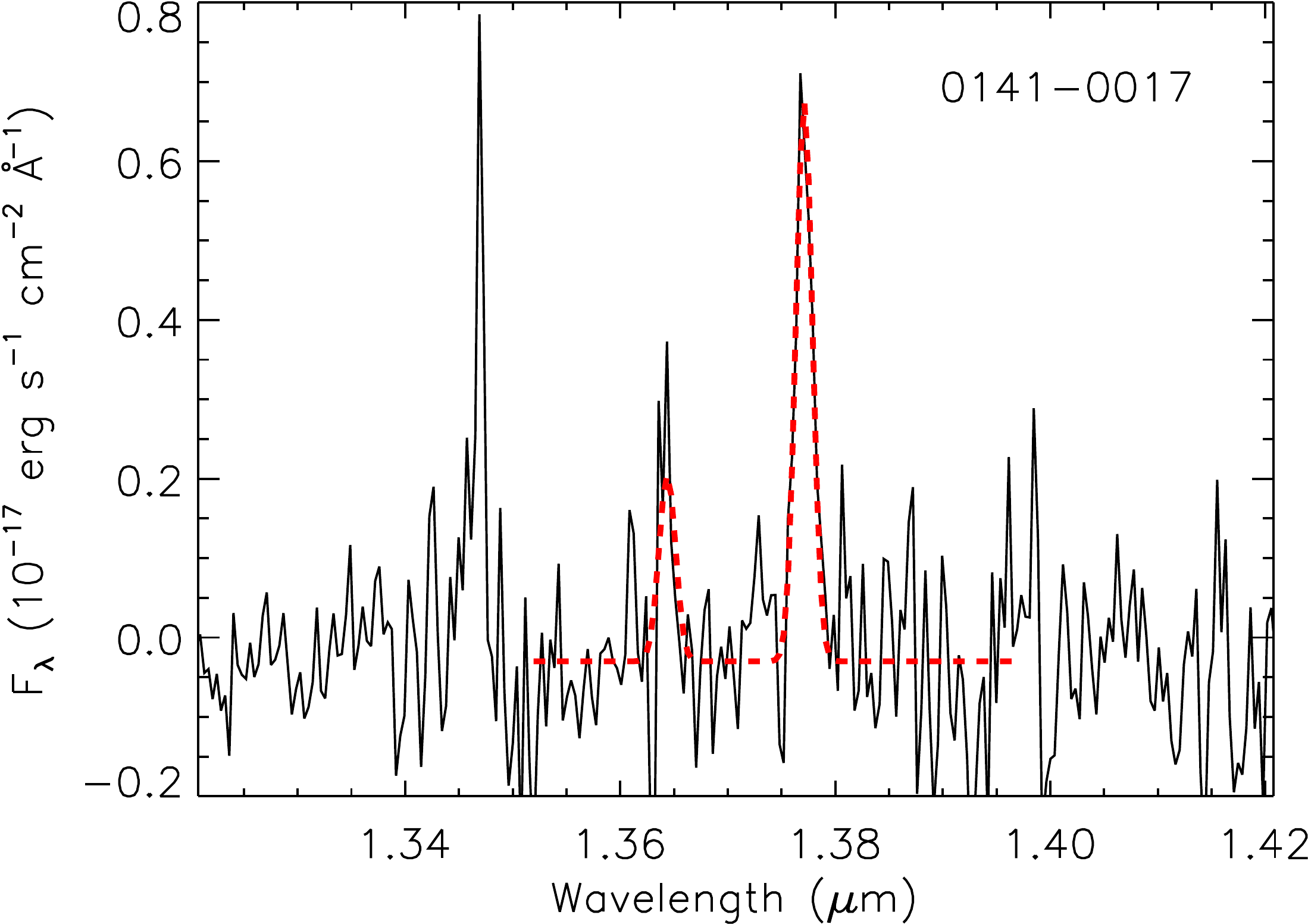}
  \caption{\label{oiii_fits} Close-up of [\ion{O}{3}] doublet fitted with a two-component Gaussian model, with the width of the lines tied together and the amplitude of the 4959\AA\ line frozen to a third of the 5007\AA\ line. We included an additional Gaussian component to fit the blue wing to the [\ion{O}{3}] doublet in S82X 0242+0005 ({\it left}), which is blueshifted with respect to the narrow component by $\Delta v = -400$ km s$^{-1}$, suggestive of outflowing gas. Spectra are in the observed frame. The red dashed line indicates the sum of the emission lines and the blue dot-dashed line represents the individual narrow and broad components.}
  \end{figure*}

We corrected the emission line FWHMs for the instrumental resolution using the relation FWHM$_{\rm corrected} = \sqrt{{\rm FWHM}_{\rm observed}^2 - {\rm FWHM}_{\rm instrument}^2}$ and the instrumental resolutions listed in Section \ref{obs}. The emission line fits are shown in Figures \ref{ha_fits} and \ref{oiii_fits} and the derived redshifts and emission line properties are listed in Tables \ref{z} and \ref{fluxes}, respectively. The quoted errors represent the propagation of the returned uncertainties associated with the fitted parameters.

All the sources we observed with this program have an H$\alpha$ FWHM exceeding 1300 km s$^{-1}$. As pointed out by \citet{zakamska2003}, the FWHM dividing line between Type 1 and Type 2 AGNs is not firmly established. Some studies use a FWHM value of 2000 km s$^{-1}$ to differentiate between Type 1 and Type 2 AGNs \citep[e.g.,][]{zakamska2003,alexandroff} while others set the limit at 1000 km s$^{-1}$ \citep[e.g.,][]{weedman} or 1100 km s$^{-1}$ \citep{reyes}. \citet{hao2005} demonstrated that the distribution of H$\alpha$ FWHMs for emission line galaxies in SDSS is bimodal: all broad line sources have a minimum FWHM of 1200 km s$^{-1}$. They thus define any source with a FWHM above this value as a Type 1 AGN. Following this convention, and for consistency with previous red quasar studies which require a FWHM exceeding 1000 km s$^{-1}$ to define a source as a quasar \citep{glikman2007}, we classify all our Stripe 82X sources as Type 1 AGNs. We note, however, that the classification of the two sources with H$\alpha$ FWHMs below 2000 km s$^{-1}$, S82X 0303-0115 and S82X 2328-0028, may be ambiguous. As discussed below, all have X-ray luminosities consistent with accretion onto a supermassive black hole.

\begin{deluxetable*}{lllll}[h]
\tablewidth{0pt}
\tablecaption{\label{z} Redshifts and X-ray Characteristics of Reddened AGNs}
\tablehead{\colhead{Source} & \colhead{$z$} & \colhead{$L_{\rm X,full}$\tablenotemark{1}} & \colhead{$HR$\tablenotemark{2}} & \colhead{$N_{\rm H}$\tablenotemark{3}}\\
& & \colhead{(erg s$^{-1}$)} & & \colhead{(10$^{22}$ cm$^{-2}$)} }

\startdata

\multicolumn{4}{c}{{\textit{Palomar TripleSpec}}}\\

S82X 0242+0005 & 2.476$\pm$0.001 & 5.57$\times10^{44}$ & -0.24$^{+0.25}_{-0.19}$ & 4$^{+16}_{-4}$  \\

S82X 0302$-$0003 & 1.2574$\pm$0.0001 & 2.51$\times10^{44}$ & 0.37$^{+0.30}_{-0.26}$ & 10$^{+10}_{-4}$ \\

S82X 0303$-$0115 & 0.5909$\pm$0.0003 & 1.71$\times10^{43}$ & 0.21$^{+0.36}_{-0.34}$ & 3$^{+4}_{-2}$  \\

S82X 2328$-$0028 & 0.5859$\pm$0.0001 & 1.08$\times10^{44}$ & -0.96$^{+0.04}_{-0.04}$ & 0  \\

\multicolumn{4}{c}{{\textit{Keck NIRSPEC}}}\\

S82X 0227+0042 & 2.16$\pm$0.01 & 8.17$\times10^{44}$ & -0.90$^{+0.10}_{-0.10}$  & 0 \\

\multicolumn{4}{c}{{\textit{Gemini GNIRS}}}\\

S82X 0100+0008 & 1.49$\pm$0.01 & 2.25$\times10^{44}$ & -0.75$^{+0.12}_{-0.25}$ & 0 \\

S82X 0118+0018 & 1.103$\pm$0.003 & 1.46$\times10^{44}$ & -0.49$^{+0.17}_{-0.51}$ & $<$0.6 \\

S82X 0141$-$0017 & 1.792$\pm$0.004 & 2.68$\times10^{44}$ & -0.43$^{+0.25}_{-0.22}$ & $<$3

\enddata
\tablenotetext{1}{$k$-corrected (i.e., rest-frame), non-absorption corrected luminosities.}
\tablenotetext{2}{$HR = (H-S)/(H+S)$, where $H$ ($S$) are net counts in the hard (soft) X-ray bands and were calculated with BEHR \citep{behr}.}
\tablenotetext{3}{Gas column density ($N_{\rm H}$) implied by the $HR$ ranges.}
\end{deluxetable*}

\begin{deluxetable*}{lllllll}[h]
\tablewidth{0pt}
\tablecaption{\label{fluxes}Emission Line Properties of Targeted Reddened AGNs\tablenotemark{1}}
\tablehead{\colhead{Source} & \colhead{$f_{\rm{H\alpha,1}}$} &  \colhead{FWHM$_{\rm{H\alpha,1}}$} & \colhead{$f_{\rm{H\alpha,2}}$} & \colhead{FWHM$_{\rm{H\alpha,2}}$} & \colhead{$f_{{\rm [OIII] 5007\AA\ }}$} & \colhead{FWHM$_{{\rm [OIII] 5007\AA\ }}$} \\
& \colhead{(10$^{-17}$ erg s$^{-1}$ cm$^{-2}$)} &  \colhead{(km s$^{-1}$)} &  \colhead{(10$^{-17}$ erg s$^{-1}$ cm$^{-2}$)} & \colhead{(km s$^{-1}$)} & \colhead{(10$^{-17}$ erg s$^{-1}$ cm$^{-2}$)} & \colhead{(km s$^{-1}$)}}

\startdata

\multicolumn{7}{c}{{\textit{Palomar TripleSpec}}}\\

S82X 0242+0005 &  820$\pm$70 & 4900$\pm$200 &  190$\pm$30 &  900$\pm$90  & 300$\pm$20 & 750$\pm$20 \\

S82X 0302$-$0003 & 1370$\pm$60 & 4690$\pm$60  &  760$\pm$30 & 1720$\pm$20 & \nodata  & \nodata  \\

S82X 0303$-$0115 & 180$\pm$10  & 1430$\pm$60  &  \nodata    & \nodata     & \nodata  & \nodata  \\

S82X 2328$-$0028 & 290$\pm$10  & 1350$\pm$20  &  \nodata    & \nodata     & \nodata  & \nodata  \\

\multicolumn{7}{c}{{\textit{Keck NIRSPEC}}}\\

S82X 0227+0042 & 70$\pm$60  & 3400$\pm$1300 & \nodata & \nodata & \nodata & \nodata \\

\multicolumn{7}{c}{{\textit{Gemini GNIRS}}}\\

S82X 0100+0008 & 60$\pm$50 & 3500$\pm$1200 & \nodata & \nodata & \nodata & \nodata \\

S82X 0118+0018 & 90$\pm$40 & 2300$\pm$300  & \nodata & \nodata & \nodata & \nodata \\

S82X 0141$-$0017 & 150$\pm$60 & 3200$\pm$600 & \nodata & \nodata & $<$39\tablenotemark{2} & 400$\pm$200

\enddata
\tablenotetext{1}{Two Gaussian components were needed to fit the H$\alpha$ emission profile for two sources (S82X 0242+0005 and S82X 0302-0003). If a second H$\alpha$ component was needed, the flux and FWHM of this feature is reported as $f_{\rm{H\alpha,2}}$ and FWHM$_{\rm{H\alpha,2}}$, respectively.  $f_{\rm{H\alpha,1}}$ and FWHM$_{\rm{H\alpha,1}}$ represent the flux and FWHM of the broader H$\alpha$ component, if two Gaussian profiles are needed to fit the spectrum, or the single Gaussian component for the remaining sources. The narrow component of the [\ion{O}{3}] 5007 \AA\ line is reported for S82X 0242+0005; the broad wing to the [\ion{O}{3}] doublet is recorded in Table \ref{outflow_fit} and discussed in Section \ref{outflows}.}
\tablenotetext{2}{3$\sigma$ upper limit.}
\end{deluxetable*}

\subsection{X-ray Properties of AGNs Targeted with Infrared Spectroscopy}
From the redshifts measured above, we calculated the $k$-corrected, observed (non-absorption corrected) full band X-ray luminosities ($L_{\rm X,full}$, where $f_{\rm k-corr} = f_{\rm observed} \times (1+z)^{\Gamma - 2}$ and $\Gamma$, the slope of the AGN continuum power law, is 1.7; \citealt{s82x1,s82x2,s82x3}). To estimate the approximate X-ray absorption, we calculated the hardness ratio: $HR = (H-S)/(H+S)$, where $H$ is the net number of counts in the hard band and $S$ is the net number of counts in the soft band. We note that while the soft range is 0.5-2 keV for both {\it Chandra} and {\it XMM-Newton}, the hard band is 2-10 keV for {\it XMM-Newton} and 2-7 keV for {\it Chandra}. For this calculation, we used the Bayesian Estimation of Hardness Ratios \citep[BEHR;][]{behr} which provides robust estimates of $HR$ in the low-count regime and in the case of non-detections in either band. BEHR takes as input the total counts in the soft and hard bands within user-defined source and background regions and the ratio of areas between the source and background regions. While {\it XMM-Newton} has three detectors, we extracted the net counts from only the MOS1 detector for a straightforward comparison with model $HR$ values.

Gas column density estimates ($N_{\rm H}$) derived from hardness ratios are redshift dependent: at higher redshifts, the higher energy photons ($>$7-10 keV), which are less attenuated by absorption, are shifted into the observed bandpass. Assuming an absorbed powerlaw, we calculated a grid of hardness ratios for various $N_{\rm H}$ values over a range of redshifts in bins of 0.05 for both {\it Chandra} and the MOS1 detector on {\it XMM-Newton}. Using the redshift of the source and range of hardness ratios returned by BEHR, we determined the implied $N_{\rm H}$ and report these values in Table \ref{z}. Of the eight sources, the hardness ratios for three are consistent with an unabsorbed X-ray source and three have lower limits on $N_{\rm H}$ of 0; we do not correct for Galactic absorption since such low column densities \citep[$N_{\rm H} \sim 3\times 10^{20}$ cm$^{-2}$;][]{dickey} have little impact on the X-ray spectrum.

Since S82X 0242+0005 is detected in hard X-rays (2-10 keV; $F_{\rm 2-10 keV} = (1.56 \pm 0.34) \times 10^{-14}$ erg s$^{-1}$) and the [\ion{O}{3}] 5007 \AA\ line is measured, we can independently assess the X-ray obscuration by $F_{\rm 2-10 keV}$/$F_{\rm [OIII]}$: because [\ion{O}{3}] forms in the AGN narrow line region, it is unaffected by circumnuclear obscuration that attenuates the X-ray emission and thus serves as a proxy of the intrinsic AGN luminosity \citep[e.g.,][]{bassani,heckman2004,me2010}. The ratio of the hard X-ray to [\ion{O}{3}] flux can then indicate whether the X-ray emission is heavily absorbed \citep{bassani,panessa,lansbury2014,lansbury2015}. We find log($F_{\rm 2-10 keV}$/$F_{\rm [OIII]}$) = 0.72$\pm$0.09 dex, which is significantly less than the mean value for unabsorbed Type 1 AGNs \citep[1.59$\pm$0.48 dex][]{heckman}, but higher than the most heavily obscured, Compton-thick systems \citep[e.g.,][]{me2009,me2011}. Both $F_{\rm 2-10 keV}$/$F_{\rm [OIII]}$ and the implied $N_{\rm H}$ from the hardness ratio are consistent with a moderately obscured AGN. However, the narrow line region can suffer extinction, which would translate into more luminous intrinsic [\ion{O}{3}] emission, causing the true $F_{\rm 2-10 keV}$/$F_{\rm [OIII]}$ to decline. The implied X-ray column density can thus be higher.

We caution that the theoretical $HR$-$N_{\rm H}$ conversion assumes a simple absorbed power law while observed X-ray spectra of obscured AGN are generally more complex, with scattered emission, or leakage through a patchy obscuring medium, that can boost the observed soft X-ray flux compared with the model assumed here. Furthermore, hardness ratios provide no information about the global distribution, or global column density, of obscuration around the AGN. Indeed, AGNs can have significantly different line-of-sight and global column densities \citep{lamassa2014,yaqoob,lamassa2016c}. Though the implied column densities for some of these AGNs are consistent with column densities of FIRST-2MASS selected reddened quasars \citep{glikman2012} derived via X-ray spectral fitting \citep{lamassa2016c}, the reported $N_{\rm H}$ ranges should be considered approximate.

\subsection{\label{supp_samp} Bright Sources with SDSS Spectra}
\begin{figure}[ht]
  \centering
{\includegraphics[scale=0.4,angle=90]{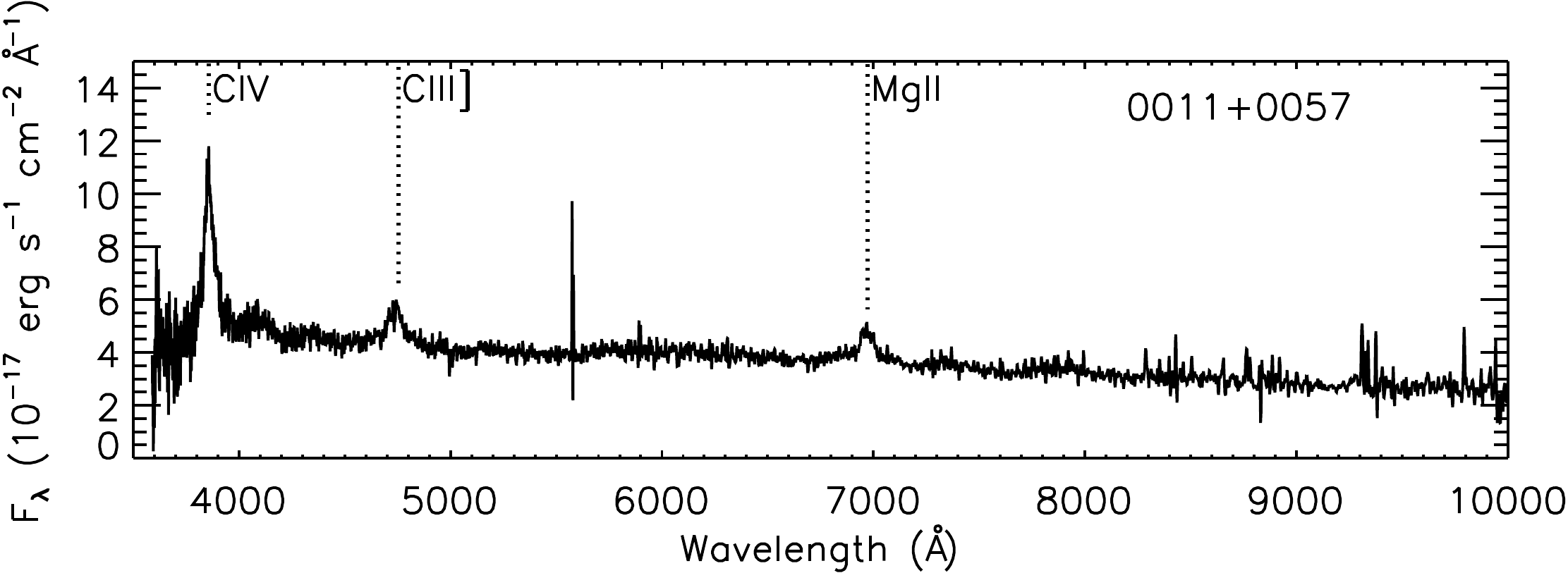}}
{\includegraphics[scale=0.4,angle=90]{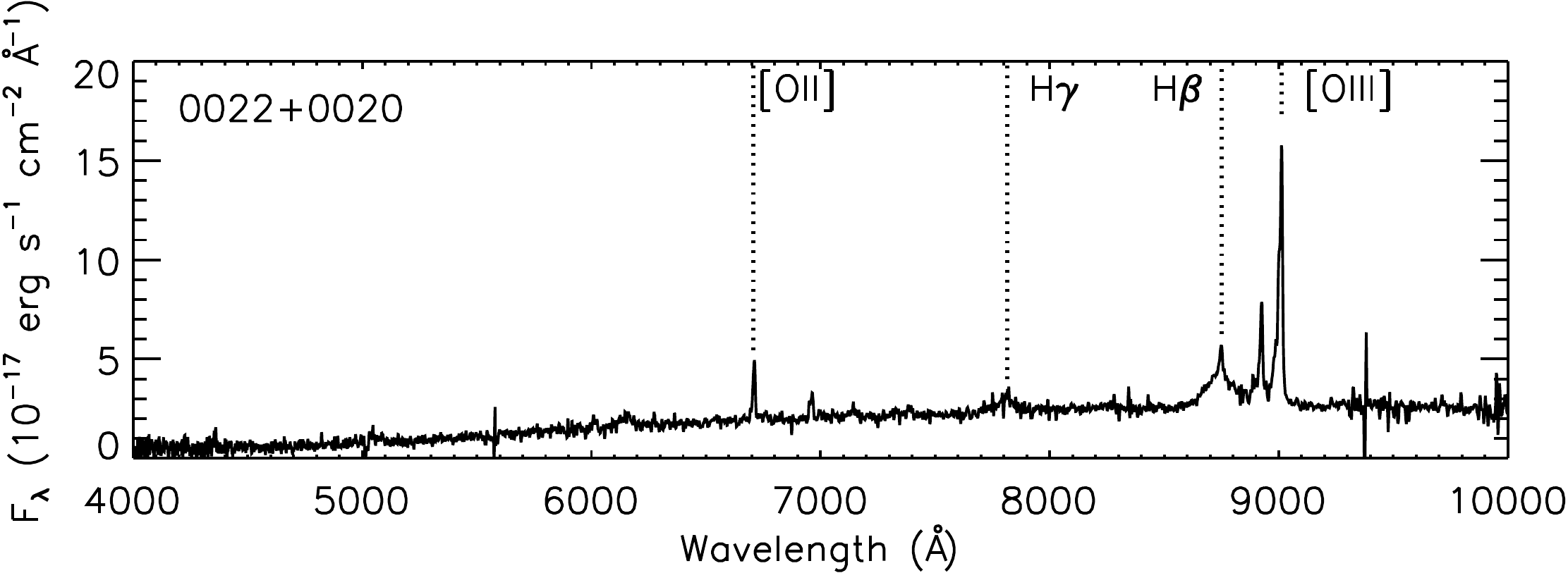}}
{\includegraphics[scale=0.4,angle=90]{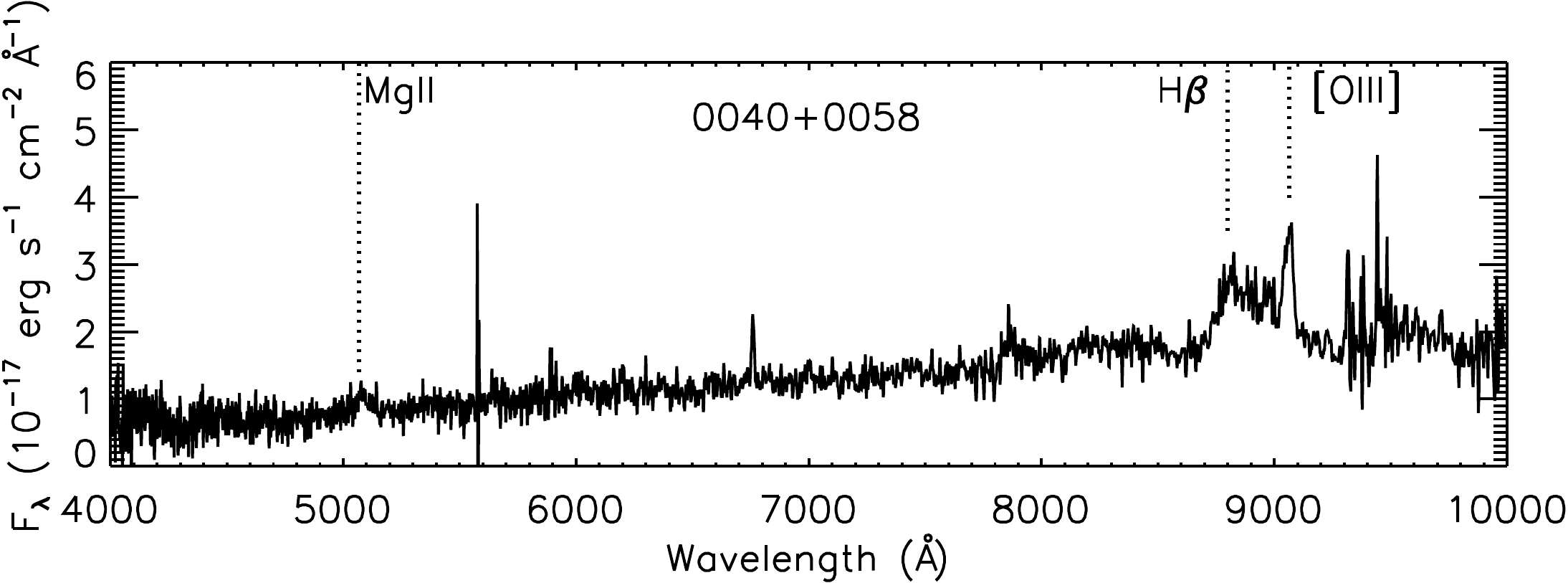}}
{\includegraphics[scale=0.4,angle=90]{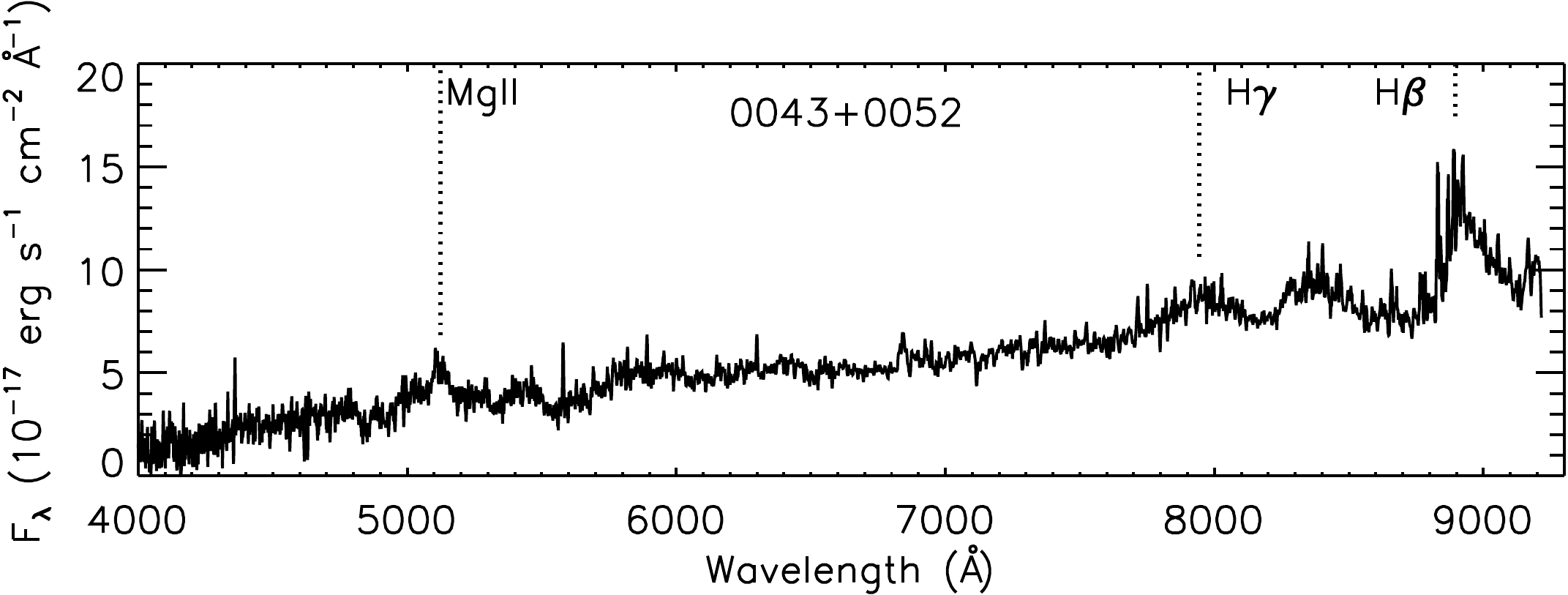}}
\caption{\label{sdss_supp}SDSS spectra of extragalactic Stripe 82X sources that meet the selection criteria of our bright NIR Stripe 82X sample, i.e., $R-K > 4$, $X/O > 0$, and $K \leq 16$.}
\end{figure}

Four Stripe 82X sources with $R-K >4$ and $X/O > 0$ have existing SDSS spectra and obey the quality control cuts and magnitude limits applied to our target list for the bright NIR sample. The optical, infrared, and X-ray properties of these sources are presented in Table \ref{sdss_samp}. As can be seen by their SDSS spectra in Figure \ref{sdss_supp}, they are all Type 1 AGN.

We also calculated BEHR-derived hardness ratios for these sources to estimate their column densities. While one source has only an upper limit on the implied column density, the other three objects have hardness ratios consistent with non-zero absorption. Two of these sources, S82X 0022+0020 and S82X 0040+0058, are detected in hard X-rays and are at sufficiently low redshift that [\ion{O}{3}] 5007 \AA\ is observed in the SDSS spectrum. Based on our fits to the optical spectra (see Section \ref{supp_sdss_fit}), we find rest-frame [\ion{O}{3}] 5007 \AA\ flux values of (2.06 $\pm$ 0.17) $\times 10^{-15}$ erg s$^{-1}$ cm$^{-2}$ and (3.2 $\pm$ 0.2) $\times 10^{-16}$ erg s$^{-1}$ cm$^{-2}$ for S82X 0022+0020 and S82X 0040+0058, respectively. With observed hard X-ray fluxes of (1.7 $\pm$ 0.2) $\times 10^{-14}$ erg s$^{-1}$ cm$^{-2}$ and (1.5 $\pm 0.2$) $\times 10^{-13}$ erg s$^{-1}$ cm$^{-2}$, we obtain log($F_{\rm 2-10keV}$/$F_{\rm [OIII]}$) values of 0.92 $\pm$ 0.06 dex and 2.68 $\pm$ 0.07 dex for S82X 0022+0020 and S82X 0040+0058. These values are consistent with the hardness ratios: S82X 0022+0020 is moderately X-ray obscured while S82X 0040+0058 is X-ray unobscured.

While this SDSS sample is optically brighter than the sources we identified with our spectroscopic campaign, their X-ray luminosities span a similar range. Below, we consider these four sources alongside the four we targeted with Palomar TripleSpec when we discuss the bright NIR $R-K$ versus $X/O$ Stripe 82X sample.

\begin{deluxetable*}{lrllllllll}
\tablewidth{0pt}
\tablecaption{\label{sdss_samp}Bright NIR $R-K$ versus $X/O$ Sample from SDSS}
\tablehead{\colhead{Stripe 82X Name} & \colhead{X-ray ID\tablenotemark{1}} & \colhead{$r$} & \colhead{$K$} & \colhead{$R-K$} & \colhead{$X/O$} & \colhead{$z$} & \colhead{$L_{\rm X,full}$\tablenotemark{2}} & \colhead{$HR$\tablenotemark{3}} & \colhead{$N_{\rm H}$\tablenotemark{4}} \\
 & & \colhead{(AB)} & \colhead{(Vega)} & \colhead{(Vega)} & & & \colhead{(erg s$^{-1}$)} & & \colhead{(10$^{22}$ cm$^{-2}$)}}

\startdata

S82X 001130.21+005751.5 & 111X   & 20.65 & 15.56 & 4.83 & 1.34 & 1.491 & 2.87$\times10^{45}$ & -0.30$^{+0.06}_{-0.06}$ & 1$^{+1}_{-0.6}$ \\

S82X 002255.06+002055.7 & 34598C & 20.51 & 15.84 & 4.25 & 0.03 & 0.799 & 3.79$\times10^{43}$ & 0.72$^{+0.12}_{-0.11}$ & 20$^{+0}_{-10}$ \\

S82X 004003.87+005853.9 & 287X   & 20.62 & 15.92 & 4.24 & 1.37 & 0.811 & 7.66$\times10^{44}$ & -0.47$^{+0.54}_{-0.53}$ & $<$3 \\

S82X 004341.18+005253.2 & 367X   & 19.45 & 14.34 & 4.65 & 0.33 & 0.828 & 2.17$\times10^{44}$ & -0.09$^{+0.14}_{-0.17}$ & 2$^{+1}_{-1.4}$ 

\enddata
\tablenotetext{1}{If the X-ray ID number is followed by a ``C'', this indicates the {\it Chandra} MSID number from the {\it Chandra} Source Catalog \citep{evans}. If the X-ray ID number is appended by an ``X'', this denotes the {\it XMM-Newton} record number introduced in the Stripe 82X survey \citep{s82x2,s82x3}.}
\tablenotetext{2}{$k$-corrected (i.e., rest-frame), non-absorption corrected luminosities.}
\tablenotetext{3}{$HR = (H-S)/(H+S)$, where $H$ ($S$) are net counts in the hard (soft) X-ray bands and were calculated with BEHR.}
\tablenotetext{4}{Gas column density ($N_{\rm H}$) implied by the $HR$ ranges.}

\end{deluxetable*}

\subsection{Radio Properties}\label{radio}
As noted above, red colors can be induced by synchrotron emission from jets along the line-of-sight that boost the $K$-band flux \citep[e.g.,][]{serjeant}. The FIRST survey covers the full Stripe 82 region \citep{helfand}, and only two of our 12 sources are detected by FIRST: S82X 0011+0057 and S82X 0302-0003. Following the prescription to calculate radio loudness from \citet{ivezic}, we first define an AB radio magnitude based on the integrated FIRST flux density at 20 cm:
\begin{equation}
  t = -2.5 {\rm log}\left(\frac{F_{\rm int}}{3631 {\rm Jy}}\right).
\end{equation}

The FIRST flux densities are 156 mJy and 0.46 mJy for S82X 0011+0057 and S82X 0302-0003, respectively, corresponding to $t$ = 10.9 and 17.2. We then calculate the radio loudness by taking the ratio of the radio and optical flux:
\begin{equation}
  R = {\rm log}\left(\frac{F_{\rm radio}}{F_{\rm optical}}\right) = 0.4\ (m_r - t),
\end{equation}
where $m_r$ is the SDSS $r$-band magnitude. We note that the $r$-band magnitude is not corrected for extinction, making our $R$ values upper limits. We find $R$ = 3.9 and 1.7 for S82X 0011+0057 and S82X 0302-0003. While S82X 0302-0003 can be classified as radio-intermediate \citep[$1 < R < 2$;][]{miller}, S82X 0011+0057 is radio loud. The prominent radio jet in this source might contribute to the $K$-band flux, which may enhance the red $R-K$ color.

\subsection{Spectral Energy Distribution Analysis}
Using any available photometric data from ultraviolet to mid-infrared wavelengths, we constructed Spectral Energy Distributions (SEDs) of these sources \citep{ananna}. The ultraviolet data are from the {\it GALEX} Medium Imaging Survey \citep{morrissey}. Due to the optical faintness of our sources, only one object, S82X 0011+0057, is detected by {\it GALEX}, and only in the near-ultraviolet band. The optical data were culled from the coadded Stripe 82 catalog from \citet{fliri}, if available,  otherwise from \citet{jiang}; one source, S82X 0141+0017, was not detected at any optical wavelength. For the NIR data, we used VHS magnitudes where available, or UKIDSS for filters that did not have a detection in VHS.\footnote{We only used both VHS and UKIDSS magnitudes if the magnitudes in filters in common between both observatories were consistent. For S82X 0303-0015, the lack of common filters between VHS and UKIDSS is due to non-coverage in the VHS $K$-band and UKIDSS $J$-band; the source was at the edge of the detector in the $J$-band in UKIDSS, precluding the UKIDSS pipeline from measuring a $J$-band magnitude. Here, we combined the UKIDSS and VHS magnitudes.} The optical and NIR magnitudes were corrected for Galactic extinction. 

In the case of non-detections, we use the upper limits reported in the various multiwavelength catalogs: $m_{\rm FUV,AB}$ = 22.6, $m_{\rm NUV,AB}$ = 22.7 \citep[{\it GALEX} 5$\sigma$ limit;][]{morrissey}; $m_{\rm u, AB}$ = 24.2, $m_{\rm g, AB}$ = 25.2, $m_{\rm r, AB}$ = 24.7, $m_{\rm i, AB}$ = 24.3, $m_{\rm z, AB}$ = 23.0 \citep[3$\sigma$ level for 50\% completeness in the SDSS co-added catalog;][]{fliri}; $m_{\rm J, AB}$ = 21.5, $m_{\rm H, AB}$ = 21.2, $m_{\rm K, AB}$ = 20.4 \citep[5$\sigma$ detection limit for point sources;][]{mcmahon}; and $F_{\rm \nu, W1}$ = 0.08 mJy, $F_{\rm \nu, W2}$ = 0.11 mJy, $F_{\rm \nu, W3}$ = 1 mJy, and $F_{\rm \nu, W4}$ = 6 mJy \citep[5$\sigma$ limit for point sources;][]{wright}. 

We fit the SEDs with AGNFitter \citep{calistro} to estimate the bolometric AGN luminosities ($L_{\rm bol}$) and reddening (E(B-V)). This algorithm employs a Bayesian Markov Chain Monte Carlo (MCMC) method, assuming a flat prior on the parameters listed in Table 1 of \citet{calistro}, and fits the following templates to the SED: accretion disk emission, which is a modified version of the \citet{richards} template, extended to wavelengths redward of 1$\mu$m assuming $F_\nu \propto \nu^{-2}$; hot dust emission from the putative torus using models from \citet{silva}; host galaxy emission using the stellar population models from \citet{bruzual}; and cold dust emission associated with star formation using templates from \citet{chary} and \citet{dale}. AGNFitter accounts for upper limits by creating a fictitious data point at half the value of the upper limit ($F_{\rm UL}$), with an error bar of $\pm$0.5 $F_{\rm UL}$, such that the upper limit data point spans the range from 0 to $F_{\rm UL}$. Inclusion of upper limits allows the MCMC sampling to accept models that lie within the bounds defined by the upper limits.

The fitted parameter space is 10-dimensional \citep[see][for details]{calistro} which is on the order of or larger than the number of photometric detections used in the fitting, so we {\it caution that our results from this exercise are approximate}. However, it is the best we are able to do with our data and does provide a sense of the bolometric AGN luminosity and reddening.

We ran AGNFitter with two burn-in sets, with 4000 steps per set and 100 chains per set: after each burn-in, the starting point in the parameter space of the subsequent MCMC chains is that of the highest likelihood of the previous chains. After the burn-in sets, the MCMC chain is run with 10,000 steps, where all sampled areas of the parameter space are used in the calculation of the posterior probability distribution functions (PDFs).

\begin{figure*}[ht]
  \centering
{\includegraphics[scale=0.5]{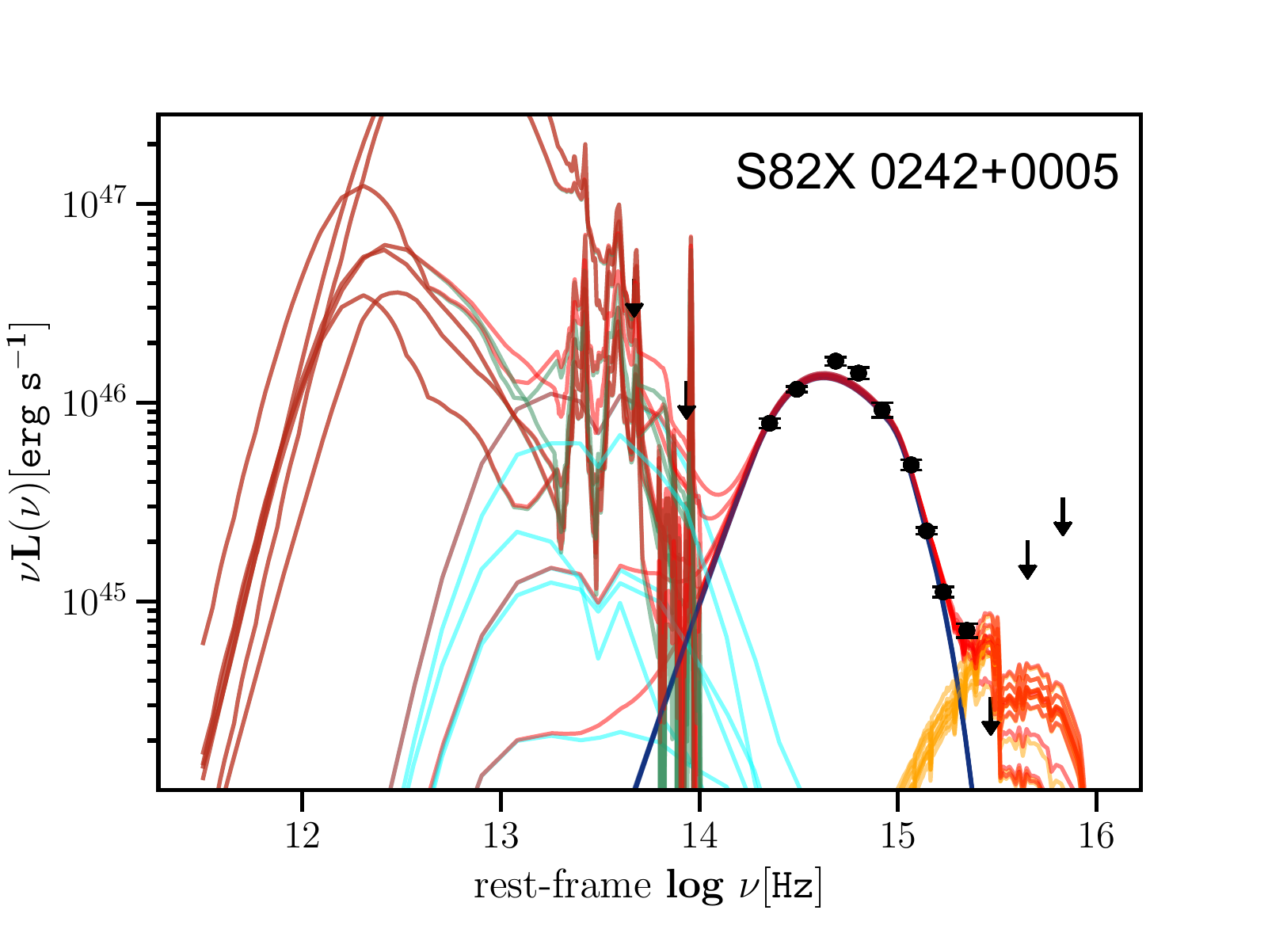}}~
{\includegraphics[scale=0.5]{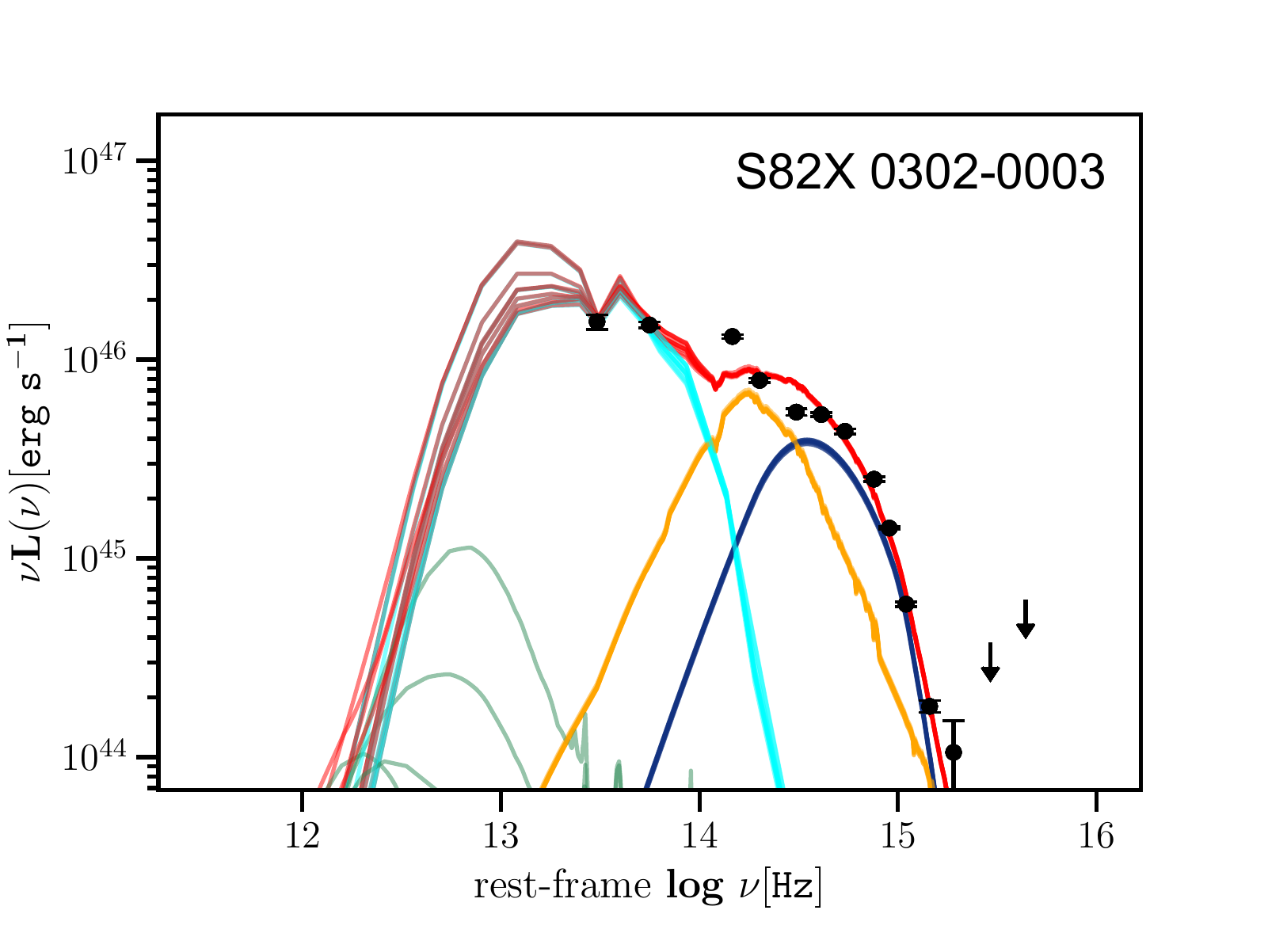}}
{\includegraphics[scale=0.5]{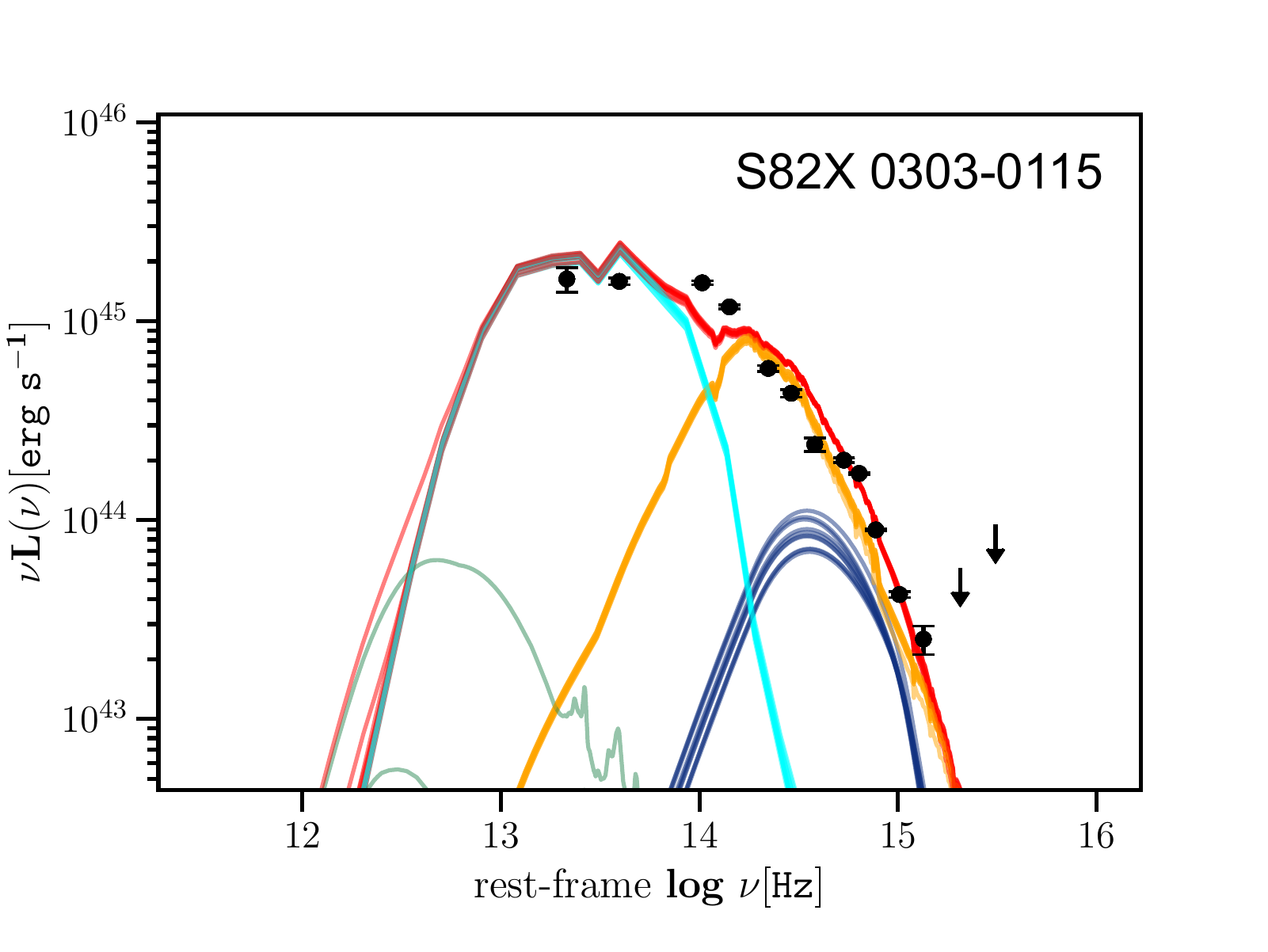}}~
{\includegraphics[scale=0.5]{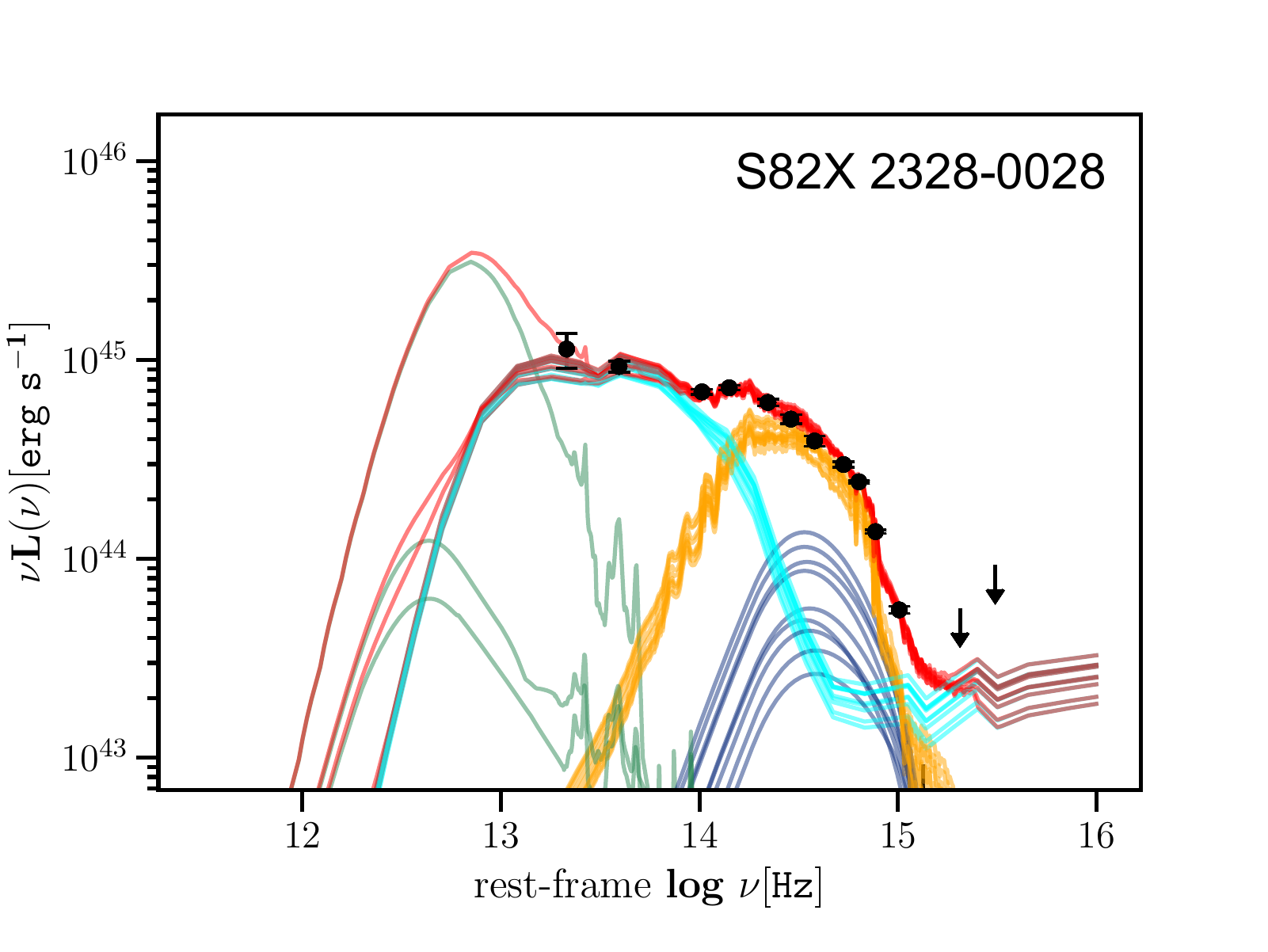}}
\caption{\label{seds_bright} SEDs of our bright NIR $R-K$ versus $X/O$ selected Stripe 82X sample fitted with AGNFitter \citep{calistro}. The {\it black data points} are the ultraviolet to mid-infrared photometric detections ({\it circles}) and upper limits ({\it arrows}) . Overplotted on the SEDs are ten random realizations from the MCMC fit of the individual templates: accretion disk emission ({\it dark blue lines}), host galaxy emission ({\it orange lines}), AGN-heated dust emission ({\it cyan lines}), and cold dust emission associated with star formation ({\it green lines}). The sum of the individual emission components, averaged over all MCMC runs, is shown by the {\it red lines}. The AGN bolometric luminosity is based on the integrated luminosity from the de-reddened accretion disk template (i.e., {\it dark blue lines}) integrated from 0.03$\mu$m to 1$\mu$m.}
\end{figure*}

\begin{figure*}[ht]
  \centering
{\includegraphics[scale=0.5]{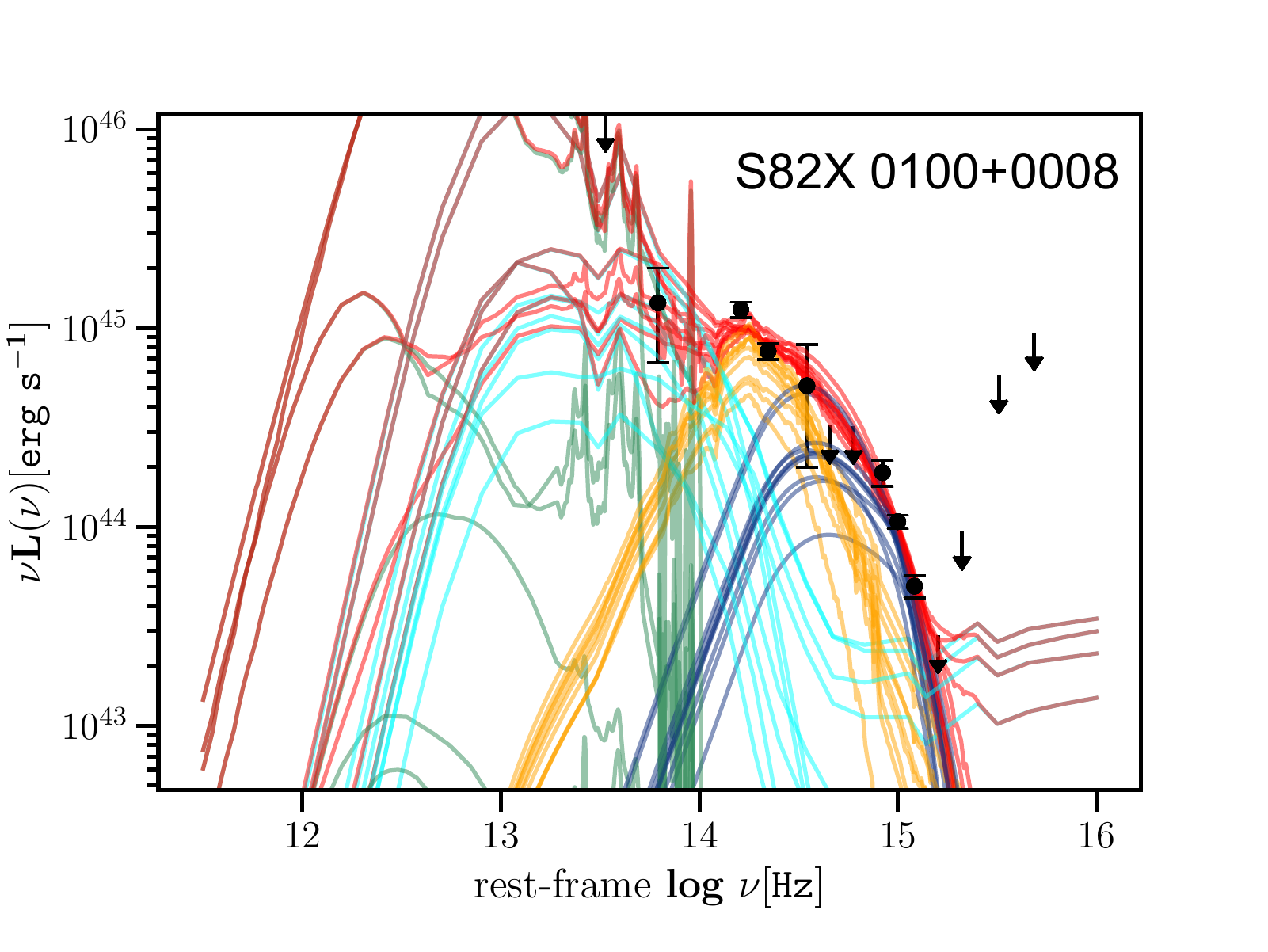}}~
{\includegraphics[scale=0.5]{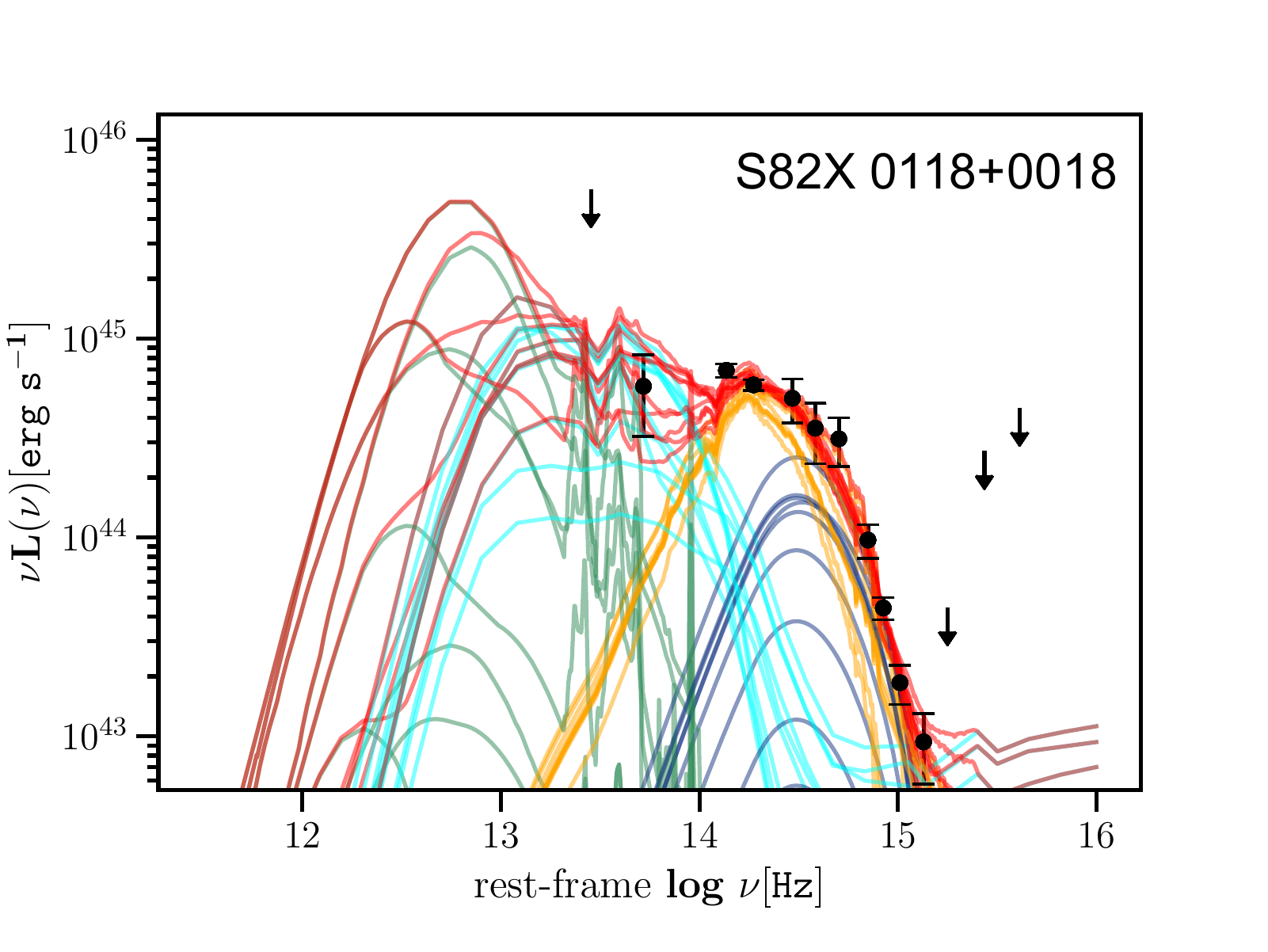}}
{\includegraphics[scale=0.5]{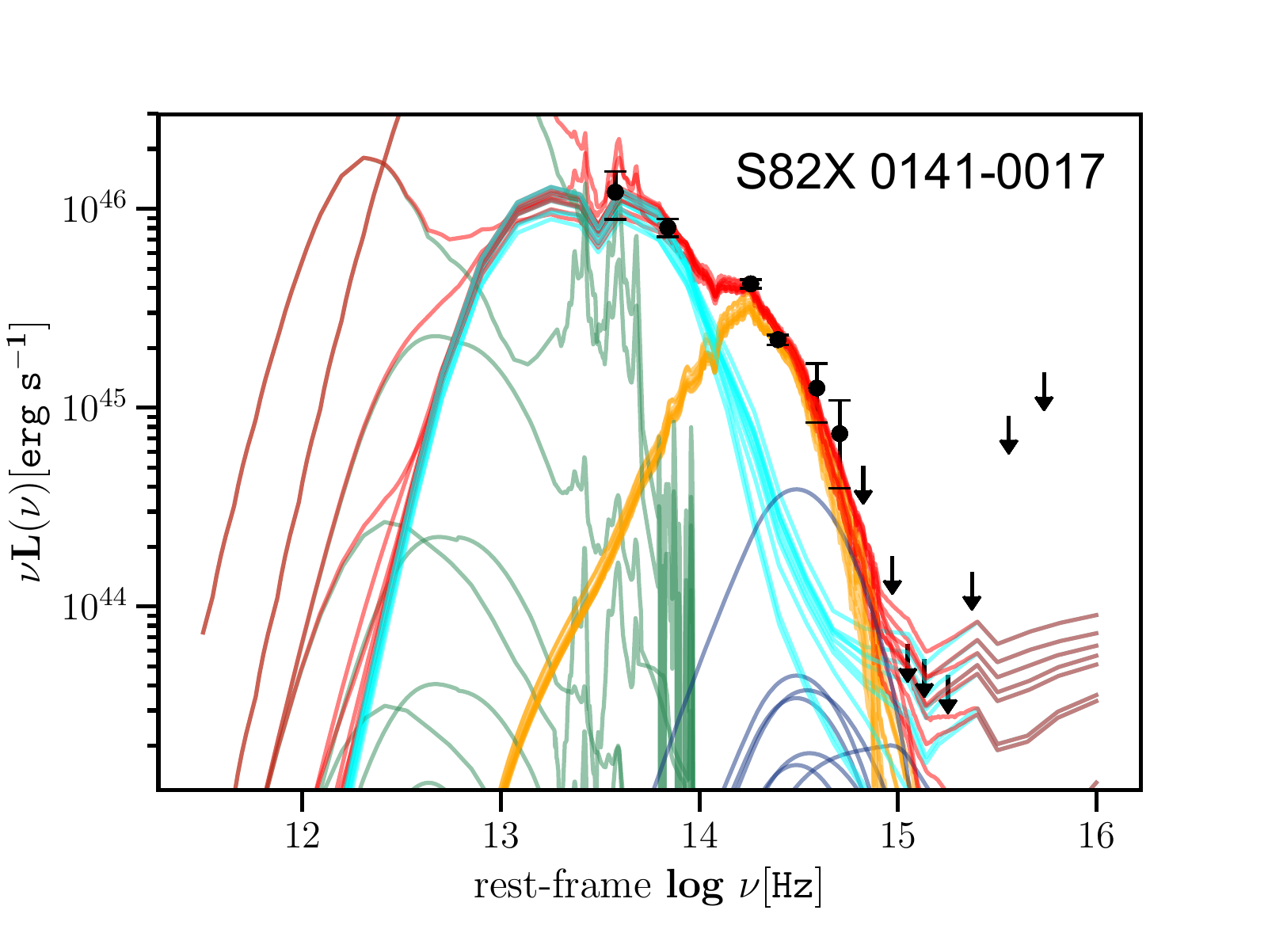}}~
{\includegraphics[scale=0.5]{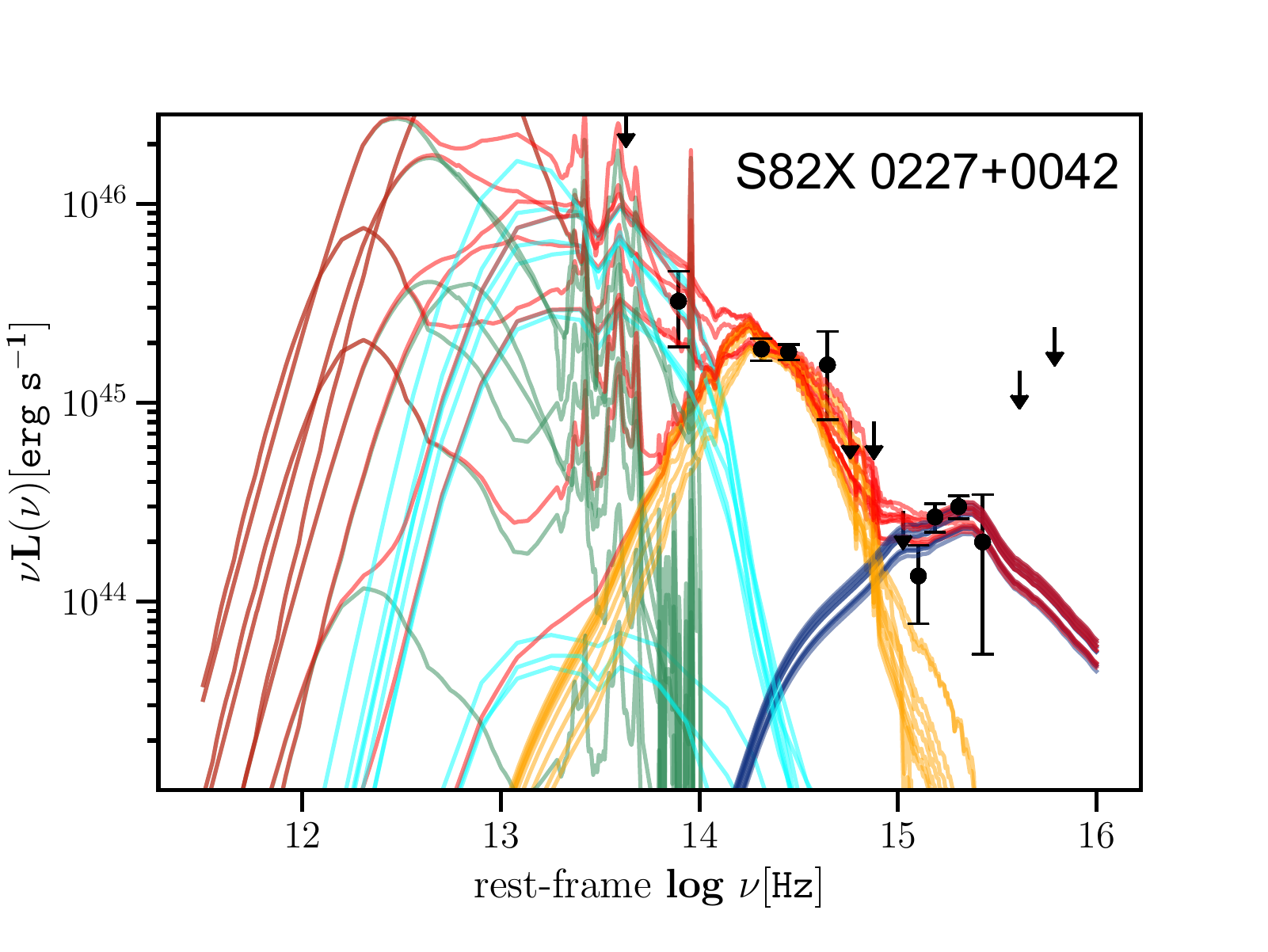}}
\caption{\label{seds_faint} SEDs of our faint NIR {\it WISE}-selected optical dropout Stripe 82X sample fitted with AGNFitter \citep{calistro}. Colors are the same as in Figure \ref{seds_bright}.}
\end{figure*}

\begin{figure*}[ht]
  \centering
{\includegraphics[scale=0.5]{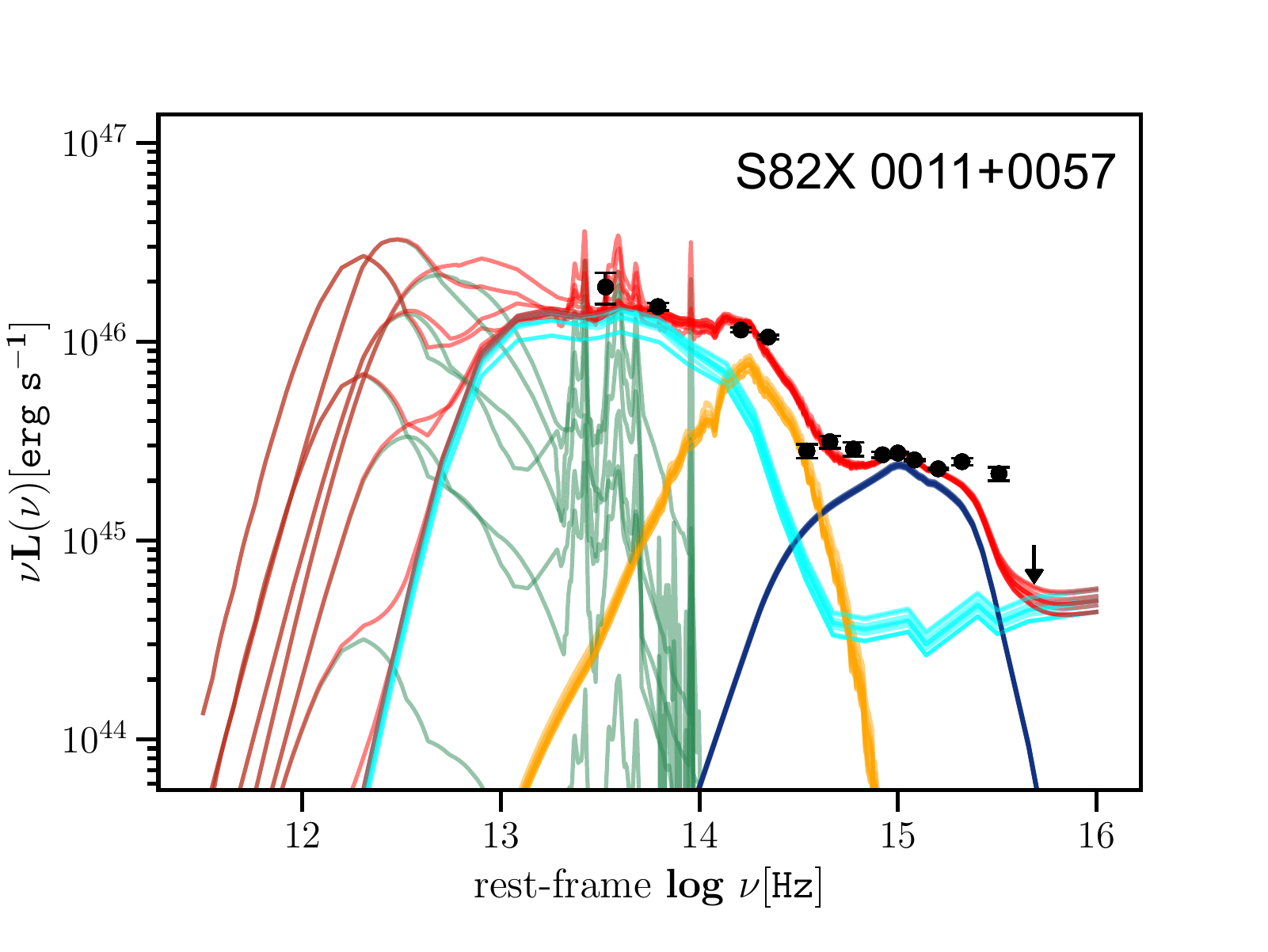}}~
{\includegraphics[scale=0.5]{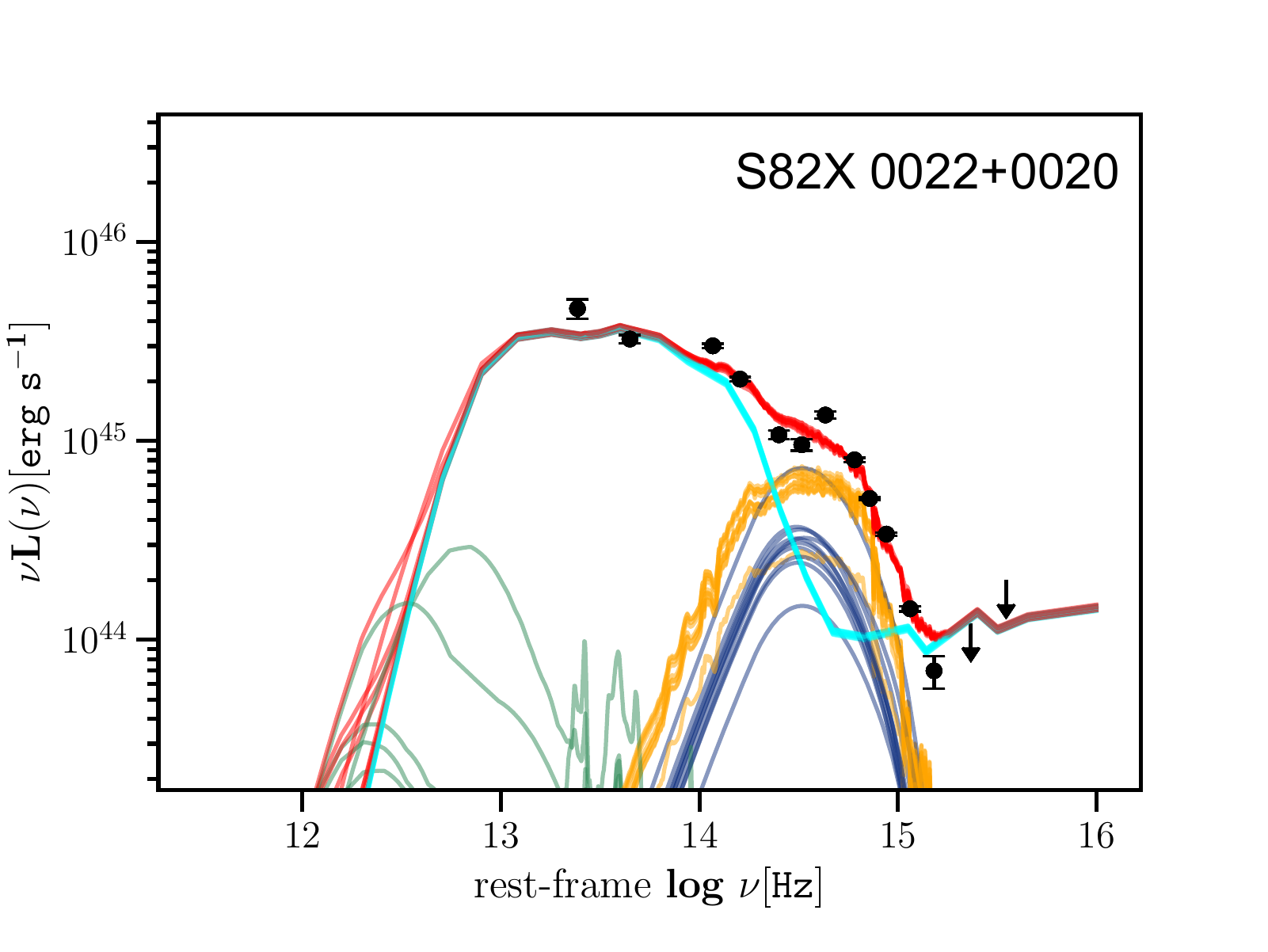}}
{\includegraphics[scale=0.5]{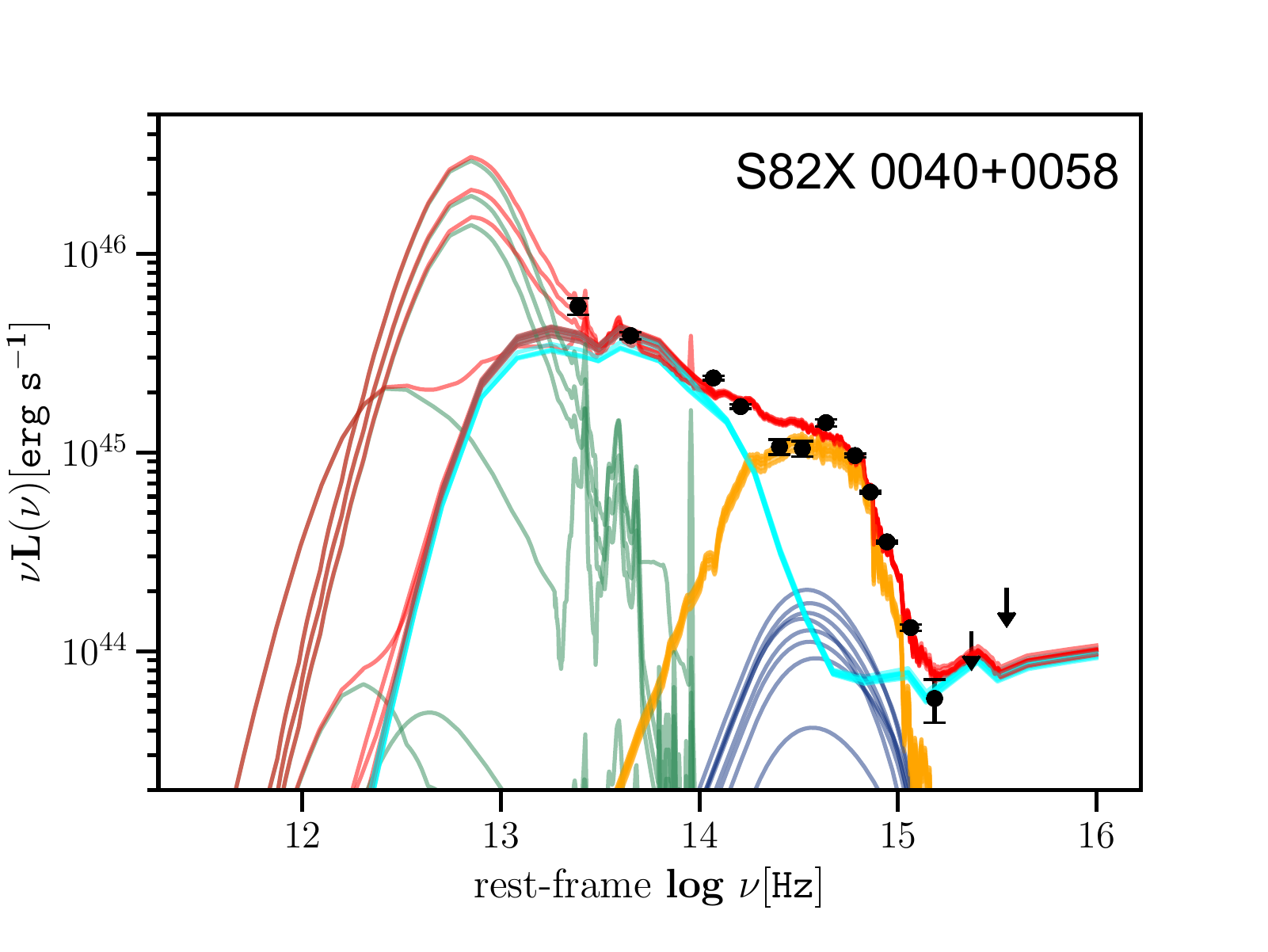}}~
{\includegraphics[scale=0.5]{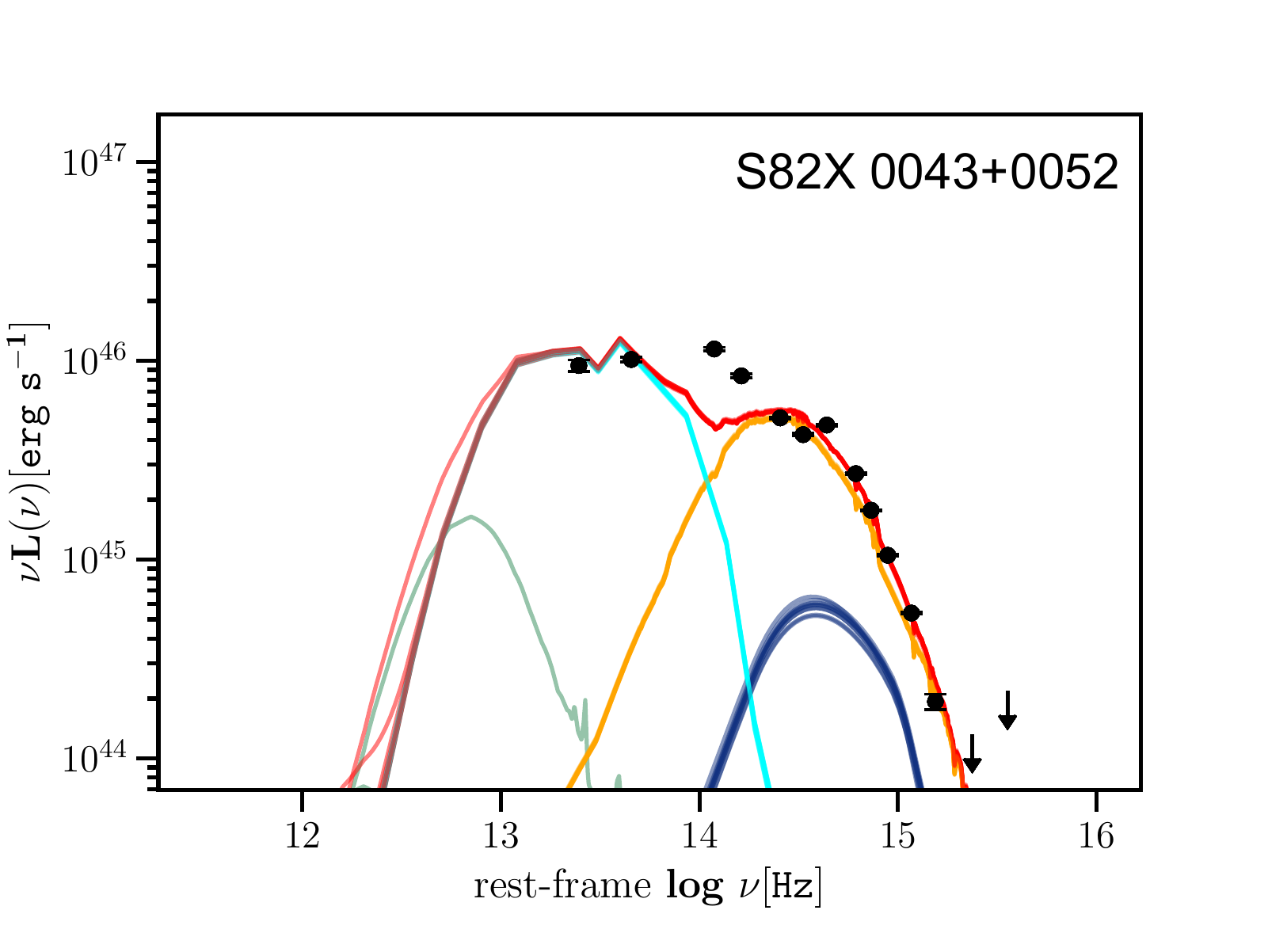}}
\caption{\label{seds_suppl} SEDs of our Stripe 82X NIR supplementary sample fitted with AGNFitter \citep{calistro}. Colors are the same as in Figure \ref{seds_bright}.}
\end{figure*}

We show the fitted SEDs in Figures \ref{seds_bright} - \ref{seds_suppl}, where the black circles represent our photometric data points. Ten random realizations from the MCMC chain are plotted. The sum of the individual templates from these realizations are shown by the solid red line, with the individual templates denoted by the other colored lines as indicated in the caption of Figure \ref{seds_bright}. In Table \ref{agn_params}, we list results from the SED fitting: the AGN and host galaxy reddening values, E(B-V)$_{\rm AGN}$ and E(B-V)$_{\rm Galaxy}$, respectively; and $L_{\rm bol}$, found by integrating the luminosity from the {\it de-reddened} accretion disk template from 0.03 $\mu$m - 1 $\mu$m. We note that since AGNFitter does not include the X-ray emission when modeling the SEDs, the AGN bolometric luminosity derived from the accretion disk template is underestimated \citep[but see][for a discussion about how the geometry of the X-ray emitting corona determines whether inclusion of $>$2 keV X-ray emission in the AGN bolometric luminosity amounts to ``double counting'' when calculating the intrinsic AGN luminosity]{krawczyk}.

To estimate black hole masses (see below), we calculate monochromatic luminosities at 5100 \AA\ ($\lambda L_{\rm 5100}$) and 3000 \AA\ ($\lambda L_{\rm 3000}$) from $L_{\rm bol}$ assuming a bolometric correction of $8.1\pm0.4$ and $5.2\pm0.2$, respectively \citep{runnoe}, and list these values, where appropriate, in Table \ref{agn_params}. The reported values derived from the SED fitting represent the median of the PDFs, with the lower and higher error bars indicating the bounds for the 16th and 84th percentiles of the PDFs.

\begin{deluxetable*}{llllllll}
\tablewidth{0pt}
\tablecaption{\label{agn_params}AGN Parameters Derived from SED \& Spectral Fitting}
\tablehead{\colhead{Stripe 82X Name} & \colhead{E(B-V)$_{\rm AGN}$} & \colhead{E(B-V)$_{\rm Galaxy}$} & \colhead{Log($L_{\rm bol}$)\tablenotemark{1}} &  \colhead{Log ($\lambda L_{\rm 5100}$)\tablenotemark{2}} &  \colhead{Log ($\lambda L_{\rm 3000}$)\tablenotemark{3}} &  \colhead{Log($M_{\rm BH}$)} & \colhead{$\lambda_{\rm Edd}$} \\
 & & & \colhead{(erg s$^{-1}$)} & \colhead{(erg s$^{-1}$)} & \colhead{(erg s$^{-1}$)} & \colhead{(M$_{\sun}$)}  }

\startdata

\multicolumn{8}{c}{\textit{Bright NIR $R-K$ versus $X/O$ Selected Sample}}\\

S82X 0242+0005   & 0.45$\pm$0.02 & -0.03$\pm$0.05            & 47.24$\pm$0.01 & 46.34$\pm$0.02 & \nodata & 9.62$^{+0.05}_{-0.06}$ & 0.32$\pm$0.04 \\

S82X 0302$-$0003 & 0.70$\pm$0.02 & 1.52$^{+0.33}_{-0.40}$       & 46.85$\pm$0.01 & 45.94$\pm$0.02 & \nodata & 9.38$\pm$0.04 & 0.23$\pm$0.02 \\

S82X 0303$-$0115 & 0.69$^{+0.03}_{-0.04}$ & 1.48$\pm0.38$       & 45.17$^{+0.05}_{-0.08}$ & 44.26$^{+0.06}_{-0.07}$ & \nodata & 7.44$^{+0.06}_{-0.07}$ & 0.41$\pm$0.08 \\

S82X 2328$-$0028 & 0.71$^{+0.20}_{-0.12}$ & 0.40$^{+0.11}_{-0.06}$ & 45.08$^{+0.23}_{-0.42}$ & 44.17$^{+0.22}_{-0.48}$ & \nodata & 7.34$^{+0.13}_{-0.19}$ & 0.42$\pm$0.32 \\

\multicolumn{8}{c}{\textit{Faint NIR {\it WISE}-Selected Optical Dropout Sample}}\\

S82X 0100+0008   & 0.54$\pm$0.09 & 1.50$^{+0.38}_{-0.44}$       & 45.52$\pm$0.22      & 44.61$^{+0.18}_{-0.32}$ & \nodata & 8.43$^{+0.25}_{-0.62}$ & 0.10$\pm$0.09 \\

S82X 0118+0018   & 1.18$^{+0.54}_{-0.41}$ & 1.20$^{+0.55}_{-0.49}$ & 45.41$^{+0.24}_{-0.99}$ & 44.50$^{+0.26}_{-0.76}$ & \nodata & 7.99$^{+0.18}_{-0.31}$ & \nodata \\

S82X 0141$-$0017\tablenotemark{4} & 1.21$^{+0.51}_{-0.65}$ & 1.38$^{+0.41}_{-0.31}$ & 44.59$^{+0.67}_{-0.44}$ & \nodata & \nodata & \nodata & \nodata \\

S82X 0227+0042\tablenotemark{4}   & -0.03$\pm$0.05 & 0.97$^{+0.70}_{-0.52}$      & 44.70$^{+0.06}_{-0.09}$ & \nodata & \nodata & \nodata & \nodata \\

\multicolumn{8}{c}{\textit{Bright NIR $R-K$ versus $X/O$ Selected Sample from SDSS}} \\

S82X 0011+0057\tablenotemark{5} & 0.10$\pm0.02$ & 1.43$^{+0.39}_{-0.37}$       & 46.02$\pm$0.01 & \nodata & 45.31$\pm$0.02 & 8.85$^{+0.13}_{-0.18}$ & 0.11$\pm$0.04 \\

S82X 0022+0020 & 0.85$^{+0.47}_{-0.07}$ & 0.03$^{+0.04}_{-0.08}$ & 45.85$^{+0.15}_{-0.34}$ & 44.94$^{+0.17}_{-0.29}$ & \nodata & 8.75$^{+0.13}_{-0.18}$ & 0.10$\pm$0.06 \\

S82X 0040+0058\tablenotemark{6} & 0.63$^{+0.09}_{-0.13}$ & -0.04$\pm$0.04     & 45.12$^{+0.37}_{-0.43}$  & \nodata & 44.41$^{+0.30}_{-1.72}$ &  $<$7.3 & \nodata \\

S82X 0043+0052\tablenotemark{5} & 0.55$\pm$0.02 & 1.42$^{+0.39}_{-0.32}$      & 45.93$^{+0.04}_{-0.03}$  & \nodata & 45.21$\pm$0.04 & 8.84$^{+0.10}_{-0.13}$ & 0.09$\pm$0.03

\enddata
\tablenotetext{1}{$L_{\rm bol}$ is the AGN bolometric luminosity found by decomposing the SED in AGNFitter and integrating the {\it de-reddened} accretion disk luminosity from 0.03$\mu$m - 1.0$\mu$m.}
\tablenotetext{2}{$\lambda L_{\rm 5100}$ is the monochromatic continuum luminosity at 5100 \AA, calculated from $L_{\rm bol}$ and assuming a bolometric correction of 8.1 $\pm$ 0.4 \citep{runnoe}, that we use to estimate $M_{\rm BH}$ for sources where H$\alpha$ or H$\beta$ is detected (and in the case of H$\beta$, not blended with the [\ion{O}{3}] doublet).}
\tablenotetext{3}{$\lambda L_{\rm 3000}$ is the monochromatic continuum luminosity at 3000 \AA, calculated from $L_{\rm bol}$ and assuming a bolometric correction of 5.2 $\pm$ 0.2 \citep{runnoe}, that we use to estimate $M_{\rm BH}$ along with the \ion{Mg}{2} FWHM for sources where H$\alpha$ and H$\beta$ are not detected or where H$\beta$ is blended.}
\tablenotetext{4}{$L_{\rm bol}$ is on the order of or lower than the observed X-ray luminosity, indicating potential errors in the AGN and galaxy decomposition in the SED fitting. We therefore refrain from calculating $M_{\rm BH}$ and $\lambda_{\rm Edd}$, and caution that the E(B-V) values may be unreliable.}
\tablenotetext{5}{$M_{\rm BH}$ is estimated using the \ion{Mg}{2} FWHM reported in \citet{shen2011} and $\lambda L_{\rm 3000}$ calculated from our SED decomposition (i.e., $\lambda L_{\rm 3000} = L_{\rm bol}/(5.2 \pm 0.2)$).}
\tablenotetext{6}{Due to uncertainties in $\lambda L_{\rm 3000}$ and the \ion{Mg}{2} FWHM, we report the 3$\sigma$ upper limit on $M_{\rm BH}$.}

\end{deluxetable*}

\subsection{Black Hole Masses and Eddington Ratios}
We emphasize that due to the limited number of photometric detections used in the SED decomposition, the monochromatic and bolometric luminosities are uncertain, which propagate to uncertainties in the estimated black hole masses ($M_{\rm BH}$) and implied accretion rate as measured by the Eddington parameter ($\lambda_{\rm Edd} = L_{\rm bol}/L_{\rm Edd}$). We estimate $M_{\rm BH}$ and $\lambda_{\rm Edd}$ using the best available data, but caution that these values should be considered approximate.

\subsubsection{Targeted Sources}
For the sources that we targeted with Palomar, Keck, and Gemini, we use the estimated de-reddened $\lambda L_{\rm 5100}$ and measured H$\alpha$ FWHM values to derive $M_{\rm BH}$ using:
\begin{equation}
  \begin{split}
  M_{\rm BH} = (9.7 \pm 0.5) \times 10^6 \left [\frac{{\rm FWHM (H\alpha)}}{1000\ {\rm km\ s^{-1}}}\right]^{2.06 \pm 0.06}
  \times \\
  \left [\frac{\lambda L_{\rm 5100}}{10^{44}\ {\rm erg\ s^{-1}}} \right ]^{0.519 \pm 0.07} M_{\sun},
  \end{split}
\end{equation}
from \citet{greene}. For the two sources where two Gaussian components were needed to fit the H$\alpha$ profile (S82X 0242+0005 and S82X 0302-0003), we use the broader H$\alpha$ FWHM to estimate $M_{\rm BH}$ since this component arises from gas closer to the black hole.

From these $M_{\rm BH}$ values, we estimate the Eddington luminosity \citep[$L_{\rm Edd} = 1.3 \times 10^{38} M_{\rm BH}/M_{\sun}$ erg s$^{-1}$;][]{frank} and $\lambda_{\rm Edd}$. These values are listed in Table \ref{agn_params}. The errors represent the propagation of the statistical measurement errors of the individual parameters and the errors associated with the bolometric corrections and virial $M_{\rm BH}$ relations. We note that comparisons of black hole masses derived via single epoch measurements, as calculated here, with those determined from reverberation mapping studies show a sample dispersion of $\sim$0.5 dex \citep[e.g.,][]{vestergaard}, which is an additional uncertainty to our $M_{\rm BH}$ values beyond the formal errors that we report.

We caution that the view to the broad line region is likely obscured, such that we are not getting an unbiased view of the gas kinematics near the black hole. If we are indeed viewing just the outer photosphere of the broad line region, the FWHM of H$\alpha$, and other broad emission lines we use below to calculate $M_{\rm BH}$, are systematically lower than those observed in unobscured AGN that were used to derive virial relations to calculate $M_{\rm BH}$. Our estimated black hole masses may thus be lower limits to the true value.

\subsubsection{\label{supp_sdss_fit} SDSS Sample}
Since H$\alpha$ is not covered in the optical spectra of the sources in the bright NIR $R-K$ SDSS sample, we estimate black hole masses, and associated Eddington ratios, using the H$\beta$ or \ion{Mg}{2} emission lines. Two of these sources have published black hole masses in the \citet{shen2011} catalog (S82X 0011+0057 and S82X 0043+0052) which were calculated using the FWHM of the \ion{Mg}{2} line and $\lambda L_{\rm 3000}$. For consistency with our targeted sample, we use the $\lambda L_{\rm 3000}$ we calculated from our SED fitting ($\lambda L_{\rm 3000}$ = $ L_{\rm Bol}$/(5.2$\pm$0.2)) along with the reported \ion{Mg}{2} FWHM in \citet{shen2011} to estimate $M_{\rm BH}$ given the relation published in \citet{trakhtenbrot}:

\begin{equation}
  \begin{split}
  M_{\rm BH} = 5.6 \times 10^{6} \left [\frac{{\rm FWHM (MgII)}}{1000\ {\rm km\ s^{-1}}}\right]^{2}
  \times \\
  \left [\frac{\lambda L_{\rm 3000}}{10^{44}\ {\rm erg\ s^{-1}}} \right ]^{0.62} {\rm M_{\sun}}.
  \end{split}
\end{equation}
The black hole masses and Eddington ratios are reported in Table \ref{agn_params}. 

For the remaining two sources, we fitted the SDSS spectra using the IRAF package \textsc{specfit} \citep{kriss} to obtain emission line FWHMs. This routine uses a $\chi^2$ minimization technique to find the best fit to the input model parameters, which consists of: 1) AGN powerlaw continuum, 2) star formation templates that span an age range from 56 Myr to 10 Gyr (S. Charlot \& G. Bruzual, private communication), 3) a \citet{cardelli} dust extinction that attenuates the AGN power law continuum with $R$=3.1, and 4) Gaussian components to fit emission lines. The spectra were corrected for Galactic reddening, shifted to the rest-frame, and the flux was multiplied by $(1+z)$ to preserve the observed integrated line flux. During the fitting procedure, all narrow emission lines are forced to have the same FWHM, and the [\ion{O}{3}] 4959 \AA\ intensity was fixed to a third of the [\ion{O}{3}] 5007 \AA\ flux.

\begin{figure*}[ht]
  \centering
      {\includegraphics[scale=0.65,angle=90]{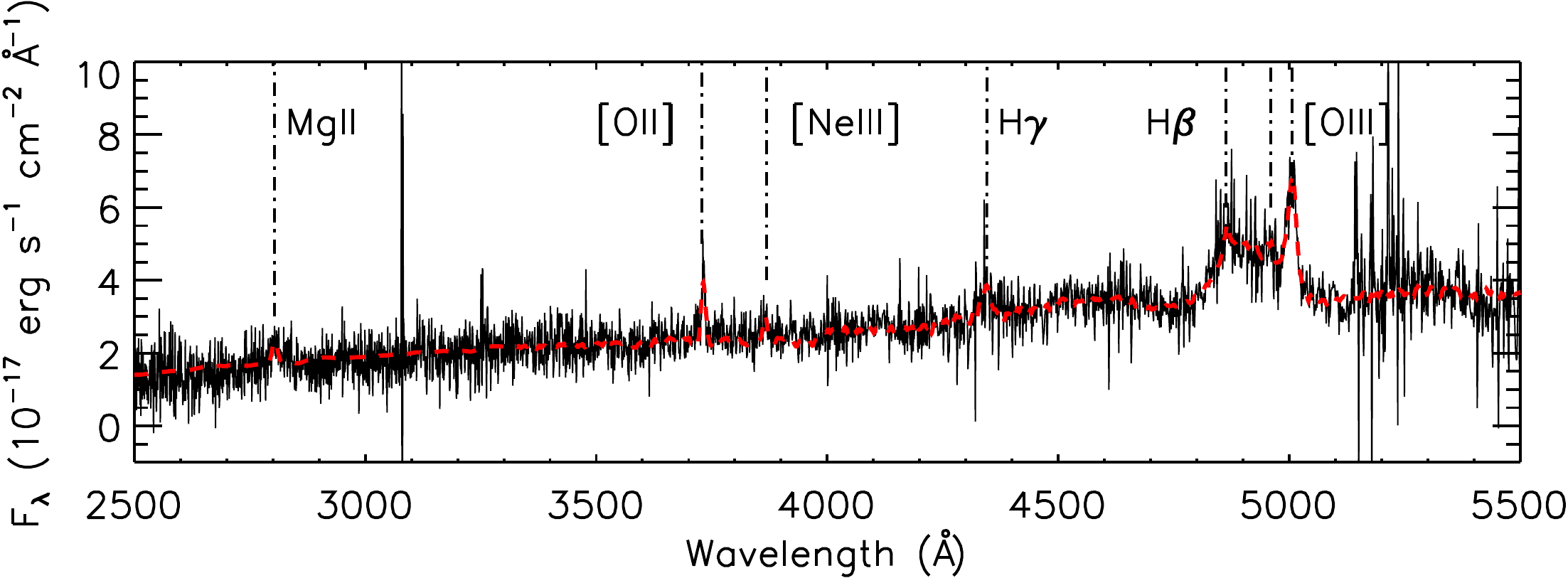}}
      {\includegraphics[scale=0.4]{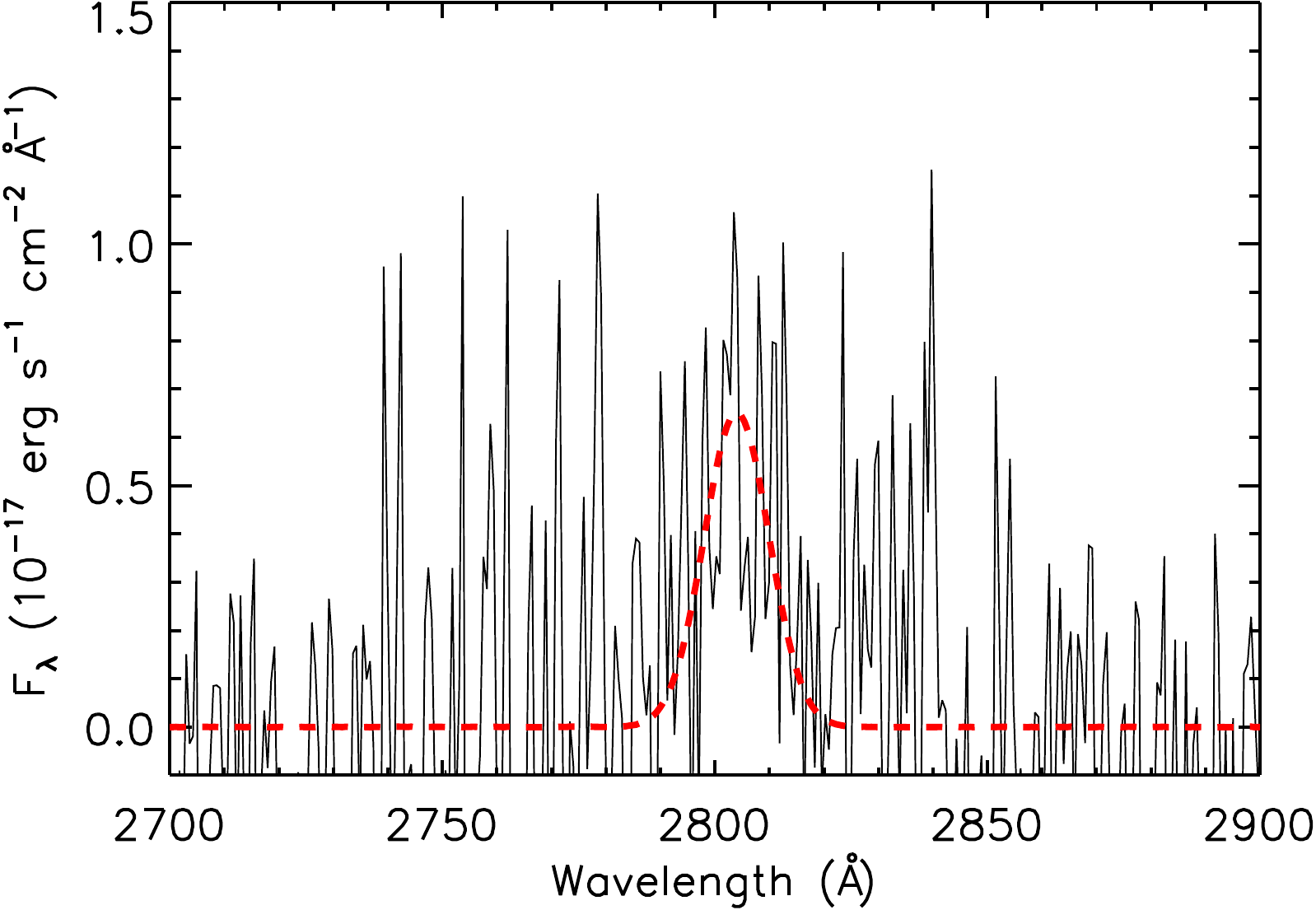}}~
      {\includegraphics[scale=0.4]{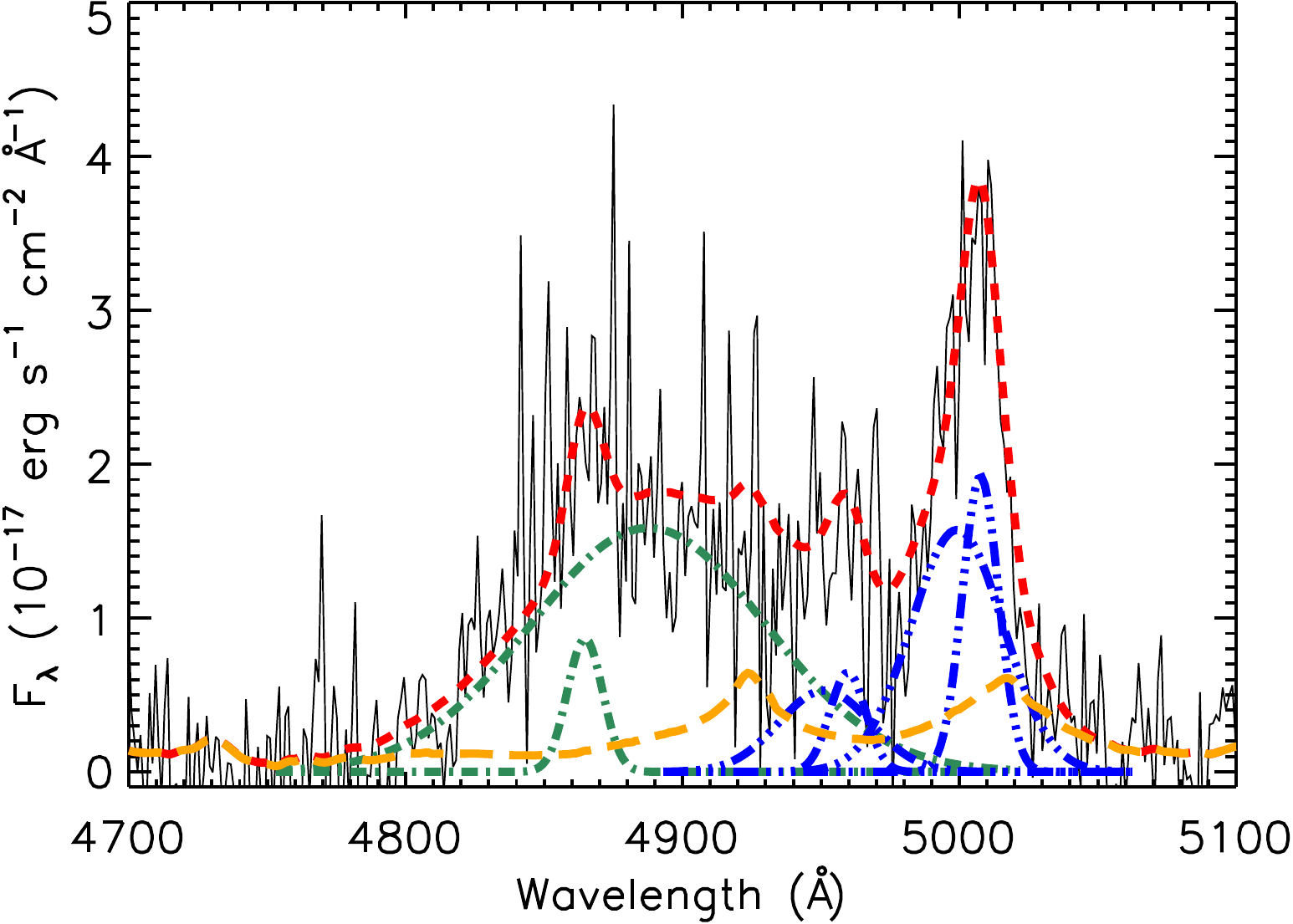}}
      \caption{\label{0040_spec} {\it Top}: Rest-frame SDSS spectrum of S82X 0040+0058 with our best fit model from IRAF \textsc{specfit} overplotted ({\it red dashed line}). Fitted emission lines are marked. {\it Bottom}: Continuum-subtracted spectra around ({\it left}) the \ion{Mg}{2} emission line, which we use to derive $M_{\rm BH}$. The H$\beta$-[\ion{O}{3}] complex is shown on the {\it right}, where we model the emission lines with an \ion{Fe}{2} optical template \citep[{\it long dashed orange lines};][]{boronson} and Gaussian profiles for the narrow and broad, redshifted ($\Delta v = 1400$ km s$^{1}$) H$\beta$ lines ({\it green dot-dash lines}), and narrow and broad, blueshifted ($\Delta v = -500$ km s$^{-1}$) [\ion{O}{3}] lines ({\it dot-dot-dot dash blue line}). The {\it red dashed line} shows the sum of these emission features.}
\end{figure*}

Since H$\beta$ is blended with the [\ion{O}{3}] doublet and \ion{Fe}{2} emission in S82X 0040+0058, we use the \ion{Mg}{2} FWHM (1500$\pm$500 km s$^{-1}$; see Figure \ref{0040_spec}) and the virial relation above to estimate $M_{\rm BH}$. However, due to the large errors in $\lambda L_{\rm 3000}$ and \ion{Mg}{2} FWHM, we are only able to estimate a 3$\sigma$ upper limit on the black hole mass.

\begin{figure*}[ht]
  \centering
      {\includegraphics[scale=0.65,angle=90]{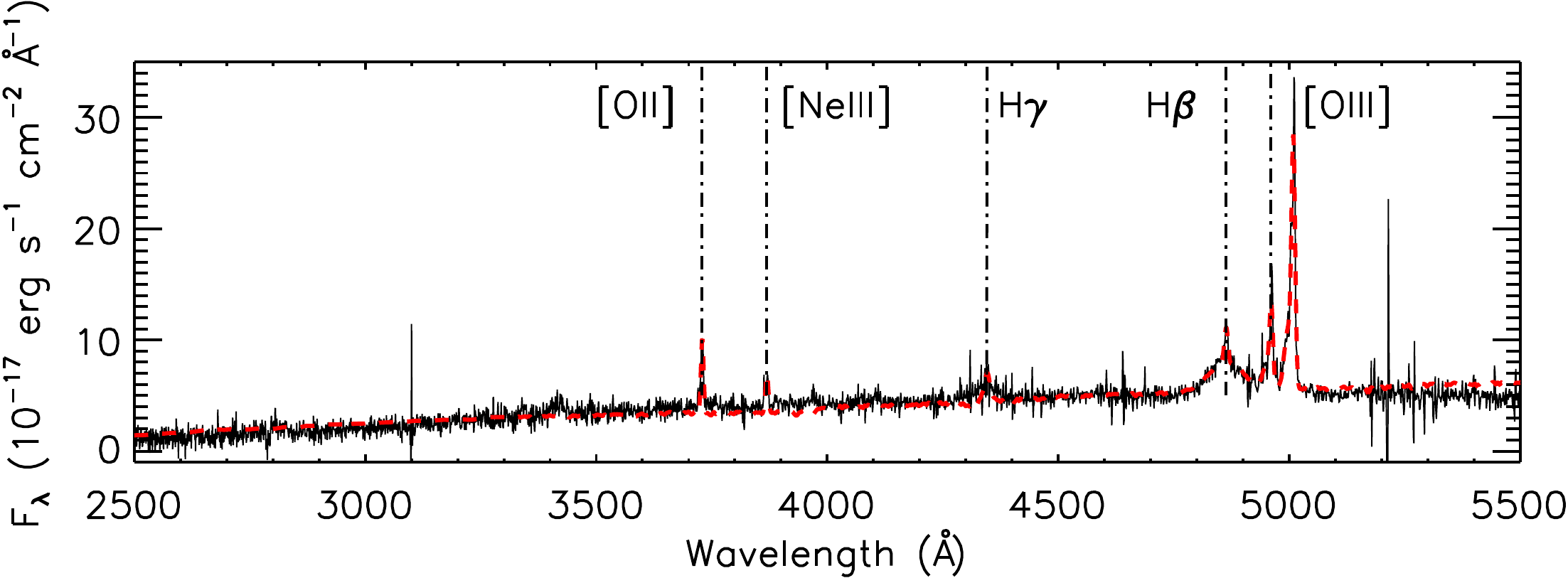}}
      {\includegraphics[scale=0.4]{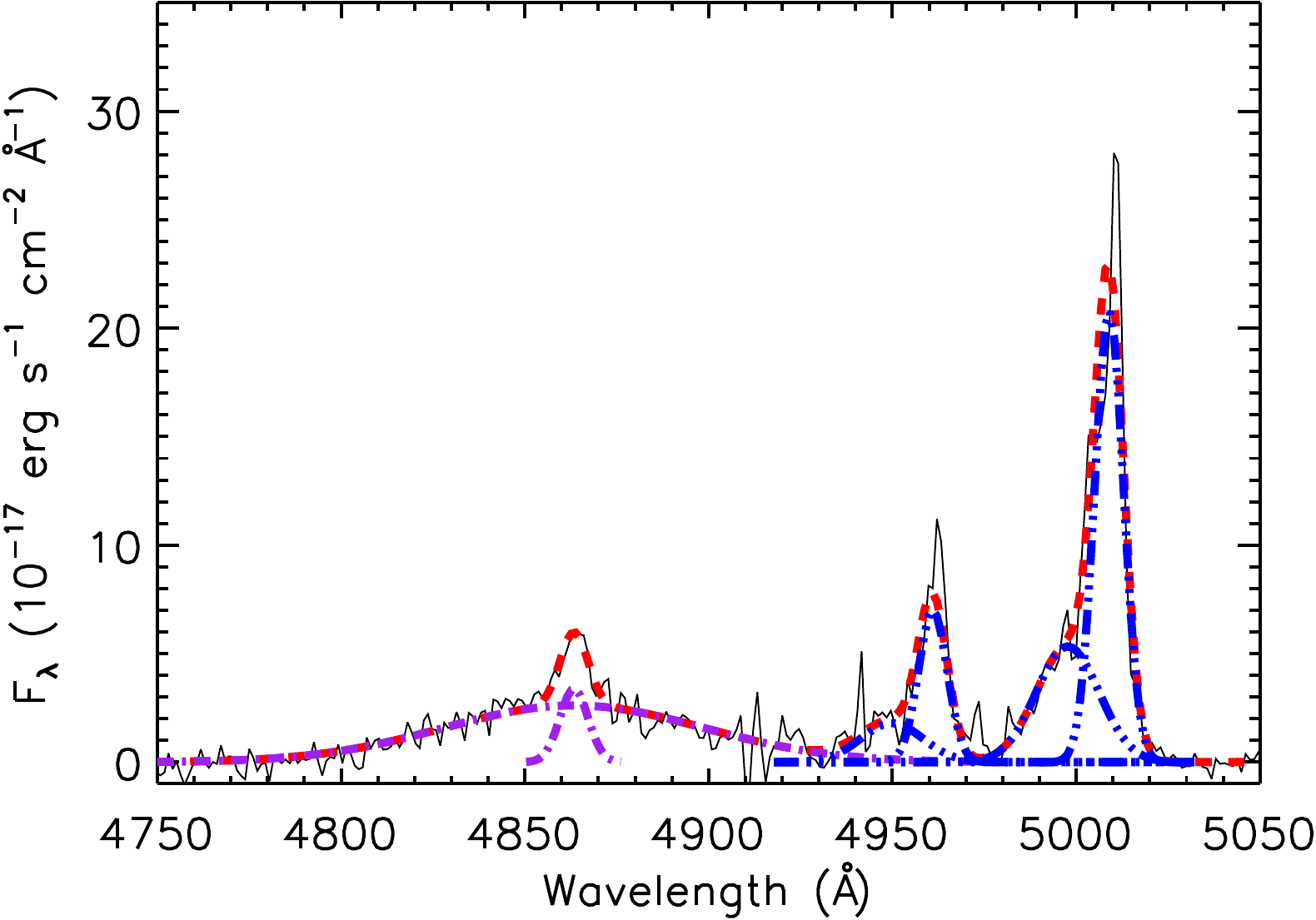}}
      \caption{\label{0022_spec} {\it Top}: Rest-frame SDSS spectrum of S82X 0022+0020 with our best fit model from IRAF \textsc{specfit} overplotted (red dashed line). Fitted emission lines are marked. {\it Bottom}: Close-up of the H$\beta$ and [\ion{O}{3}] complex for S82X 0022+0020. Here, the continuum (from AGN and host galaxy) has been subtracted off. Overplotted are the sum of the fitted emission lines ({\it red dashed line}): broad and narrow H$\beta$ emission lines ({\it dot-dash purple line}) and [\ion{O}{3}] emission ({\it dot-dot-dot-dash blue line}), including the narrow and broad broad (FWHM = 1100$\pm$100 km s$^{-1}$), blueshifted ($\Delta v = -740\pm50$ km s$^{-1}$) [\ion{O}{3}] components. This blue wing to the [\ion{O}{3}] doublet is likely a signature of an AGN outflow.}
\end{figure*}

We use the H$\beta$ FWHM to derive a black hole mass for S82X 0022+0020. With a fitted FWHM of 5100$\pm$300 km s$^{-1}$ (see Figure \ref{0022_spec}, top), and using:
\begin{equation}
  \begin{split}
  M_{\rm BH} = 1.05\times10^{8} \left [\frac{{\rm FWHM (H\beta)}}{1000\ {\rm km\ s^{-1}}}\right]^{2}
  \times \\
  \left [\frac{\lambda L_{\rm 5100}}{10^{46}\ {\rm erg\ s^{-1}}} \right ]^{0.65} {\rm M_{\sun}}
  \end{split}
\end{equation}
from \citet{trakhtenbrot}, we find $M_{\rm BH}$ = 8.75$^{+0.13}_{-0.18}$ M$_{\sun}$, with an associated $\lambda_{\rm Edd}$ of 0.10$\pm$0.06.

\citet{trakhtenbrot} do not provide formal errors on the $M_{\rm BH}$-\ion{Mg}{2} or $M_{\rm BH}$-H$\beta$ virial relations. Typical standard deviations in the samples used for their calibrations range from $\sim$0.13 - 0.15 dex. When estimating black hole masses using the \ion{Mg}{2} and H$\beta$ emission FWHMs, we only propagated formal errors on the fit parameters and bolometric corrections, and note that there is likely an additional uncertainty of up to $\sim$0.15 dex as well as an $\sim$0.5 dex uncertainty associated with single-epoch black hole mass measurments \citep{vestergaard}.

\subsection{Absorption Line System \& Asymmetric Line Profiles: Indications of Outflows?}\label{outflows}
Half of the sources from the bright NIR $R-K$ versus $X/O$ selected sample have spectroscopic signatures of narrow line region kinematics, with either absorption line troughs or broadend [\ion{O}{3}] emission: S82X 0043+0052 (Figure \ref{sdss_supp}, bottom), S82X 0022+0020 (Figure \ref{0022_spec}), S82X 0040+0058 (Figure \ref{0040_spec}), and S82X 0242+0005 (Figure \ref{oiii_fits}). None of the spectra from the faint NIR {\it WISE}-selected optical dropout sample show any sign of outflowing gas, though this sample is only $\sim12$\% complete.

S82X 0043+0052 was identified as a \ion{Mg}{2} quasar narrow absorption line (FWHM $\leq$ 500 km s$^{-1}$) system in \citet{lundgren}. Such absorption line systems can be associated with quasar outflows or from gas within the quasar environment \citep{weymann,vandenberk}. For the remaining three sources, we found asymmetries in the [\ion{O}{3}] 5007 \AA\ line from our own fits to the spectra, as detailed below and summarized in Table \ref{outflow_fit}. In all cases, the widths of the [\ion{O}{3}] doublet lines were tied together, the flux of the [\ion{O}{3}] 4959 \AA\ line was fixed to 1/3 of the [\ion{O}{3}] 5007 \AA\ line, and the central wavelength of the [\ion{O}{3}] 4959 \AA\ line was fixed to 0.99 of the [\ion{O}{3}] 5007 \AA\ line.

In S82X 0040+0058, H$\beta$, \ion{Fe}{2} emission, and the [\ion{O}{3}] doublet are blended (Figure \ref{0040_spec}). To fit the spectrum, we include an optical \ion{Fe}{2} emission template \citep{boronson} as well as broad components to the [\ion{O}{3}] doublet. The broad H$\beta$ emission (FWHM = 6100$\pm$600 km s$^{-1}$) is redshifted with respect to the narrow component ($\Delta v = 1400$ km s$^{-1}$), while the broad [\ion{O}{3}] component (FWHM = 2400$\pm$200 km s$^{-1}$) is blueshifted compared to the fitted wavelength of the narrow component ($\Delta v = -500$ km s$^{-1}$). Shifted broad H$\beta$ emission is sometimes observed in double-peaked emitters with asymmetric line profiles \citep{eracleous94,eracleous2003,barrows}. This feature is typically explained by asymmetries in a Keplarian accretion disk. We note that similar signatures, i.e., high velocity shifts in the broad H$\beta$ line, can also be produced by supermassive black hole binaries \citep{eracleous2012,runnoe2015} and rapidly recoiling black holes \citep{bonning,komossa}.

An apparent blue wing to the [\ion{O}{3}] doublet is present in S82X 0022+0020, which we are able to fit with broad Gaussian components in addition to narrow Gaussians to fit the narrow line doublet (Figure \ref{0022_spec}). The broad component of the [\ion{O}{3}] line has a FWHM of 1200$\pm$200 km s$^{-1}$, and is blueshifted with respect to the narrow component (FWHM = 560$\pm$20 km s$^{-1}$) by $\Delta v = -700$ km s$^{-1}$.

The Palomar spectrum of S82X 0242+0005 shows a blue wing to the [\ion{O}{3}] doublet. As shown in Figure \ref{oiii_fits},  additional Gaussian components, with FWHM = $2300\pm200$ km s$^{-1}$, accommodates this additional emission. It is blueshifted by $\Delta v = -400$ km s$^{-1}$ compared with the narrow component of the line.

We note that these [\ion{O}{3}] FHWM values and velocities are on the order of those observed in {\it XMM}-COSMOS reddened quasars \citep{brusa2015b}, but less extreme than the SDSS-selected luminous reddened ($r_{\rm AB} - W4_{\rm Vega} > 14$) quasars \citep{ross} presented in \citet{zakamska}.

\begin{deluxetable}{llll}
\tablewidth{0pt}
\tablecaption{\label{outflow_fit}Asymmetric [\ion{O}{3}] Line Profiles}
\tablehead{\colhead{Stripe 82X Name} & \colhead{FWHM} & \colhead{$\Delta v$\tablenotemark{1}} & \colhead{Spectrum} \\
 & \colhead{(km s$^{-1}$)} & \colhead{(km s$^{-1}$)} }

\startdata

S82X 0022+0020 &  $1200\pm200$ &  $-700$   & SDSS \\

S82X 0040+0058 & $2400\pm200$ & $-500$ & SDSS \\

S82X 0242+0005 & $2300\pm200$ & $-400$ & Palomar TSpec 

\enddata
\tablenotetext{1}{$\Delta v$ is measured between the fitted wavelengths of the broad and narrow components of the [\ion{O}{3}] 5007 \AA\ line.}
\end{deluxetable}

\section{Discussion}\label{disc}

\subsection{AGN Properties Derived from Spectral Analysis and SED Fitting}
Two of the sources from the NIR faint optical drop-out sample, S82X 0141-0017 and S82X 0227+0042, have estimated bolometric luminosities on the order of, or lower than, the observed full-band X-ray luminosity. This apparent inconsistency points to limitations in the SED decomposition due to the relatively few photometric detections for these sources. We therefore refrain from estimating their black hole masses and Eddington parameters, and note that their E(B-V)$_{\rm AGN}$ and E(B-V)$_{\rm Galaxy}$ values may also be unreliable. We therefore discard these objects when considering the AGN properties derived from SED fitting below.

Based on the fitted E(B-V) values from the SED decomposition, nine out of the remaining ten sources are ``reddened'' Type 1 AGNs, with E(B-V) $\geq$ 0.45. We note that the blue source is S82X 0011+0057, which we showed to be radio loud (Section \ref{radio}). The SDSS spectrum for this source (Figure \ref{sdss_supp}) also shows a blue powerlaw slope, consistent with an unobscured quasar. Thus we conclude that the red $R-K$ color for S82X 0011+0057 is due to synchrotron emission boosting the $K$-band flux and the low E(B-V) value is to be expected. We point out, however, that S82X 0302-0003, which is radio-intermediate appears to be truly reddened based on the E(B-V) values derived from SED decomposition. For the reddened sources, the extinction is along the line-of-sight to the AGN.

The black hole masses and Eddington ratios span a range of values, with similar $M_{\rm BH}$ - $\lambda_{\rm Edd}$ relationships as unobscured quasars from SDSS \citep{trakhtenbrot}. However, we reiterate that the bolometric and monochromatic luminosities from which we derive these values are approximate. Furthermore, obscuration in the broad line region can skew the emission line FWHM which results in systematically lower $M_{\rm BH}$ estimates compared with unobscured quasars, such that a comparison between both populations is not straightforward.

\subsection{Bright NIR Reddened AGN Are Less Numerous and More Luminous Than Blue Type 1 AGNs}
\citet{glikman2007,glikman2012,glikman2013} analyzed properties of radio-selected reddened quasars, finding that their observed surface space density was $\sim$17-21\% lower than a matched sample of radio-selected blue quasars. They also reported that after correcting the $K$-band magnitude for reddening, these red quasars were more luminous than their unobscured counterparts. Similarly, \citet{banerji2015} and \citet{assef2015} found that reddened quasars selected on the basis of red near-infrared colors ($J-K > 2.5$, $K < 16.5$, Vega) and red {\it WISE} colors, respectively, have higher bolometric luminosities than blue Type 1 AGNs culled from SDSS. 

Since our bright NIR $R-K$ sample is $\sim$89\% complete (only one source that fits our selection criteria lacks a spectroscopic redshift; Figure \ref{rkxo_selection}), we compare the properties of these reddened AGNs with a matched sample of X-ray selected blue ($R-K < 3$) Type 1 AGNs, also drawn from Stripe 82X. This comparison sample obeys the same infrared and optical magnitude cuts as the $R-K$ versus $X/O$ sample: $X/O > 0$, $K < 16$ (Vega). We discard all sources spectroscopically identified as stars or galaxies (i.e., they lack broad lines in their optical spectra). For sources that lacks a redshift, we discarded objects that lie along the $R-K$ versus $R-W1$ stellar locus \citep[Figure 6][]{rw1}, so that the comparison sample is made up of likely extragalactic sources. There are 62 such blue sources for comparison, 56 of which have spectra and are confirmed Type 1 AGNs.

\begin{figure}[ht]
  \centering
  \includegraphics[scale=0.35,angle=90]{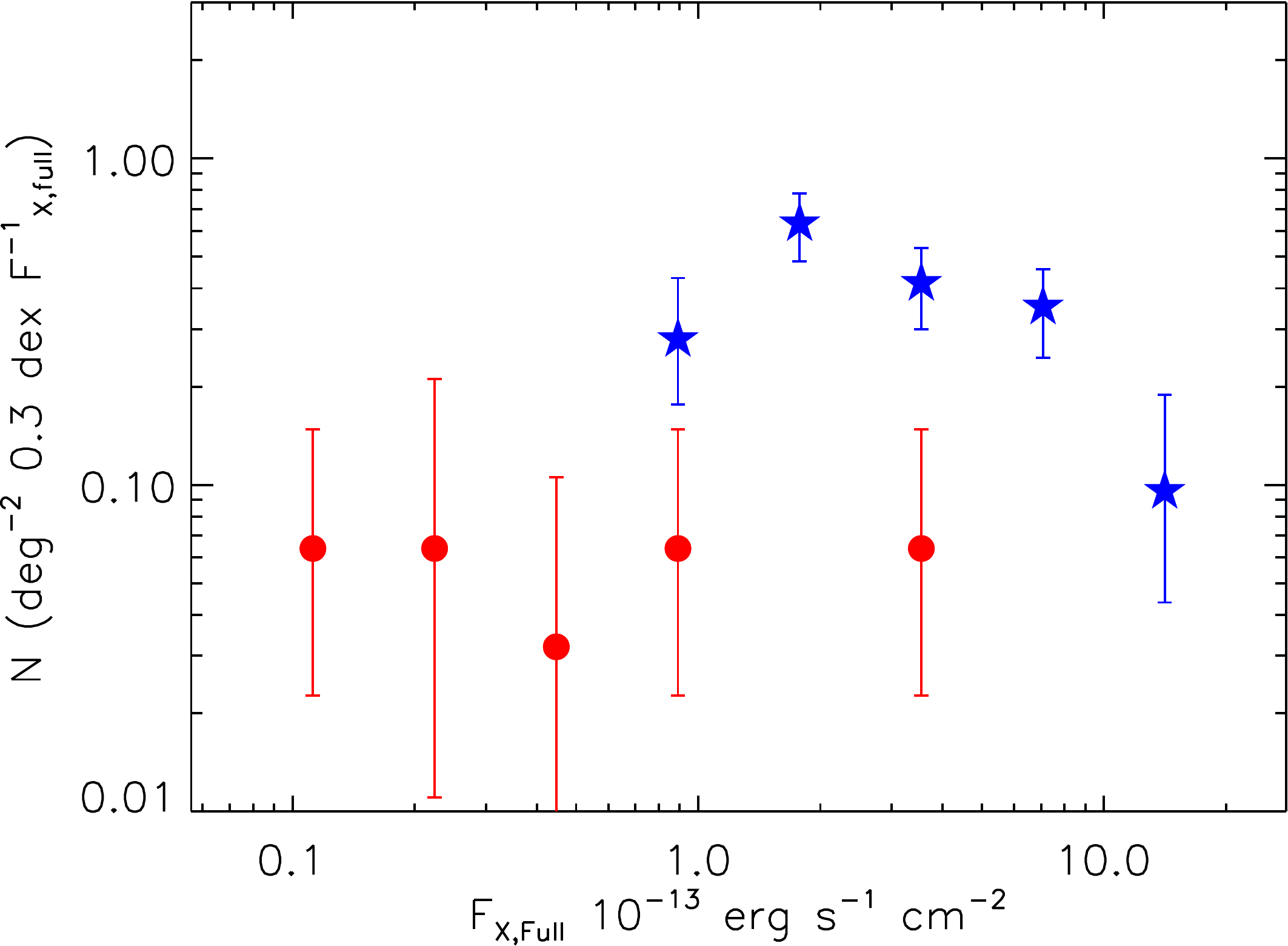}
\caption{\label{space_dens} Observed surface space density in bins of 0.3 dex of full-band X-ray flux ($F_{\rm X,full}$) of reddened AGNs from the nearly complete bright NIR $R-K$ versus $X/O$ selected sample (red circles) compared with a matched sample of X-ray selected blue ($R - K < 3$) Type 1 AGNs (blue stars). While the space density of the reddened AGN show is relatively constant with X-ray flux, no blue Type 1 AGNs are found at fluxes under 10$^{-13}$ erg s$^{-1}$ cm$^{-2}$. }
\end{figure}

First, we calculated the observed surface density for both the reddened and comparison blue AGN samples in X-ray flux bins with a width of 0.3 dex. To account for spectroscopic incompleteness, we multiplied the observed space density (i.e., $N$/31.3 deg$^2$) within each bin by the fraction of sources spectroscopically identified in that bin. Figure \ref{space_dens} shows the observed space density for the reddened and blue AGN, where the errors are Poissonian ($\sqrt{N}$) (if there are 10 or more sources in the bin) or are derived from \citet{gehrels}. From this exercise, we find that blue Type 1 AGNs have a higher space density than the reddened AGNs, and that they have higher X-ray fluxes than the reddened population. There are no blue Type 1 AGNs at X-ray fluxes below 10$^{-13}$ erg s$^{-1}$ cm$^{-2}$, while the reddened AGNs have a roughly constant space density ($\sim$0.06 deg$^{-2}$) as a function of observed X-ray flux.

\begin{figure*}[ht]
  \centering
  \includegraphics[scale=0.35,angle=90]{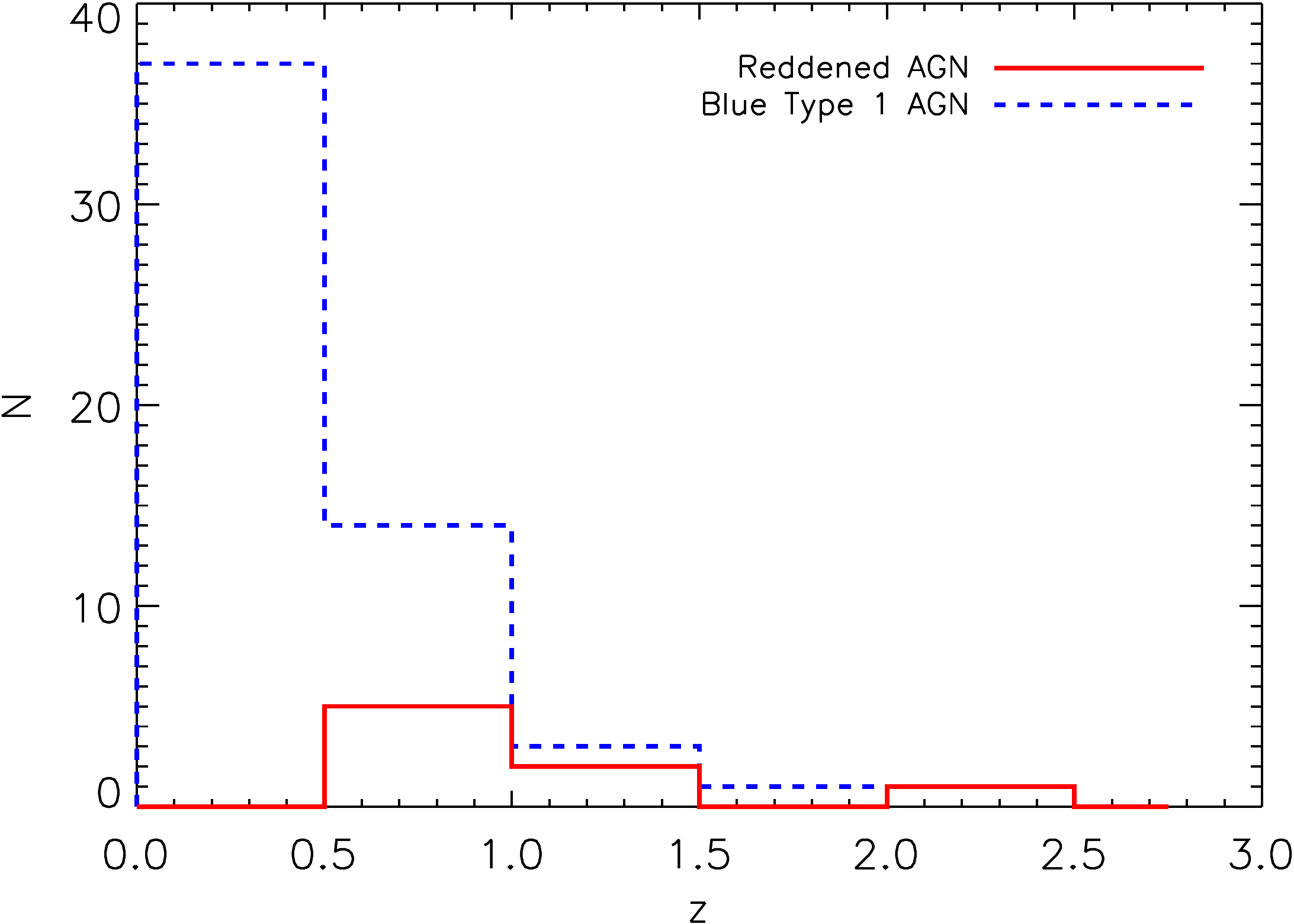}~
  \includegraphics[scale=0.35,angle=90]{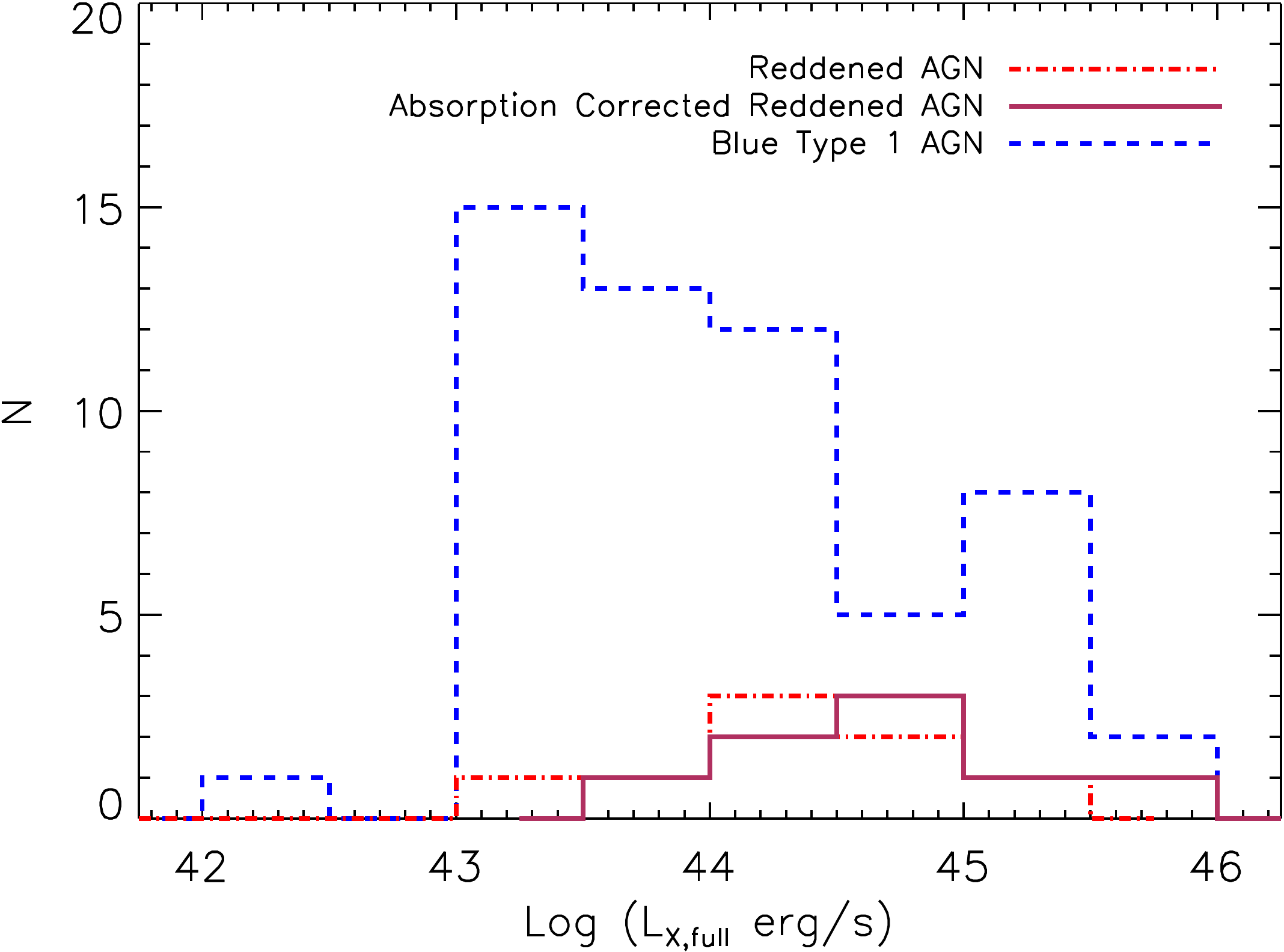}
\caption{\label{z_lum_hist} {\it Left}: Redshift distribution of our bright NIR $R-K$ versus $X/O$ selected reddened AGNs compared with blue ($R-K < 3$, Vega) Type 1  X-ray selected AGNs from the Stripe 82X survey at similar magnitude limits (i.e., $K < 16$, Vega). The blue AGNs are predominantly at lower redshift ($z < 0.5$) compared with the reddened AGNs. {\it Right}: Luminosity distribution for the reddened and blue AGNs samples, where the {\it estimated} intrinsic luminosities, as implied by the hardness ratios, for the reddened AGNs are shown for reference. Though the average luminosities are similar between the reddened and blue populations, a higher fraction of reddened AGNs have X-ray luminosities exceeding $10^{44}$ erg s$^{-1}$ than the blue AGNs. }
\end{figure*}

The blue AGNs extend to brighter X-ray fluxes because they are predominantly nearby compared with the reddened population. As shown in Figure \ref{z_lum_hist} (left), most of the blue AGNs (66\%) reside at $z < 0.5$ while all the reddened AGNs are more distant. Furthermore, the X-ray luminosities of the reddened population are drawn from the higher end of that observed in the blue AGN population, as illustrated in Figure \ref{z_lum_hist} (right), where we also show the {\it estimated} instrinsic X-ray luminosity for the reddened AGNs for reference. The mean X-ray luminosities of both the reddened (log($L_{\rm X,full}$/erg s$^{-1}$) = 44.7 $\pm$0.4 (observed); 45.0$\pm$0.4 (intrinsic)) and blue (log($L_{\rm X,full}$/erg s$^{-1}$) = 44.8 $\pm$0.5) AGNs are consistent. However, under half of the blue AGNs have observed X-ray luminosities exceeding 10$^{44}$ erg s$^{-1}$ while 67\% of the reddened AGNs are at these high X-ray luminosites.

Focusing on reddened AGNs in a flux-limited X-ray sample favors detection of AGNs that are more distant, and more luminous, than their unreddened counterparts. This bias is induced by the red $R-K$ criterion and the $K$-band flux limit: sources that are more reddened are those where the AGN dominates over the host galaxy, which are preferentially high-luminosity AGN since lower luminosity AGN would fall below the $K$-band flux limit. Furthermore, wide-area coverage is required to identify this luminous population at a relatively bright near-infrared flux limit. We find no X-ray AGNs (i.e., $L_{\rm X,full} > 10^{42}$ erg s$^{-1}$) from the smaller 2.2 deg$^2$ {\it Chandra} COSMOS Legacy survey \citep{civano,marchesi} with the same colors ($R-K > 4$, Vega; $X/O > 0$) at the same magnitude limit (i.e., $K <$ 16, Vega).\footnote{To match the magnitude system used in this study, we converted the COSMOS $r$-band magnitude from Subaru SuprimeCam reported in the PSF-homogeneized photometric catalog of \citet{laigle} to the SDSS $r$-band filter using the formula in \citet{capak}. We then transformed to the Bessel $R$ bandpass and converted to the Vega magnitude system using Eqs. 1 \& 2 above.}

\subsection{Stripe 82X Reddened AGNs Compared with Those Previously Known}
We compare our Stripe 82X reddened quasars with samples from the literature selected based on radio emission and red optical-infrared colors \citep[$R-K > 4$, $J -K > 1.7$, Vega;]{glikman2007,glikman2012,glikman2013}, near-infrared colors \citep[$J-K > 2.5$, $K < 16.5$, Vega;][]{banerji2012,banerji2013,banerji2015}, mid-infrared colors \citep[$W1W2$ drop-outs;][]{eisenhardt,assef2015}, and reddened ($R - K > 4-4.5$, Vega) X-ray selected AGNs presented in \citet{bongiorno}.

Though these previous samples of reddened AGNs have been selected via independent methods, there are several traits that many of these sources have in common: the extinction ranges from moderate \citep[i.e., 0.1 $< E$($B-V$)$<$1.55;][]{glikman2007,glikman2012,glikman2013,banerji2012,banerji2013,banerji2015} to extreme \citep[i.e., 2.5 $< E$($B-V$)$<$ 21.5;][]{assef2015}, the black holes are massive \citep[$M_{\rm BH} > 10^9$ M$_{\sun}$;][]{banerji2015,bongiorno,wu2017}, the AGNs are generally distributed beyond $z > 1$ \citep{bongiorno} and $z > 2$ \citep{banerji2012,banerji2013,banerji2015,assef2015}, and they tend to be more luminous than blue Type 1 quasars at comparable redshifts \citep{glikman2007,glikman2012,glikman2013,banerji2015,assef2015}. Most of the AGNs have broad H$\alpha$ emission, and are not narrow-line only Type 2 AGNs, with the exception of the $W1W2$ drop-outs detected by {\it WISE} which are a mixture of Type 1 and Type 2 AGNs \citep{eisenhardt,assef2015}.

\citet{glikman2012} presented a sample of 120 reddened quasar candidates selected from the FIRST radio and 2MASS near-infrared surveys that have red colors. An analogous sample of radio-selected quasar candidates was presented in \citet{glikman2013}, with similar color selection, but pushed down to lower near-infrared flux limits, using sources detected in the deeper UKIDSS survey. We find that our Stripe 82X sources span similar redshift ranges ($0.6 < z < 2.5$) as those in \citet{glikman2007,glikman2012,glikman2013}, $0.13 < z < 3.1$. We also obtain similar AGN reddening values, where ours have a range of $0.45 < E$($B-V$) $< 1.18$ (after excluding the radio loud AGN and the two optical dropout AGN with inconsistent bolometric and X-ray luminosities) compared with $0.1 < E$($B-V$)$< 1.55$; our $E$($B-V$) values are derived from SED fitting while those from \citet{glikman2007,glikman2012,glikman2013} are measured from fitting a reddened quasar template to the optical and/or near-infrared spectra. Only two of our Stripe 82X sources are detected in the radio by FIRST, indicating that the orthogonal axis of X-ray selection aids in recovering reddened AGNs not detected by radio surveys.

Similar to \citet{banerji2015}, we estimated $M_{\rm BH}$ using the FWHM of the broad H$\alpha$ emission line, where the continuum and bolometric luminosities were calculated via SED fitting. Both samples are subjected to similar broad line region obscuration biases that could potentially affect emission line FWHMs. Compared with the 38 $z > 2$ red quasars presented in \citet{banerji2012,banerji2013,banerji2015}, where log($M_{\rm BH}/M_{\sun}$)=9.7$\pm$0.46 and log($L_{\rm bol}$/erg s$^{-1}$) = 47.1$\pm$0.4 \citep{banerji2015}, our X-ray selected reddened AGNs have lower black hole masses (log($M_{\rm BH}/M_{\sun}$) = 9.0$\pm$ 0.8) and bolometric luminosities (log($L_{\rm bol}$/erg s$^{-1}$)= 46.5$\pm$0.8), though there is a wide spread on these values for the Stripe 82X AGN. Additionally, as the faint NIR optical dropout sample is only $\sim$12\% complete, and we are unable to derive estimates of $M_{\rm BH}$ for half of the sample we have observed, more observations are needed to test whether the pilot sample observed thus far is representative of the parent sample. The measured AGN reddening in the \citet{banerji2015} sample ($0.5 < E(B-V) < 1.5$) spans a similar range to the values calculated in our Stripe 82X sample.

The most luminous, reddened quasars yet identified were selected based on their mid-infrared colors in {\it WISE}: these $W1W2$ dropouts are weak or undetected in {\it WISE} bands $W1$ and $W2$ but are bright in bands $W3$ and $W4$ \citep{eisenhardt}. \citet{assef2015} analyzed the SEDs of 52 $W1W2$ drop-outs at $z > 1$ and $W4 < 7.2$ (Vega) that have {\it Spitzer} IRAC data. The reddening in these objects are much more extreme (i.e., $\langle E$($B-V$)$\rangle$ = 6.8) than what we observe in the Stripe 82X reddened AGNs presented here and seen in other reddened AGN samples. The typical bolometric luminosities of the $W1W2$ drop-outs, 10$^{47}$ - 10$^{48}$ erg s$^{-1}$, are also much higher than the Stripe 82X sample. Three of these sources have been followed up with X-ray observations, with {\it XMM-Newton} and {\it NuSTAR}, and were found to be X-ray faint, consistent with Compton-thick levels of obscuration \citep{stern2014}. With our X-ray-optical-infrared selection of reddened quasars, we appear to be selecting an AGN population that is less extreme than the {\it WISE} $W1W2$ drop-outs, which at a space density of $\sim$1/30 deg$^2$, are more rare than reddened AGN selected via other diagnostics.

Finally, we compare our Stripe 82X reddened AGNs with the 21 reddened ($R - K > 4.5$, Vega) AGNs from \citet{bongiorno} that were selected from small-to-moderate area X-ray surveys: the original {\it Chandra} Deep Field South \citep[0.109 deg$^2$;][]{giacconi}, {\it XMM}-COSMOS \citep[2 deg$^2$;][]{hasinger,cappelluti,brusa}, and the literature \citep{alexander,sarria,melbourne,delmoro}. In this sample, the X-ray emission was mildly absorbed ($N_{\rm H} > 10^{21} - 10^{22}$ cm$^{-2}$, as implied by X-ray spectral analysis or hardness ratios), similar to the implied obscuration of our bright $R-K$ sample. Our faint NIR optical dropout sample, which largely has hardness ratios consistent with no X-ray absorption, spans a similar redshift range as the \citet{bongiorno} sample ($1.2 < z < 2.6$). The \citet{bongiorno} sample has a similar average black hole mass (log($M_{\rm BH}/M_{\sun}$)=9.3$\pm$0.5) as the Stripe 82X reddened AGNs. They used the FWHM of the H$\alpha$ line in conjunction with the intrinsic hard X-ray (2-10 keV) luminosity as a proxy for the AGN continuum luminosity \citep[using the relationship between $\lambda L_{\rm 5100}$ and hard X-ray luminosity found in][]{maiolino} to estimate $M_{\rm BH}$:
\begin{equation}
  \begin{split} 
  M_{\rm BH} = 10^{7.11} \left [\frac{{\rm FWHM (H\alpha)}}{1000\ {\rm km\ s^{-1}}}\right]^{2.06}
  \times \\
    \left [\frac{\lambda L_{\rm 2-10keV, intrinsic}}{10^{44}\ {\rm erg\ s^{-1}}} \right ]^{0.693} {\rm M_{\sun}}.
  \end{split}
\end{equation}
     \citet{bongiorno} do not provide errors on the parameters in this virial relation, but note that there is a scatter of about 0.1 dex in the normalization.

     For reference, in Table \ref{bhmass_xray} we list the black hole masses we obtain using this scaling relation for the four sources that have hard band X-ray detections and H$\alpha$ coverage. We propagate the errors on the H$\alpha$ FWHM and $L_{\rm 2-10keV, intrinsic}$ values and note that there is likely an additional $\sim$0.1 dex uncertainty in $M_{\rm BH}$ that is associated with this virial relation as well as a general $\sim$0.5 dex uncertainty that is found for single-epoch measurements, as discussed above \citep{vestergaard}. We obtain similar black hole masses compared with what we calculated using $\lambda L_{\rm 51000}$ as the continuum luminosity, though the intrinsic hard X-ray luminosities are based on column densities derived from hardness ratios which are a very crude measure of absorption.

\begin{deluxetable}{lll}
\tablewidth{0pt}
\tablecaption{\label{bhmass_xray}Black Hole Masses Estimated from $L_{\rm 2-10keV, intrinsic}$ and H$\alpha$ FWHM}
\tablehead{\colhead{Stripe 82X Name} & \colhead{Log ($L_{\rm 2-10keV, intrinsic}$)} & \colhead{$M_{\rm BH}$} \\
 & \colhead{(erg s$^{-1}$ cm$^{-2}$)} & \colhead{(M$_{\Sun}$)} }

\startdata

S82X 0242+0005   & 45.05$^{+0.27}_{-0.31}$ & 9.26$^{+0.17}_{-0.29}$  \\

S82X 0302$-$0003 & 44.84$^{+0.14}_{-0.08}$ & 9.07$^{+0.08}_{-0.09}$  \\

S82X 0303$-$0115 & 43.77$^{+0.11}_{-0.10}$ & 7.29$^{+0.07}_{-0.09}$  \\

S82X 0118+0018\tablenotemark{1}   & 44.12$^{+0.14}_{-0.00}$ & 7.98$^{+0.16}_{-0.26}$ 

\enddata
\tablenotetext{1}{Since we only have an upper limit on the estimated $N_{\rm H}$, the lower limit on the intrinsic luminosity is the observed luminosity.}
\end{deluxetable}

\section{Conclusions}
We presented the results of a ground-based, near-infrared spectroscopic campaign to follow up reddened AGN candidates in the wide-area (31 deg$^2$) Stripe 82 X-ray survey \citep{s82x1,s82x2,s82x3}. Our bright NIR sample selected on the basis of red $R-K$ colors ($>$4, Vega) and $X/O > 0$ \citep[c.f.,][]{brusa} consists of 9 sources, four of which had existing spectroscopy in SDSS and five of which we targeted with Palomar TripleSpec (Figure \ref{rkxo_selection}); four of the targeted sources were identified via spectroscopic redshifts (Figure \ref{tspec_spec}). This sample is 89\% complete to a magnitude limit of $K=16$ (Vega).

We also presented a pilot program to follow-up sources that are not detected in the single-epoch SDSS imaging, yet have {\it WISE} colors consistent with quasars \citep[Figure \ref{wiseqso_selection};][]{wright}. The spectra of these four sources were obtained with Keck NIRSPEC and Gemini GNIRS since 8-10m class telescopes are required to spectroscopically identify sources at these faint NIR magnitudes (i.e., $K > 17$, Vega; Figure \ref{keck_gem_spec}).

All sources have at least one permitted emission line with FWHM exceeding 1300 km s$^{-1}$ in their optical or infrared spectra, and can thus be classified as Type 1 AGNs \citep{hao2005, glikman2007}. The bright NIR sample spans a range of redshifts, $0.59 < z < 2.5$, while faint optical dropout AGNs all lie beyond a redshift of 1.

We used AGNFitter \citep{calistro} to fit the SEDs and decompose AGN and galaxy emission (Figures \ref{seds_bright} - \ref{seds_suppl}), obtaining estimates of the reddening and AGN bolometric luminosity for each source. Two sources from the optical dropout sample, S82X 0141-0017 and S82X 0027+0042, have estimated $L_{\rm bol}$ values on the order of or less than the observed X-ray luminosity, suggesting limitations in the SED decomposition for these sources. All but one of the remaining AGNs are reddened, with 0.45 $<$ E(B-V)$_{\rm AGN}$ $<$ 1.18. The blue source is radio loud, such that its red $R-K$ color is likely due to jet-dominated synchrotron emission.

Half the sources in the bright NIR sample have features in their optical spectra indicative of outflows (Figures \ref{oiii_fits} left, \ref{sdss_supp} bottom, \ref{0040_spec} bottom right, and \ref{0022_spec} bottom). Since many quasars host outflows \citep[e.g.,][]{ganguly}, follow-up high-resolution imaging of the host galaxies would be necessary to search for morphological signatures of mergers to test whether these features signify feedback predicted by the major merger AGN evolution paradigm \citep{sanders,hopkins}.

Because the bright NIR sample is nearly complete, we compared the characteristics of these AGNs with blue ($R-K < 3$, Vega) Type 1 AGNs selected from the Stripe 82X survey. While the blue Type 1 AGNs have systematically higher X-ray fluxes (Figure \ref{space_dens}), they are predominantly at low redshift ($z<0.5$), with a greater percentage at lower X-ray luminosities compared with the reddened AGNs (Figure \ref{z_lum_hist}). Hence, focusing on reddened populations in shallow X-ray surveys, like Stripe 82X, for follow-up will likely unveil more distant and more luminous AGNs than blue AGN at similar X-ray, optical, and infrared flux limits.

Compared with reddened AGNs selected on the basis of their radio, optical, near-infrared, and/or mid-infrared emission, the Stripe 82X red quasars have similar reddening \citep[though less extreme than $W1W2$ drop-outs which are extreme sources that have the highest levels of extinction compared to all samples of reddened quasars;][]{assef2015}, and a range of estimated black hole masses and Eddington parameters. 

Our pilot sample of {\it WISE}-selected optical dropouts is only $\sim$12\% complete, precluding us from drawing any firm conclusions about this population as a whole. Our program does demonstrate proof-of-concept for using this selection technique to recover reddened quasars at $z > 1$ that are missed by optical surveys like SDSS.

We highlight that Stripe 82X complements other X-ray surveys by discovering reddened AGNs at relatively bright NIR magnitudes (i.e., $K < 16$, Vega) that are missed entirely by smaller-area X-ray surveys, like the 2.2 deg$^2$ {\it Chandra} COSMOS Legacy \citep{civano,marchesi}. As these sources are rare, potentially due to the reddened stage being a short-lived AGN evolutionary phase, wide-area X-ray surveys like Stripe 82X, {\it XMM}-XXL \citep{pierre}, XBo\"otes \citep{kenter,murray}, the upcoming {\it eROSITA} mission \citep{merloni}, and serendipitous surveys/catalogs like ChaMP \citep{kim}, the {\it Chandra} Source Catalog \citep{evans}, and the {\it XMM} Serendipitous catalog \citep{rosen}, are necessary to reveal this missing tier of luminous, obscured black hole growth at the brightest fluxes.

\acknowledgements We thank the anonymous referee for a careful reading of this manuscript and providing helpful comments. Most of this work was completed while S.M.L was supported by an appointment to the NASA Postdoctoral Program at the NASA Goddard Space Flight Center, administered by Universities Space Research Association under contract with NASA. S.M.L. thanks A.-N. Chene for support when running the Gemini GNIRS reduction pipeline and G. Calistro Rivera for guidance in using AGFNitter. Palomar and Keck observations were obtained through guaranteed Yale time on these facilities. E.G acknowledges the generous support of the Cottrell College Award through the Research Corporation for Science Advancement. The work of D.S. was carried out at the Jet Propulsion Laboratory, California Institute of Technology, under a contract with NASA.

Some of the data presented herein were obtained at the W. M. Keck Observatory, which is operated as a scientific partnership among the California Institute of Technology, the University of California and the National Aeronautics and Space Administration. The Observatory was made possible by the generous financial support of the W. M. Keck Foundation. The authors wish to recognize and acknowledge the very significant cultural role and reverence that the summit of Maunakea has always had within the indigenous Hawaiian community.  We are most fortunate to have the opportunity to conduct observations from this mountain.

\facility{XMM}, \facility{CXC}, \facility{Sloan}, \facility{Hale (TSPEC)}, \facility{Gemini:Gillet (GNIRS)}, \facility{Keck:II (NIRSPEC)}

\end{document}